\documentclass[sn-apa]{sn-jnl}

\usepackage{natbib}

\usepackage{graphicx}%
\usepackage{multirow}%
\usepackage{amsmath,amssymb,amsfonts}%
\usepackage{amsthm}%
\usepackage{mathrsfs}%
\usepackage[title]{appendix}%
\usepackage{xcolor}%
\usepackage{textcomp}%
\usepackage{manyfoot}%
\usepackage{booktabs}%
\usepackage{algorithm}%
\usepackage{algorithmicx}%
\usepackage{algpseudocode}%
\usepackage{listings}%
\usepackage{dsfont}
\newcommand{\bm}[1]{\mbox{\boldmath{$#1$}}}
\newcommand{\1}{\mathds{1}}

\usepackage{booktabs,subcaption,dcolumn}

\theoremstyle{thmstyleone}%
\newtheorem{theorem}{Theorem}[section]
\newtheorem{proposition}[theorem]{Proposition}%

\numberwithin{equation}{section}

\begin{document}

\title[B-Spline Copula with Penalized Pseudo-Likelihood]{EM Estimation of the B-Spline Copula with Penalized Pseudo-Likelihood Functions}

\author*[1,2]{\fnm{Xiaoling} \sur{Dou}}\email{xiaolingd@fc.jwu.ac.jp}

\author[2]{\fnm{Satoshi} \sur{Kuriki}}\email{kuriki@ism.ac.jp}
\equalcont{These authors contributed equally to this work.}

\author[3]{\fnm{Gwo Dong} \sur{Lin}}\email{gdlin@stat.sinica.edu.tw}
\equalcont{These authors contributed equally to this work.}

\author[4]{\fnm{Donald} \sur{Richards}}\email{richards@stat.psu.edu}
\equalcont{These authors contributed equally to this work.}

\affil*[1]{\orgdiv{Faculty of Science}, \orgname{Japan Women's University}, \orgaddress{\street{2-8-1}, \city{Bunkyo-ku}, \postcode{112-8681}, \state{Tokyo}, \country{Japan}}}

\affil[2]{
\orgname{The Institute of Statistical Mathematics}, \orgaddress{\street{10-3 Midoricho}, \city{Tachikawa}, \postcode{190-8562}, \state{Tokyo}, \country{Japan}}}

\affil[3]{\orgdiv{Institute of Statistical Science}, \orgname{Academia Sinica}, \orgaddress{\street{128, Section 2, Academia Rd}, \city{Taipei}, \postcode{11529}, \state{Taiwan}, \country{R.O.C}}}

\affil[4]{\orgdiv{Department of Statistics}, \orgname{Pennsylvania State University}, \city{University Park}, \state{Pennsylvania 16802}, \country{U.S.A}}

\abstract{The B-spline copula function is defined by a linear combination of elements of the normalized B-spline basis.  We develop a modified EM algorithm, to maximize the penalized pseudo-likelihood function, wherein we use the smoothly clipped absolute deviation (SCAD) penalty function for the penalization term.  We conduct simulation studies to demonstrate the stability of the proposed numerical procedure, show that penalization yields estimates with smaller mean-square errors when the true parameter matrix is sparse, and provide methods for determining tuning parameters and for model selection.  We analyze as an example a data set consisting of birth and death rates from 237 countries, available at the website, ``Our World in Data,'' and we estimate the marginal density and distribution functions of those rates together with all parameters of our B-spline copula model.}




\maketitle

\section{Introduction}\label{sec1}

A copula is a multivariate probability distribution function for which each univariate marginal distribution is the uniform distribution on the interval $[0,1]$ \citep{Nelsen06, Sklar59}. 
Copulas are widely used to describe the dependence structure of a collection of jointly distributed random variables, and in estimating a multivariate distribution we infer the copula function and the marginal distributions separately.

To date, numerous copulas have been developed. These include the well-known Gaussian, Frank, Clayton, Gumbel-Hougaard, and many other parametric copulas, each of which enjoys distinctive properties and uses.

\cite{Sancetta-Satchell04} defined the Bernstein copula, a notable nonparametric copula based on the Bernstein polynomials. 
It is known that, with uniform marginals on the unit interval $[0,1]$, multivariate distributions constructed with order statistics are Bernstein copulas \citep{Baker08, Dou-etal13}. When the degrees of the Bernstein polynomials are equal to the sample size, the Bernstein copula becomes the empirical beta copula \citep{Segers-etal16} which is constructed with respect to the ranks of the data. \cite{Kojadinovic-Yi22} also introduced recently a rank-based approach to constructing a smooth class of empirical beta copulas.  

We comment in particular on the B-spline copula, a focus of the present article and a generalization of the Bernstein copula.  The B-spline copula, introduced by \cite{Shen-etal08}, comprises a copula constructed from linear B-spline functions.  \cite{Dou-etal21} subsequently introduced a B-spline copula that generalized the linear B-spline copula by allowing the degree of the B-spline basis functions to be any positive integer. 
As a method of estimating the B-spline copula has not been developed, one main objective of the present article is to construct an EM algorithm with penalized log-likelihood function to estimate the B-spline copula.

To date, many interesting ideas for estimating copulas have been proposed.  \cite{CaiWang14} 
applied a penalized likelihood method to select a mixed copula model from a large number of candidate copulas, thereby capturing the dependence structure.  
Cai and Wang estimated the weights of the candidate copulas by applying a \textit{smoothly clipped absolute deviation} (SCAD) penalty function to the likelihood function, discarding copulas with small weights, and then treating the remaining weights as the component elements of the mixed copula; in this approach, an EM algorithm is used to estimate the component weights and the parameters of the copulas, and the tuning parameters in the SCAD penalty function are selected by cross-validation.
In related work, \cite{Kauermann-etal13} developed a hierarchical linear B-spline method, with an $L^2$ penalty function, that used general optimization routines for parameter estimation.
\cite{KauermannSchellhase14} also used penalized linear B-spline functions to estimate copulas, their penalty function is a difference matrix of certain coefficients for smoothness of the estimation procedure, and they applied a quadratic programming algorithm to estimate pairwise copulas for multivariate distributions.

In this paper, we focus on the optimization aspects of estimating copulas.  In a previous paper \citep{Dou-etal16}, we developed an EM algorithm approach to estimating the Bernstein copula, a special case of the B-spline copula.  Here, we extend our earlier EM algorithm for the B-spline copula by attaching a penalty term, the new EM algorithm to be developed being in the sense of \cite{Green90}, and the penalty function to be employed being the SCAD penalty.  

The contents of the article are organized as follows.  Section \ref{sec_prelim} provides a review of the B-spline copula.  In Section \ref{sec_em_algorithm}, we propose the new EM algorithm for the penalized pseudo-log-likelihood to estimate B-spline copula and we establish its convergence properties; further, we provide methods for determining the tuning parameters in the penalty function and for choosing the size of the parameter matrix.  
In Section \ref{sec_simulation}, we conduct simulation studies to illustrate the stability of the proposed numerical procedure and demonstrate that penalization yields estimates with smaller mean-square errors when the true parameter matrix is sparse.  
We also assess the difference between the estimated joint densities obtained using the Bernstein and the B-spline copulas for a simulated $3$-dimensional data set.  
In Section \ref{sec_example}, we analyze as an example a data set consisting of birth and death rates from 237 countries, available at the website, ``Our World in Data,'' and we estimate the marginal density and distribution functions of those rates together with all parameters of our B-spline copula model.
The contributions of the paper are discussed in Section \ref{sec_discussion} and, finally, the proofs of propositions, an algorithm for generating random numbers from the B-spline copula, and small-sample simulations for the pseudo-MLE are given in Appendices \ref{sec_appendix}, \ref{sec_sampling}, and \ref{sec_small_pseudo_mle}, respectively.

\section{A review of the B-spline copula}\label{sec_prelim}

\cite{Dou-etal21} constructed the B-spline copula with B-spline basis functions, as follows.  
For simplicity, we consider the bivariate case with random variables $X$ and $Y$.
Let $d$ be the degree of the B-spline basis functions \citep{deBoor72, deBoor01}.  For a positive integer $p$ and a set of interior knots $t_i \in [0,1]$, $i=1, \ldots, p-1$, define 
\begin{equation}
\label{knots}
\underbrace{t_{-d} = \cdots = t_{-1}}_d = t_0 = 0 \le t_1 \le \cdots
\le t_{p-1} \le 1 = t_p = \underbrace{t_{p+1} = \cdots = t_{p+d}}_d.
\end{equation}
Given the $m=p+d$ B-spline basis
functions $\{N^d_{k-d-1}:k=1,2,\ldots,m\},$ we define the
quantities $q_{k,m}$ and the functions $\phi_{k,m}$ and $\Phi_{k,m}$ for $X$ by 
\begin{equation}
\label{q_phi}
q_{k,m} = \int_0^1 N^d_{k-d-1}(x) \, dx, \quad  \phi_{k,m}(x) = \frac{1}{q_{k,m}}N^d_{k-d-1}(x),
\end{equation}
and 
\begin{equation}
\label{Phi}
\Phi_{k,m}(x) = \int_0^x\phi_{k,m}(u)du, \quad x \in[0,1], \quad k=1,2,\ldots,m.
\end{equation}
Analogously, for a new pair $d^*$ and $p^*$, new interior knots $t^*_i \in [0,1]$, $i=1,\ldots,p^*-1$ defined similarly to \eqref{knots}, and $n=p^*+d^*,$ we define 
$$
(q^*_{\ell,n}, \psi_{\ell,n}, \Psi_{\ell,n}), \quad \ell=1,2,\ldots,n,
$$
for $Y$
by proceeding analogously to \eqref{q_phi} and \eqref{Phi}.  The general form of the bivariate B-spline copula is defined as
\begin{equation}
\label{C}
C(x,y;\!R) = \sum_{k=1}^m \sum_{\ell=1}^n r_{k,\ell} \Phi_{k,m}(x) \Psi_{\ell,n}(y), \quad x,y \in [0,1],
\end{equation}
where the $m \times n$ parameter matrix $R=(r_{k,\ell})_{1\le k\le m, \, 1\le\ell\le n}$ satisfies
\begin{equation}
r_{k,\ell} \ge 0, \quad 
\sum_{k=1}^mr_{k,\ell} = q^*_{\ell,n}, \quad 
\sum_{\ell=1}^nr_{k,\ell} = q_{k,m}, \quad
\sum_{\ell=1}^nq^*_{\ell,n} = 1, \quad 
\sum_{k=1}^mq_{k,m} = 1.
\label{R}
\end{equation}
Similar to \eqref{C}, for random variables $U, V \in [0,1]$, the density function of the B-spline copula can be written as
\begin{equation}
c(u,v;\!R) = \sum_{k=1}^m \sum_{\ell=1}^n r_{k,\ell} \phi_{k,m}(u) \psi_{\ell,n}(v), \quad u,v \in [0,1].
\label{c}
\end{equation}

It is worth noting that the normalization of the B-spline basis in (\ref{q_phi}) guarantees that the marginal distribution is the uniform distribution on $[0,1]$.  This follows from the observations that since, for all $x$, 
\[
 \sum_{k=1}^{m} N^d_{k-d-1}(x)=1,
\]
then we have
\[
\sum_{k=1}^{m} q_{k,m} \phi_{k,m}(x)=\sum_{k=1}^{m} q_{k,m} \frac{N^d_{k-d-1}(x)}{ q_{k,m}} = 1.
\]
We see that for all $x \in [0,1]$, the marginal distribution of the copula is 
\begin{align*}
C(x,1;\!R)  = &  \sum^m_{k=1}\sum^{n}_{\ell=1}r_{k,\ell} \Phi_{k, m}(x) \int^1_0 \psi_{\ell,n}(v) \ dv\\
=& \sum^m_{k=1}\sum^{n}_{\ell=1}r_{k,\ell} \Phi_{k, m}(x) \int^1_0 \frac{N^d_{\ell-d-1}(v)}{q^*_{\ell,n}}\ dv \\
=&  \sum^m_{k=1}\sum^{n}_{\ell=1}r_{k,\ell} \Phi_{k, m}(x) = \sum^m_{k=1}q_{k,m} \Phi_{k,m}(x)\\
= & \int^x_0 \bigg(\sum_{k=1}^{m} q_{k,m} \phi_{k,m}(u)\bigg) du = \int^x_0 1\  du = x.
\end{align*}
Similarly, $C(1, y; \!R)= y$, $y \in [0,1]$, can be confirmed. 

For the special case $d=d^*$, $p=p^*$, $t_i=t^*_i$, for all $i$, $n=m$, and 
$$
q_{k,m}=q^*_{k,n}:=q_k, \quad   k=1,2,\dots, n,
$$
the maximum correlation of the B-spline copula is attained when the parameter matrix is diagonal, i.e.,
$$
R = (r_{k,\ell})={\rm diag}(q_k)_{1\le k\le n}.
$$
Then by \eqref{c}, the copula density function becomes
\begin{equation}
c_n^+(u,v) = \sum_{k=1}^n q_k \phi_{k,n}(u) \phi_{k,n}(v), \quad u,v\in [0,1].
\label{c+n}
\end{equation}
We remark that \cite{Dou-etal21} showed that the B-spline copulas with equally-spaced interior knots are more flexible than the Bernstein copula.

When the component data sets $\{x_t, 1 \le t \le N\}$ and $\{y_t, 1 \le t \le N\}$ are highly correlated, it is convenient to consider the joint density as a mixture of two components: 
\begin{equation}
h(x,y; q, n) = (1-q) f_X(x) f_Y(y) + q c^+_n(F_X(x),F_Y(y)) f_X(x) f_Y(y),
\label{h_qn}
\end{equation}
where the first term on the right is intended to detect independence between $X$ and $Y$; the second term accounts for the situation in which $X$ and $Y$ are highly-correlated, with the special copula in \eqref{c+n}; and $q \in (0,1)$ is the mixture proportion.  \cite{Dou-etal16} have provided an EM algorithm 
for estimating $(q, n)$, and this approach can be used to estimate the joint density function given by the model (\ref{h_qn}); in that way, testing for independence between $X$ and $Y$ can also be performed.

\section{An EM algorithm for the penalized pseudo- likelihood function}\label{sec_em_algorithm}

From now on, for simplicity, we consider only the B-spline copulas with equally-spaced interior knots and we assume that the degree of the B-spline functions is fixed at $d$.  To estimate the joint density function, 
$$
h(x,y;\!R) = c(F_X(x),F_Y(y);\!R) \, f_X(x)f_Y(y),
$$
we assume that the marginal density functions $f_X(x)$ and $f_Y(y)$, and the marginal cumulative distribution functions $F_X(x)$ and $F_Y(y)$ can be estimated separately by other methods, e.g., kernel density estimation and empirical cumulative distribution function method, respectively.  We will focus on the estimation of the parameter matrix $R$ of the copula and propose an EM algorithm for estimating $R$ in \eqref{c}.
In the case of the B-spline copula semiparametric model, the algorithm developed in this article provides a pseudo-likelihood estimator that coincides with an estimator obtained by \cite{Genest-etal95} and \cite{Tsukahara05}.  

As regards alternative approaches to calculating the estimator, analytical (i.e., calculus-based) methods generally cannot handle cases in which the score equation has multiple solutions, and in the case of our paper, the score equations are far from the kind of explicit equations arising in classical problems (e.g., in regression models with Gaussian errors).  Hence, it appears to us that analytical methods may be generally infeasible for deriving the estimators.  

As regards numerical methods, such as the Newton-Raphson method, it is well-known that that method can be unstable, or can converge to saddle-points, local maxima, or to local minima; on the other hand, the EM-algorithm always converges to local maxima (see, e.g.,\cite{Herzet-etal06}). We acknowledge that the standard EM-algorithm may require a large number of iterations to attain convergence (see, e.g., \cite{LindstromBates88}), but in that case, the rate of convergence can be accelerated by a method of \cite{Louis82}.

\subsection{An EM algorithm for the penalized pseudo-likelihood function for general B-spline copulas}

Suppose that we have data $(x_t,y_t)$, $t=1,\ldots,N$, representing the observed values of a random sample from $(X,Y)$.  Following the approach of \cite{Genest-etal95}, we construct the rescaled empirical distribution functions
$$
\widehat F_X(x) = \frac{1}{N+1} \sum_{t=1}^N \1(x_t\le x), \quad x \in \mathbb{R}
$$
and 
$$
\widehat F_Y(y) = \frac{1}{N+1} \sum_{t=1}^N \1(y_t\le y), \quad y \in \mathbb{R},
$$
for $X$ and $Y$, respectively.
Define 
\begin{equation}
\label{eq_ut_vt}
 \widehat u_t = \widehat F_X(x_t), \qquad  \widehat v_t = \widehat F_Y(y_t)
\end{equation}
for $t=1,\ldots,N$.  
Since $u_t=F_X(x_t)$ and $v_t=F_Y(y_t)$ cannot be observed, we replace them with $\widehat{u}_t$ and $\widehat{v}_t$, respectively.  The pairs 
$(\widehat u_t,  \widehat v_t)$, $t=1,\ldots,N$, were first referred to by \cite{GhoudiRemillard04} as \textit{pseudo-observations from the copula} $C(u, v)$.  Also see \cite{Hofert-etal19}, Section 4.1.2.
Using these pseudo-observations, 
we now present an algorithm for estimating the copula density function (\ref{c}).  

To start the algorithm, we propose an initial value for $R=(r_{k,\ell})$ as
\begin{equation}
\label{r-tilde}
\widetilde r_{k,\ell} = q_{k,m} q^*_{\ell,n} \frac{1}{N} \sum_{t=1}^N \phi_{k,m}( \widehat u_t) \psi_{\ell,n}( \widehat v_t), \quad k=1,\ldots, m, \ \ \ \ell=1, \ldots, n.
\end{equation}
This is appropriate because, at least for large $N$, 
\begin{align*}
 \sum_{k=1}^m \widetilde r_{k,\ell}
 &= q^*_{\ell,n} \frac{1}{N} \sum_{t=1}^N \biggl( \sum_{k=1}^m q_{k,m} \phi_{k,m}( \widehat u_t) \biggr) \psi_{\ell,n}( \widehat v_t) \\
 &= q^*_{\ell,n} \frac{1}{N} \sum_{t=1}^N \psi_{\ell,n}( \widehat v_t) \\
 & \approx q^*_{\ell,n} \int_0^1 \psi_{\ell,n}(v) dv \\
 & = q^*_{\ell,n},
\end{align*}
and, similarly 
$$
\sum_{\ell=1}^n \widetilde r_{k,\ell} \approx q_{k,m}.
$$

Similar to \cite{Dou-etal16}, we consider in \eqref{c} a mixture distribution of $mn$ components $\phi_{k,m}(u) \psi_{\ell,n}(v)$, $k=1, \ldots, m$, $\ell=1, \ldots, n$. 
We introduce $N$ matrices of size $m\times n$, $\tau_t=(\tau_{t, k, \ell})$, $t=1, \ldots N$, which we will consider to be latent dummy variables. 
 If the $t$-th individual belongs to component $\phi_{k,m}(u) \psi_{\ell,n}(v)$,
 then we set $\tau_{t, k, \ell}=1$; otherwise, we set $\tau_{t, k, \ell}=0$.
 The pseudo-likelihood for $( \widehat u_t,  \widehat v_t, \tau_t)$, $t=1, \ldots, N$, is given by
\begin{equation}
\prod^N_{t=1}  \prod^m_{k=1} \prod^n_{ \ell=1} \left\{ r_{k, \ell} \phi_{k,m}( \widehat u_t) \psi_{\ell,n}( \widehat v_t) \right\}^{\tau_{t, k, \ell}}.
\label{l_uvt}
\end{equation}
The conditional expectation of $\tau_{t, k, \ell}$ given $( \widehat u_t,  \widehat v_t)$, $t=1, \ldots, N$, can be estimated by
\begin{align*}
{\widetilde{\tau}}_{t, k, \ell} & = E[\tau_{t, k, \ell} \vert ( \widehat u_t, \widehat v_t); R] \\
& = \frac{r_{k, \ell}   \phi_{k,m}( \widehat u_t) \psi_{\ell,n}( \widehat v_t) }{\sum_{k'=1}^m \sum_{\ell'=1}^n r_{k',\ell'} \phi_{k',m}( \widehat u_t) \psi_{\ell',n}( \widehat v_t)}.
\end{align*}

Conditional on $\tau_{t, k, \ell} = \widetilde{\tau}_{t, k, \ell}$ in (\ref{l_uvt}), the pseudo-log-likelihood divided by $N$ becomes
\begin{equation*}
\frac{1}{N} \sum^N_{t=1} \sum^m_{k=1} \sum^n_{\ell=1} \widetilde{\tau}_{t, k, \ell} \log \left( r_{k, \ell} \phi_{k,m}( \widehat u_t) \psi_{\ell,n}( \widehat v_t) \right) = \sum^m_{k=1} \sum^n_{\ell=1}  \bar{\tau}_{k, \ell} \log r_{k, \ell} + \mbox{const.},
\end{equation*}
where 
\begin{equation}
\bar{\tau}_{k,\ell}= \frac{1}{N}\sum^N_{t=1} \widetilde{\tau}_{t, k, \ell}
= \frac{1}{N} \sum^N_{t=1} \frac{ r_{k,\ell} \phi_{k,m}( \widehat u_t) \psi_{\ell,n}( \widehat v_t)}{ \sum_{k'=1}^m \sum_{\ell'=1}^n r_{k',\ell'} \phi_{k',m}( \widehat u_t) \psi_{\ell',n}( \widehat v_t)}.
\label{m_tau}
\end{equation}
This calculation constitutes the E-step of the algorithm.

For the M-step of the algorithm, since $R$ must satisfy the restrictions
\begin{equation*}
 \sum_{\ell=1}^n r_{k,\ell} = q_{k,m},  \qquad \sum_{k=1}^m r_{k,\ell} = q^*_{\ell,n},
\end{equation*}
we need to introduce Lagrange multipliers $\mu_k, \lambda_{\ell}$.
Additionally, similar to \cite{Green90}, 
we introduce into the pseudo-log-likelihood a penalty function $p(r_{k,\ell})$, and then
we maximize the average penalized pseudo-log-likelihood function
\begin{align}
L_p(R) 
& = \frac{1}{N} \sum_{t=1}^N \log \left( \sum_{k=1}^m \sum_{\ell=1}^n
 r_{k,\ell} \phi_{k,m}( \widehat u_t) \psi_{\ell,n}( \widehat v_t) \right) \nonumber \\
& \quad  - \sum_{k} \mu_{k}\biggl(\sum_{\ell}r_{k,\ell}-q_{k,m}\biggr) 
        - \sum_{\ell}\lambda_{\ell}\biggl(\sum_{k}r_{k,\ell}-q^*_{\ell,n}\biggr) 
        - \sum_{k,\ell}p(r_{k,\ell}).
\label{pll}
\end{align}

Motivated by results of \cite{CaiWang14}, on the estimation of a sparse parameter vector, we introduced the SCAD penalty function for the purpose of estimating our sparse parameter matrices, by which the mean-square error defined in (\ref{mse}) for the copula parameter estimation is expected to reduce. 
We also note that, in a general setting, \cite{Green90}
investigated the properties of the EM algorithm for penalized likelihood estimation and encouraged the use of the penalty function because the penalized algorithm can be more practical and converges at least as quickly as the unpenalized version.

The SCAD penalty function
\begin{equation*}
p(r_{k,\ell}; \alpha, \beta)  = \begin{cases}
\alpha r_{k,\ell} , & r_{k,\ell} \le \alpha   \nonumber \\
\big(2 \alpha \beta r_{k,\ell} - r^2_{k,\ell} -\alpha^2\big)/2 (\beta-1), &  \alpha < r_{k,\ell} \le \alpha \beta  \\ 
\alpha^2 (\beta+1)/2, & r_{k,\ell} > \alpha \beta \nonumber
\end{cases}
\end{equation*}
was introduced by \cite{FanLi01}.  We will show that this function provides better estimation of $R$ when $R$ is sparse.

The tuning parameters $\alpha$ and $\beta$ satisfy $\alpha \ge 0$ and $\beta > 2$ \citep{Hastie-etal09, CaiWang14}.
Note that if $\alpha = 0$ then the penalty function reduces to $p(r_{k,\ell}; 0, \beta) = 0$, and the problem of penalized maximum likelihood estimation reduces to a non-penalized problem. 
If $\alpha \ge \max\{ r_{k,\ell}\}$, we can see that the penalty is a linear combination of the elements of $R$, and it becomes constant in (\ref{pll}). 
Hence, for the cases in which $\alpha=0$ or $\alpha \ge \max\{ r_{k,\ell}\}$, the maximization problem provides the same estimate of $R$. 

Let us now denote the first term of (\ref{pll}) by
\begin{equation}
L_p^*(R) :=  \frac{1}{N} \sum_{t=1}^N \log \left( \sum_{k=1}^m \sum_{\ell=1}^n
 r_{k,\ell} \phi_{k,m}( \widehat u_t) \psi_{\ell,n}( \widehat v_t) \right).
\label{l*r}
\end{equation}
In the sequel, we will see that the function $L_p^*(R)$ plays a role in cross-validation for the tuning parameters $\alpha$ and $\beta$.  

We now differentiate (\ref{pll}) with respect to each $r_{k,\ell}$ and set the derivative equal to $0$.  Then we obtain  
\begin{multline}
\frac{1}{N} \sum^N_{t=1} \frac{  \phi_{k,m}( \widehat u_t) \psi_{\ell,n}( \widehat v_t)}{ \sum_{k'=1}^m \sum_{\ell'=1}^n
 r_{k',\ell'} \phi_{k',m}( \widehat u_t) \psi_{\ell',n}( \widehat v_t)} \\
-\mu_{k}-\lambda_{\ell}-  \frac{\partial}{\partial r_{k,\ell}} p(r_{k,\ell}; \alpha, \beta) = 0,
\label{dldr}
\end{multline}
where the derivative of the SCAD penalty function is
$$
\dot{p}(r_{k,\ell}; \alpha, \beta) := \frac{\partial}{\partial r_{k,\ell}} p(r_{k,\ell}; \alpha, \beta) = \alpha I(r_{k,\ell} \le \alpha) + \frac{(\alpha \beta -r_{k,\ell})_+}{\beta-1} I(r_{k,\ell} > \alpha).
$$
Here $t_+ = t$ or $0$ according as $t \ge 0$ or $t < 0$, respectively; and $I(\cdot)$ denotes the indicator function, so that $I(r_{k,\ell} \le \alpha) = 1$ if $r_{k,\ell} \le \alpha$, and $I(r_{k,\ell} > \alpha) = 1 - I(r_{k,\ell} \le \alpha)$.  

Multiplying (\ref{dldr}) by $r_{k,\ell}$, and solving the equation, we obtain 
\begin{equation}
\big(\mu_k + \lambda_{\ell} + \dot{p}(r_{k,\ell}; \alpha, \beta)\big) r_{k,\ell} = \frac{1}{N} \sum^N_{t=1} \dfrac{ r_{k,\ell} \phi_{k,m}( \widehat u_t) \psi_{\ell,n}( \widehat v_t)}
{\sum_{k'=1}^m \sum_{\ell'=1}^n r_{k',\ell'} \phi_{k',m}( \widehat u_t) \psi_{\ell',n}( \widehat v_t)}.
\label{mlpr}
\end{equation}
Using the notation $\bar{\tau}_{k,\ell}$ in (\ref{m_tau}), from (\ref{mlpr}) we find
\begin{equation}
r_{k,\ell} = \frac{\bar{\tau}_{k,\ell}}{\mu_k + \lambda_{\ell} + \dot{p}(r_{k,\ell}; \alpha, \beta)}.
\label{r_kl}
\end{equation}
Thus, for given values of $\bar{\tau}_{k,\ell}, \mu_k, \lambda_{\ell}$ and tuning parameters $\alpha, \beta$ we can update $r_{k, \ell}$, and this constitutes the M-step of our algorithm.

In the M-step, vectors $\bm{\lambda}=(\lambda_1, \ldots, \lambda_n)'$ and $\bm{\mu}=(\mu_1, \ldots, \mu_m)'$ can be obtained by executing the following algorithm.

\begin{algorithm}
\caption{Calculate $\bm{\lambda}$ and $\bm{\mu}$}
\label{alg:m}
\begin{algorithmic}[1]
\State 
Set $\mu^{(0)}_{k}=1/2$ and $s=0$.

\State 
For fixed $\bm{\mu}^{(s)}=\bigl(\mu_1^{(s)},\ldots,\mu_m^{(s)}\bigr)'$ and for $1\le \ell \le n$, find $\lambda_{\ell}^{(s)}$, 
a solution $\lambda_{\ell}$ of
\[
 \sum_{k=1}^m\frac{\bar{\tau}_{k,\ell}}{\mu_k^{(s)}+\lambda_{\ell} + \dot{p}(r_{k,\ell}; \alpha, \beta)}=q^*_{\ell,n}
\]
such that $\lambda_{\ell} > -\min_k\bigl(\mu_k^{(s)} + \dot{p}(r_{k,\ell}; \alpha,\beta)\bigr)$.

\State 
For fixed $\bm{\lambda}^{(s)}=\bigl(\lambda_1^{(s)},\ldots,\lambda_n^{(s)}\bigr)'$ and for $1 \le k \le m$, find $\widetilde\mu_k^{(s)}$, 
a solution $\widetilde\mu_k$ of 
\[
 \sum^n_{\ell=1}\frac{\bar{\tau}_{k,\ell}}{\widetilde\mu_k+\lambda_{\ell}^{(s)}+  \dot{p}(r_{k,\ell}; \alpha, \beta)}=q_{k,m}
\]
such that $\widetilde\mu_k>-\min_{\ell}\bigl(\lambda_{\ell}^{(s)} + \dot{p}(r_{k,\ell}; \alpha, \beta) \bigr)$.

\State 
Let
\[
\mu_{k}^{(s)} = \widetilde\mu_{k}^{(s)} - \biggl( \sum_{k=1}^m q_{k,m} \widetilde\mu_{k}^{(s)} -\sum_{k=1}^m q_{k,m} \mu^{(0)}_{k} \biggr), \quad 1 \le k \le m.
\]

\State 
Increase the counter $s$ by $1$, and repeat Steps 2--4 
until $\bm{\lambda}$ and $\bm{\mu}$ converge.
\end{algorithmic}
\end{algorithm}

Then, the EM algorithm for estimating $R$ can be summarized as follows.

\begin{algorithm}
\caption{The EM algorithm for the penalized pseudo-likelihood function}
\label{alg:em}
\begin{algorithmic}[1]
\State
Set $r_{k,\ell}$ equal to $\widetilde r_{k,\ell}$ in (\ref{r-tilde}).

\State [E-step] 
For given  $r_{k,\ell}$, calculate $\bar{\tau}_{k,\ell}$ by (\ref{m_tau}).

\State [M-step] 
For given   $\bar{\tau}_{k,\ell}$,  $\alpha$ and $\beta$, find first $\bm{\mu}$ and $\bm{\lambda}$ by Algorithm \ref{alg:m}, and then update $r_{k,\ell}$ by (\ref{r_kl}).

\State Repeat Steps 2 and 3 
until $r_{k,\ell}$ converges.
\end{algorithmic}
\end{algorithm}

\subsection{Convergence properties of the EM algorithm for the penalized pseudo-likelihood function}

As explained by \cite{Green90}, the monotonicity and convergence properties of the penalized pseudo-likelihood function are inherited from the original EM algorithm \citep[Section 3.2]{McLachlan-Krishnan08} as follows:

\begin{proposition}
\label{prop_Rinf}
The EM algorithm for the penalized pseudo-log-likelihood function converges to $R^{(\infty)}=(r^{(\infty)}_{k,\ell})$, which is a solution of 
$$
\partial L_p(R)/\partial R = 0,
$$
i.e., for $k=1,\ldots, m$ and $\ell=1,\ldots,n$, 
$$
\frac{\partial L_p(R)}{\partial r_{k,\ell}} \bigg\vert_{R=R^{(\infty)}} = 0.
$$
\end{proposition}

\begin{proposition}
\label{prop_monot}
The penalized pseudo-log-likelihood function provides monotonically increasing values under the EM algorithm: 
$$
L_p(R^{(s)}) \le L_p(R^{(s+1)}),
$$
where $R^{(s)} = (r^{(s)}_{k,\ell})$ consists of the estimated values of $r_{k,\ell}$ in the $s$-th iteration of the algorithm.
\end{proposition}

According to these properties given above, the limiting value of $R^{(\infty)}$ is at least a local maximum of the penalized pseudo-likelihood function.  The proofs of Propositions \ref{prop_Rinf} and \ref{prop_monot} are given in Appendix \ref{sec_appendix}.

\subsection{Choosing the tuning parameters}\label{sec_tuning_parameters}

The tuning parameters $(\alpha, \beta)$ in the penalty function $p(\cdot \, ; \alpha,\beta)$ can be selected by the general method of cross-validation, a method that is described in detail by \cite{Hastie-etal09, CaiWang14}.  In the context of our results, let $D$ be the full data set and let $D_1,\ldots,D_M$ be subsets of $D$ that will serve as test sets.  For $i=1,\ldots, M$ let $N_i = \#\{ (x_t, y_t) \in D_i \}$ denote the cardinality of $D_i$, and we use as training data sets the collection $D \backslash D_1,\ldots, D \backslash D_M$.  

For each pair $(\alpha, \beta)$, we use the training data sets $D \backslash D_i$ to estimate 
$\widehat{R} = (\widehat{r}_{k,\ell})$.  Next, we calculate 
\begin{equation}
L^*_i(\widehat{R}) 
= \frac{1}{N_i} \sum_{t=1}^{N_i} \log \left( \sum_{k=1}^m \sum_{\ell=1}^n \widehat{r}_{k,\ell} \phi_{k,m}( \widehat u_t) \psi_{\ell,n}( \widehat v_t)\right),
\label{L*}
\end{equation}
where $ \widehat u_t$ and $ \widehat v_t$ are defined in \eqref{eq_ut_vt}; note that $L^*_i(\widehat{R})$ is an analog of (\ref{l*r}), for the data $(x_t, y_t) \in D_i$.  Further, we define 
\begin{equation}
CV(\alpha,\beta; m,n) = \sum_{i=1}^M L^*_i(\widehat{R}),
\label{cv}
\end{equation}
and then for each fixed $(m,n)$ pair, we find $(\widehat{\alpha}, \widehat{\beta})$ that maximizes $CV(\alpha, \beta; m,n)$.  
In Section \ref{subsec:simu_ab}, we will carry out simulations to assess the performance of (\ref{cv}).

\subsection{Model selection}

The size $m \times n$ of the parameter matrix $R$ can be chosen by cross-validation or the Akaike information criterion (AIC).  
First, for each fixed pair $(\alpha, \beta)$
we use the cross-validation method to calculate $CV(\alpha, \beta; m,n)$
for numerous pairs of $(m,n)$ and, second,
we identify a pair, $(\widehat{m}, \widehat{n})$, that maximizes $CV(\alpha, \beta ; m,n)$.

The minimizer of the pseudo-AIC \citep{Akaike74} can also be considered a choice for $(m,n)$.  Here, we define the AIC-type statistic, 
\begin{align}
AIC(m,n) &= -2 \sum_{t=1}^N \log \bigg[\sum_{k=1}^m \sum_{\ell=1}^n 
\widehat{r}_{k,\ell}\, \phi_{k,m}\big(\widehat F_X(x_t)\big) \psi_{\ell,n}\big(\widehat F_Y(y_t)\big)\bigg] \nonumber \\
&\qquad + 2(m-1)(n-1),
\label{aic}
\end{align}
where the correction term, $2(m-1)(n-1)$, treats the penalty function as if it were not a parameter.  
For both methods of choosing $(m,n)$, we keep fixed the tuning parameters $\alpha$ and $\beta$.  In the simulations described in Section \ref{subsec:simu_mn}, we use the EM algorithm without penalty, i.e., $\alpha=0$.

\section{Simulation studies}
\label{sec_simulation}

\begin{figure}[htbp]
\begin{center}
\begin{tabular}{cc}
  \includegraphics[width=0.5\linewidth]{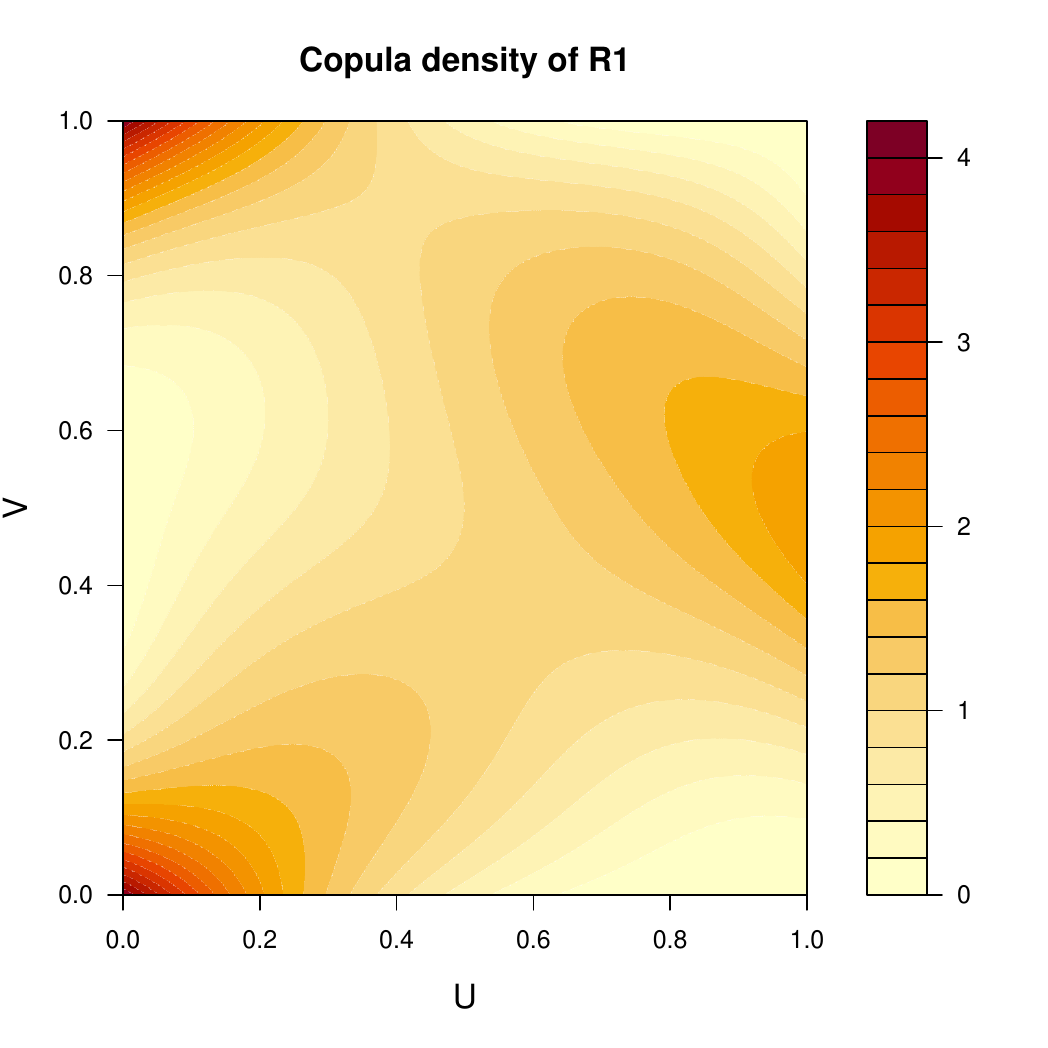} &
  \includegraphics[width=0.5\linewidth]{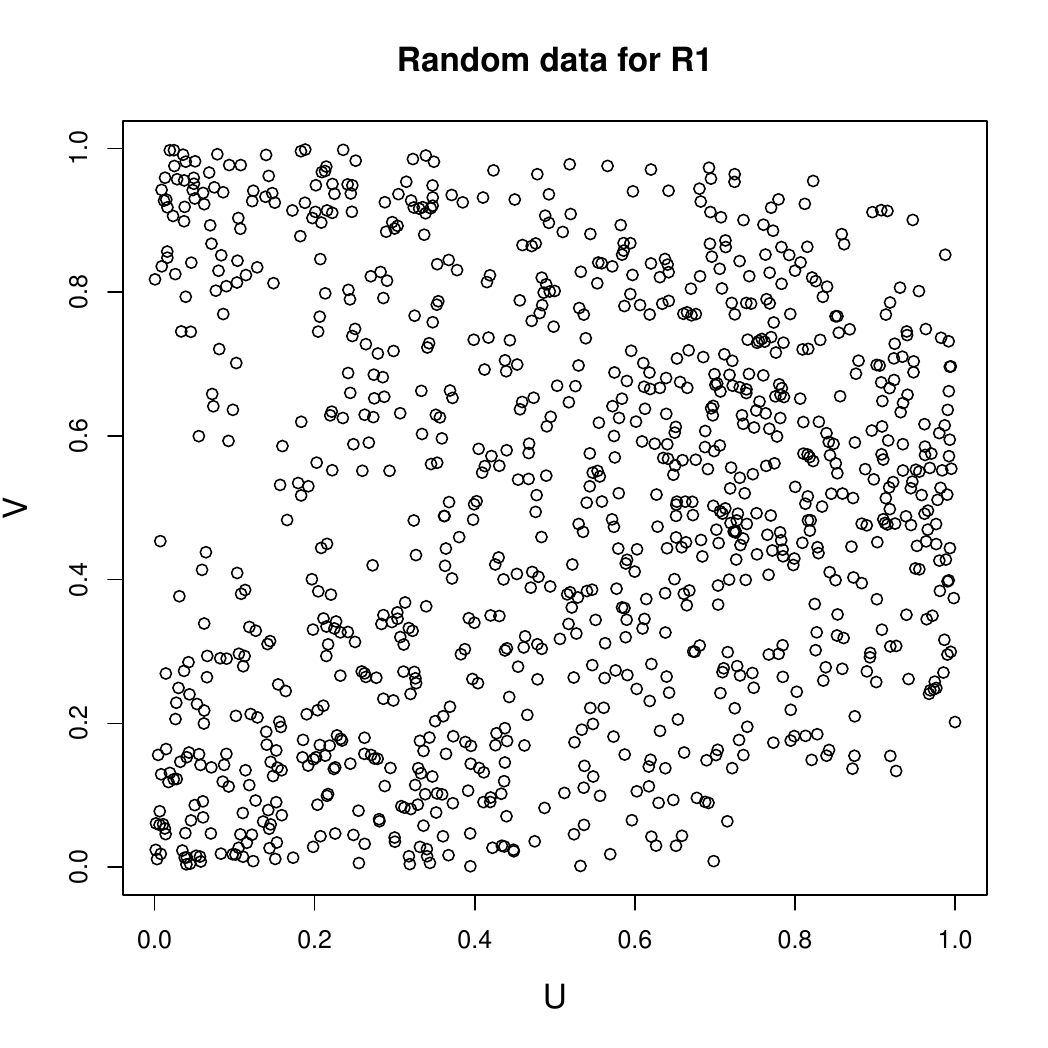}  \\
  \includegraphics[width=0.5\linewidth]{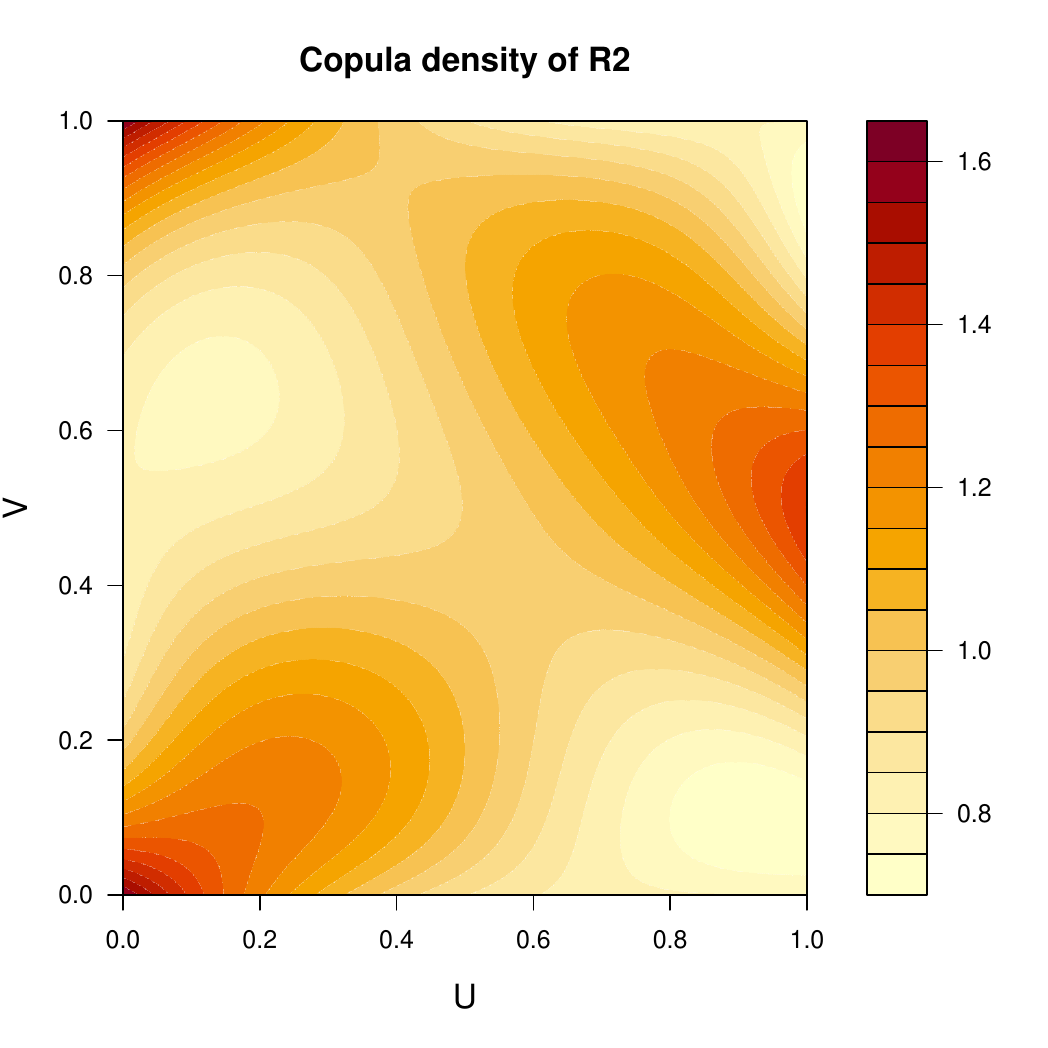} &
  \includegraphics[width=0.5\linewidth]{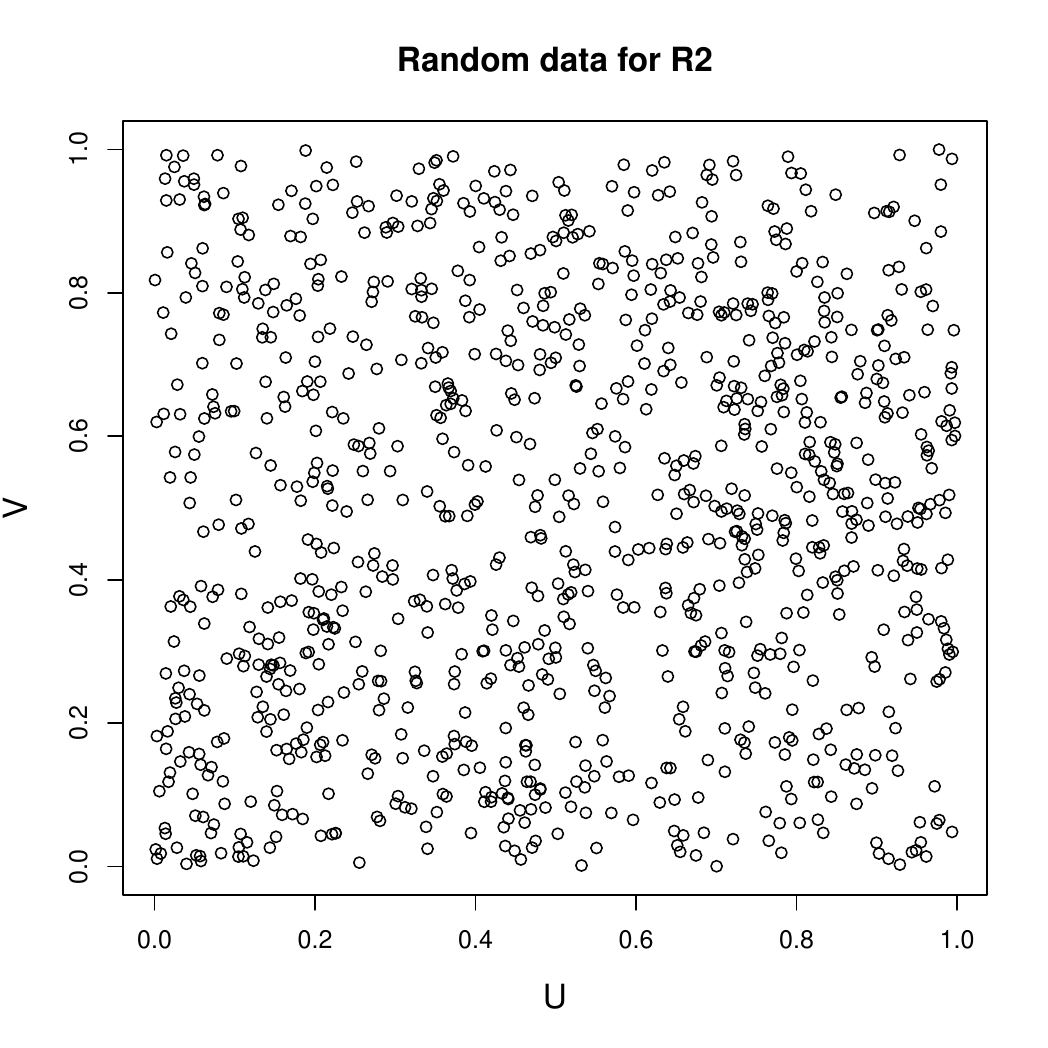} \\
  \includegraphics[width=0.5\linewidth]{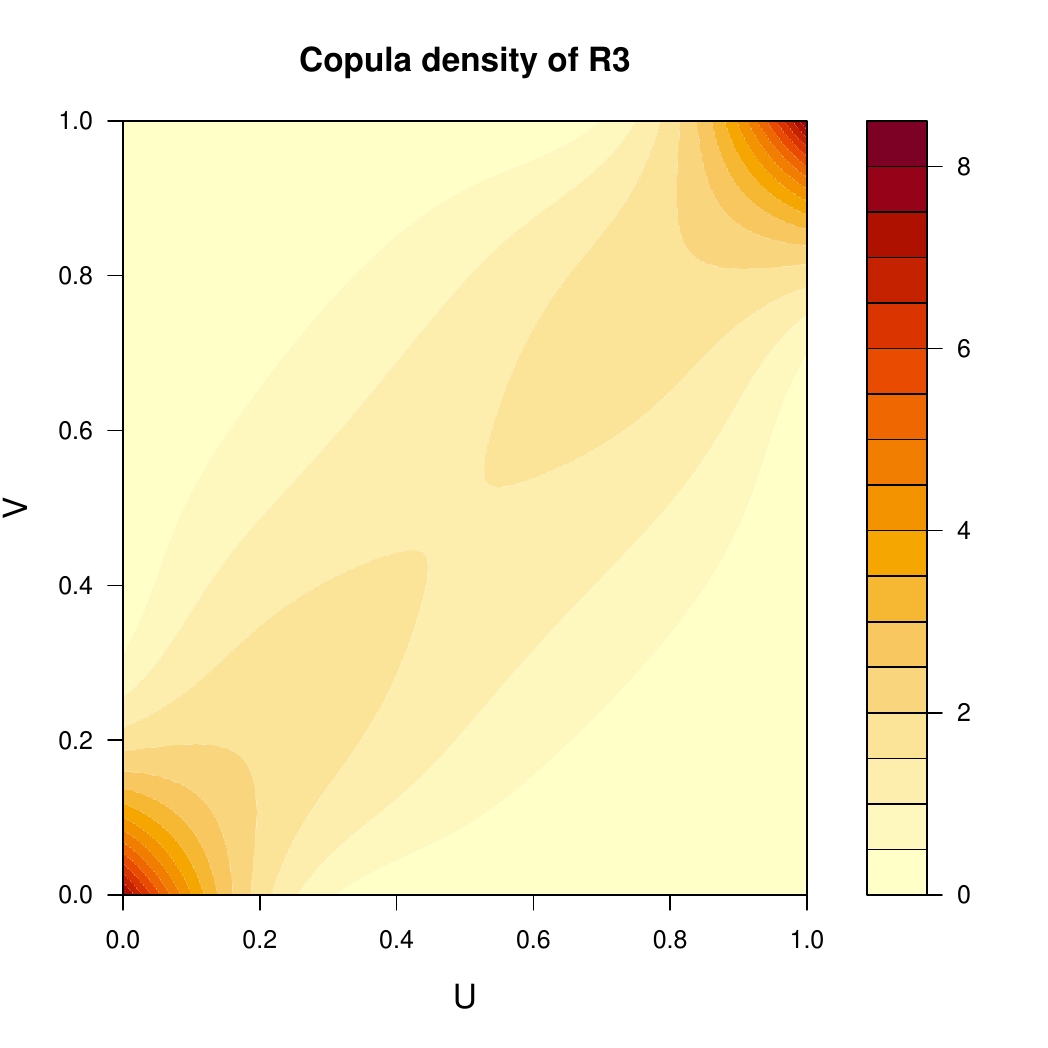} &
  \includegraphics[width=0.5\linewidth]{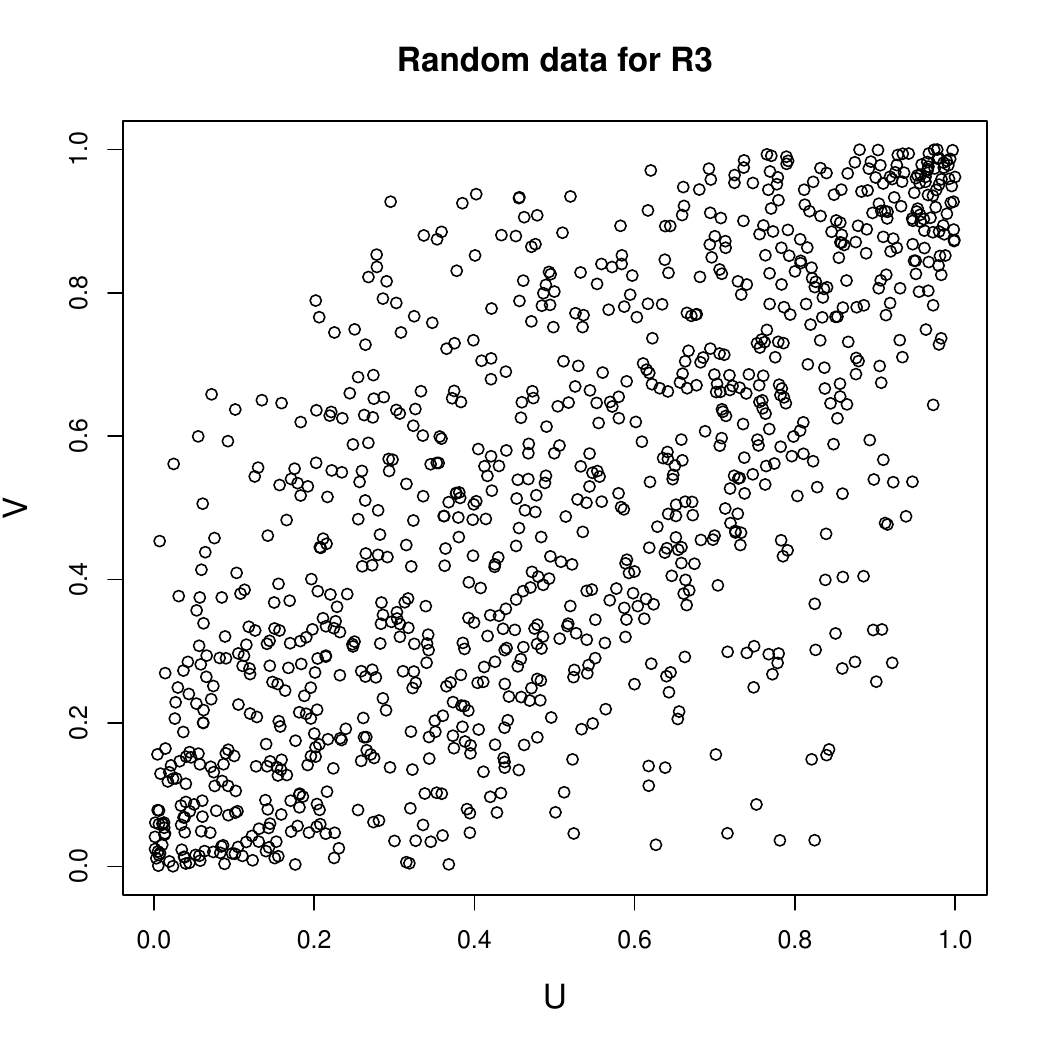}
\end{tabular}
\end{center}
\caption{The copula densities $c(u, v; R_i)$, $i=1,2,3$ 
 (left), and corresponding randomly generated data $(u_t, v_t)$, $t=1,\ldots,1000$ (right).} 
\label{fig:simc}
\end{figure}

To examine the performance of the proposed methods by simulation, we first generate random numbers using the rejection sampling method given in Appendix \ref{sec_sampling}.  In the first three simulations, we fix the degree of the B-spline basis function at $d=3$. For each of the parameter matrices, 
$$
R_1=
\begin{pmatrix}
 0.125 & 0    & 0    & 0    & 0.125 \\
 0     & 0.25 & 0    & 0    & 0     \\
 0     & 0    & 0    & 0.25 & 0     \\
 0     & 0    & 0.25 & 0    & 0 
\end{pmatrix},
$$
$$
R_2=
\begin{pmatrix}
 0.050 & 0.050 & 0.050 & 0.050 & 0.050 \\
 0.025 & 0.150 & 0.025 & 0.025 & 0.025 \\
 0.025 & 0.025 & 0.025 & 0.150 & 0.025 \\
 0.025 & 0.025 & 0.150 & 0.025 & 0.025 
\end{pmatrix}, 
$$
and 
$$
R_3=
\begin{pmatrix}
 0.12  & 0.005 & 0 & 0 & 0     \\
 0.005 & 0.245 & 0 & 0 & 0     \\
 0 & 0 & 0.24  & 0.01  & 0     \\
 0 & 0 & 0.01  & 0.24  & 0     \\
 0 & 0 & 0     & 0     & 0.125 
\end{pmatrix}, 
$$
we generate 100 sets of random values of $(U,V)$, where each set contains 1,000 pairs of values of $(u_t, v_t)$, $t=1,\ldots,1000$.  The graphs of the copula densities $c(u,v;R_i)$, $i=1,2,3$, and a scatterplot of 1,000 random data pairs $(u_t,v_t)$, $t=1,\ldots,1000$ generated from each copula are shown in Figure \ref{fig:simc}.

To examine the effectiveness of the methods in Section \ref{sec_em_algorithm}, we conduct three simulation studies. The first simulation study allows us to ascertain conditions under which the penalization is necessary and, using some examples, we also demonstrate the convergence of the algorithm for the penalized pseudo-likelihood functions. The second study illustrates the performance of the cross-validation procedure for choosing tuning parameters, and the third study compares and contrasts methods of model selection including the cross-validation and AIC methods.
Additionally, the fourth study compares the performance of Bernstein copula and B-spline copula for a 3-dimensional data.  

In the first three simulation studies, which are provided in subsections \ref{subsec_1}--\ref{subsec:simu_mn}, we use the $(u_t,v_t)$, rather than the $(\widehat{u}_t,\widehat{v}_t)$, to construct the estimates of $F_X$ and $F_Y$.  Although this ignores the effect of replacing $F_X$ and $F_Y$ with the corresponding empirical distribution functions $\widehat{F}_X$ and $\widehat{F}_Y$ we note that, by the Glivenko-Cantelli theorem, $\widehat{F}_X$ and $\widehat{F}_Y$ converge uniformly to $F_X$ and $F_Y$, respectively, as $N \to \infty$.  
Therefore,  consistent estimators for the copula density functions are obtained even when $(\widehat F_X,\widehat F_Y)$ is used instead of $(F_X,F_Y)$.  That is, the pseudo-likelihood provides consistent estimators, and indeed the consistency and inefficiency of the pseudo-likelihood estimators is established by \cite{Genest-etal95} and \cite{Tsukahara05}.  

Therefore, in the first three simulations studies, we use the $(u_t, v_t)$ rather than the pseudo-observations, and the results of the simulations can be viewed as providing assurance that the results obtained are accurate under the best-case scenarios.  Finally, we provide in Subsection 4.5 a fourth simulation study for a three-dimensional example, and we construct estimates of $F_X$, $F_Y$, and $F_Z$ using the pseudo-observations $(\widehat{F}_X, \widehat{F}_Y, \widehat{F}_Z)$.

\subsection{Conditions under which penalization is necessary (Simulation Study I)}
\label{subsec_1}

Given the true parameter matrices $R_1, R_2$, and $R_3$, and also using the generated data, we apply the EM method for penalized log-likelihood functions to estimate $\widehat{R}_1$, $\widehat{R}_2$, and $\widehat{R}_3$.  For any such estimator $\widehat{R}$ we define the mean-square error (MSE) corresponding to each pair of tuning parameters $(\alpha, \beta)$ as
\begin{equation}
\mbox{MSE}(\widehat{R}; \alpha, \beta) = \frac{1}{J}\sum^J_{j=1} \sum_{k,\ell} (\widehat{r}^{(j)}_{k,\ell}(\alpha, \beta)- r_{k,\ell})^2 , 
\label{mse}
\end{equation}
where 
$J=100$, $\alpha \in [0, 0.25]$, $\beta \in [2, 4.5]$, and each $\widehat{R}^{(j)}(\alpha, \beta) = \big( \widehat{r}^{(j)}_{k,\ell} (\alpha, \beta) \big)$ is obtained from the EM method with penalization for given $(\alpha, \beta)$ and data set $j$, where $j=1, \ldots, J$.  Following \eqref{mse}, we calculate the mean-square errors $\mbox{MSE}(\widehat{R}; \alpha, \beta)$ for 15 equally spaced values of $\alpha$ and 10 equally spaced values of $\beta$, so that 150 MSEs are obtained for each $R_i$, $i=1,2,3$.  

\begin{figure}[htbp]
\begin{center}
\begin{tabular}{c}
  \includegraphics[width=0.6\linewidth]{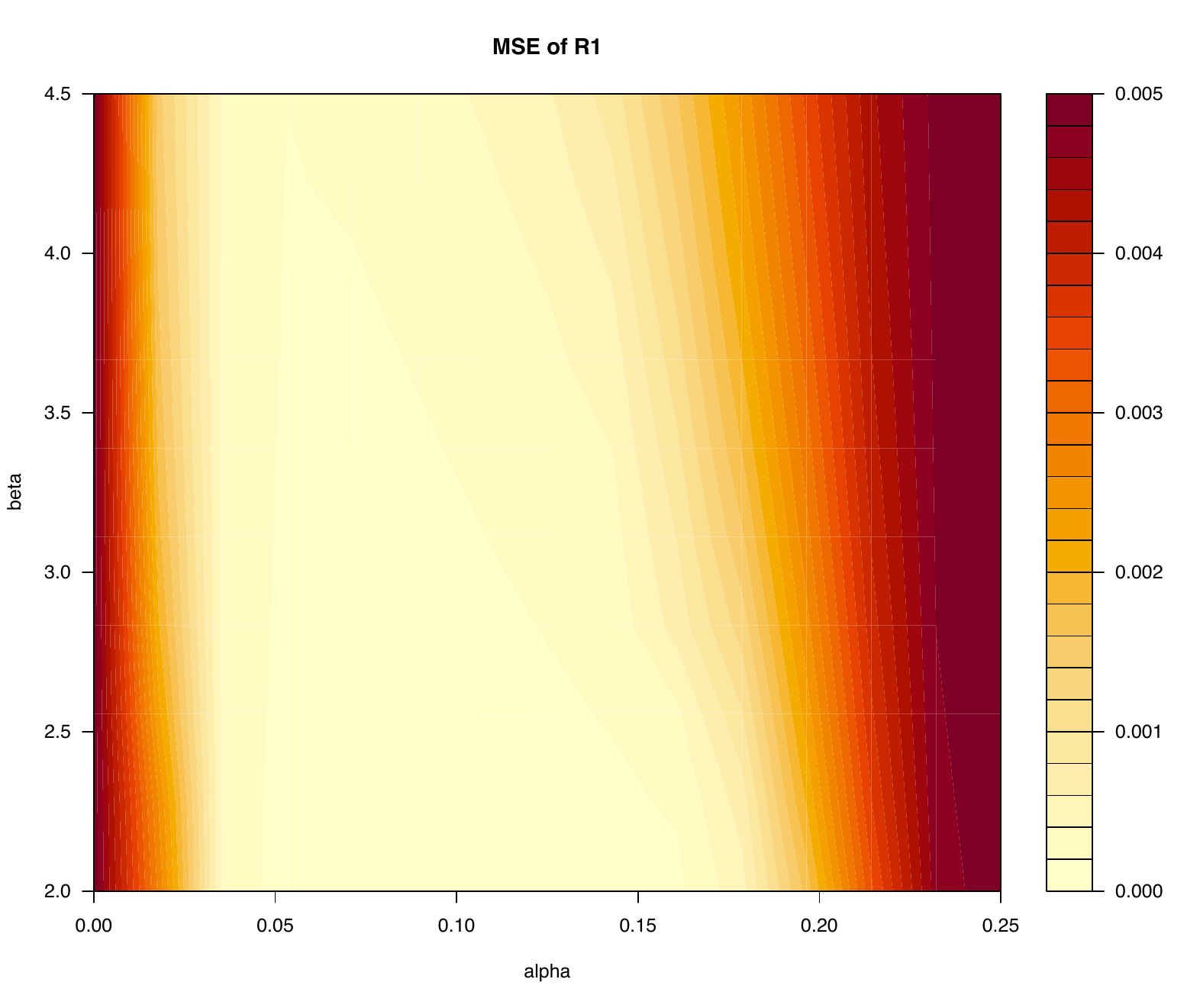} \\
\includegraphics[width=0.6\linewidth]{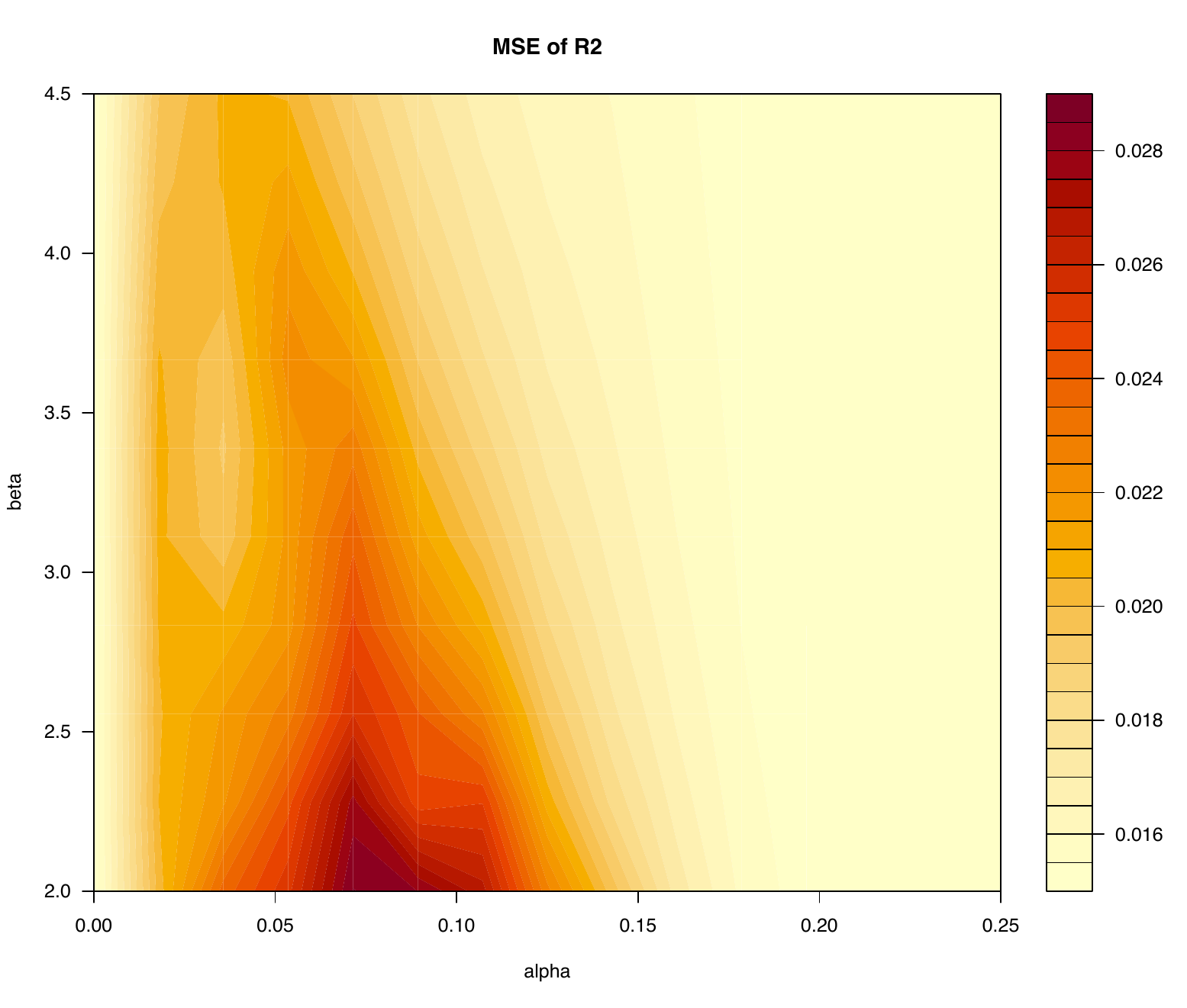} \\
  \includegraphics[width=0.6\linewidth]{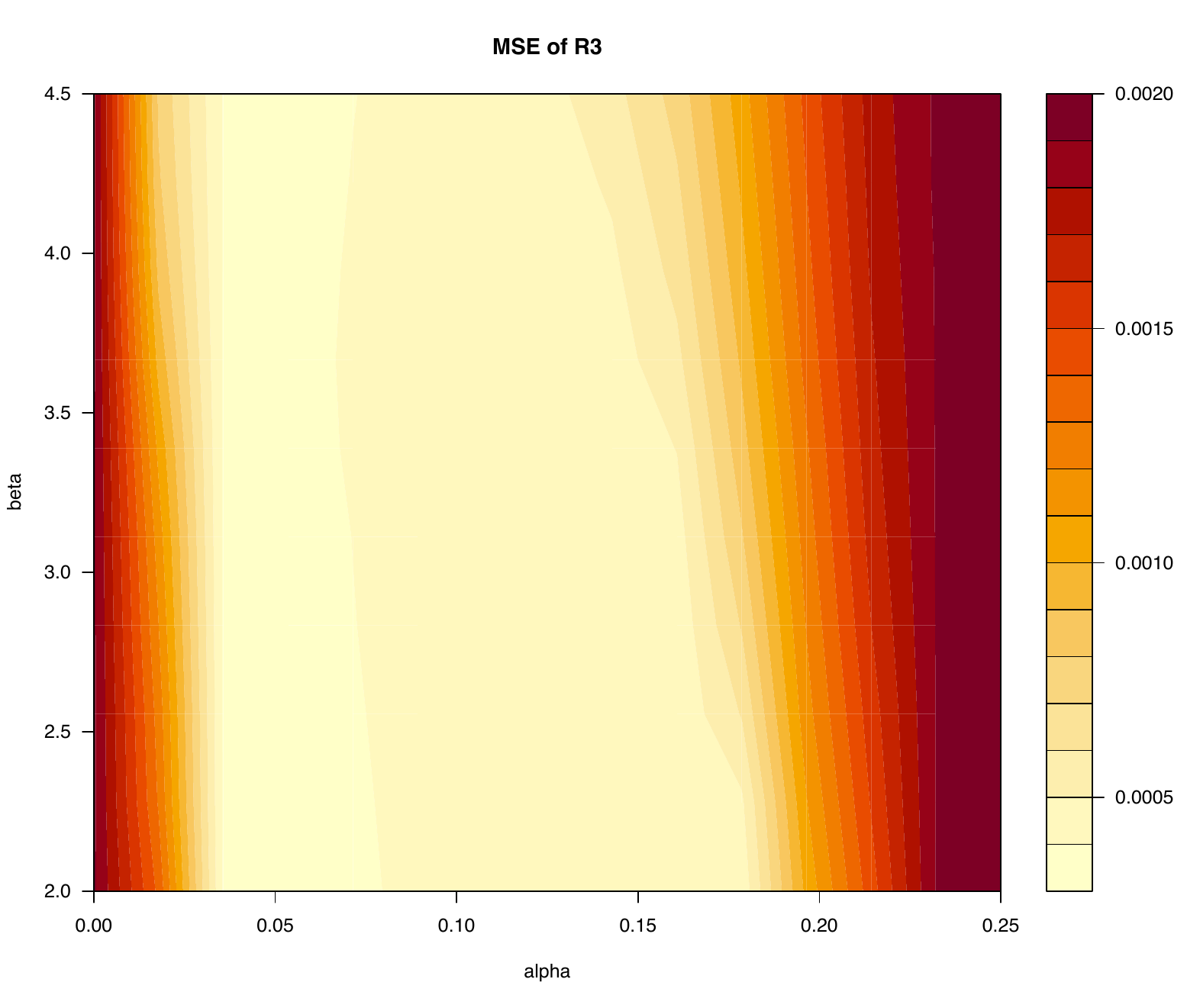} 
\end{tabular}
\end{center}
\caption{Horizontal axis is $\alpha \in [0, 0.25]$; vertical axis is $\beta \in [2, 4.5]$.  Contour plots of MSEs in (\ref{mse}) of $\widehat{R}_1$, $\widehat{R}_2$ and $\widehat{R}_3$ for tuning parameters $(\alpha, \beta)$.}
\label{fig:mse}
\end{figure}

The MSEs of $\widehat{R}_1$, $\widehat{R}_2$, and $\widehat{R}_3$ for different pairs of $(\alpha, \beta)$ are shown in Figure \ref{fig:mse} from top to bottom, respectively.  We see that the MSEs of $\widehat{R}_1$ and $\widehat{R}_3$ are large when $\alpha$ is close to $0$ or greater than $0.2$; and the MSE is small when $0.05 < \alpha < 0.2 $.  However, the MSE of $\widehat{R}_2$ shows a contrary image, which indicates that it is the extreme values of $\alpha$ that will lead to good estimation of $R_2$.  In other words, when the true parameter matrix $R$ is sparse, penalization is necessary and a properly chosen $\alpha > 0$ can lead to better estimation of $R$; on the other hand, if $R$ is not sparse then the EM algorithm without penalization is generally superior.  

We also note that the variability of the observed MSEs in the horizontal ($\alpha$) direction is larger than the variability in the vertical ($\beta$) direction.  This phenomenon is due to the fact that, in the penalty function, the parameter $\alpha$ is a more essential tuning parameter than $\beta$.

\subsection{Convergence properties of the EM algorithm (Simulation Study I)}
\label{subsec_2}

With three examples of the simulation data sets in Simulation Study I, we illustrate the performance and the convergence properties of the EM algorithm.

\begin{figure}[htbp]
\begin{center}
\begin{tabular}{cc}
    \includegraphics[width=0.5\linewidth]{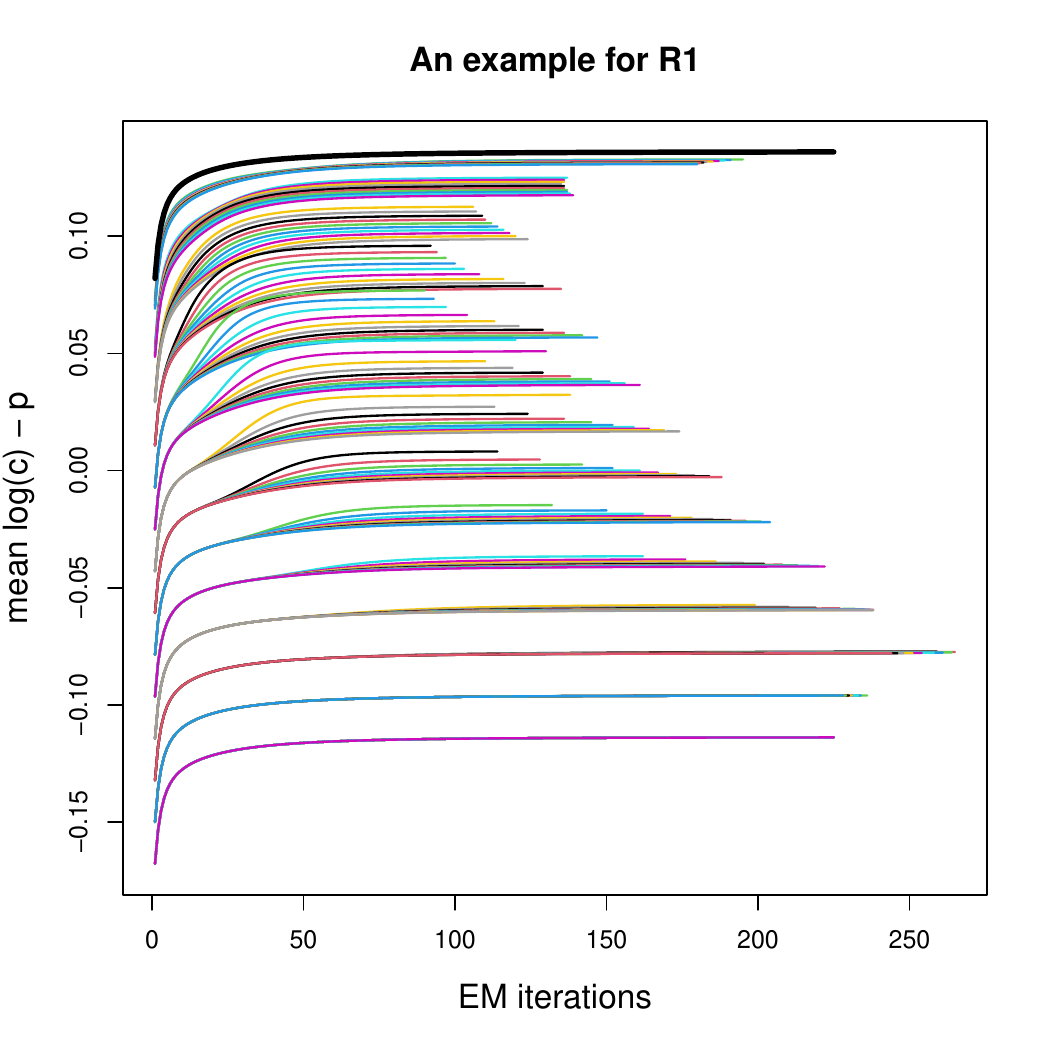}  &
      \includegraphics[width=0.5\linewidth]{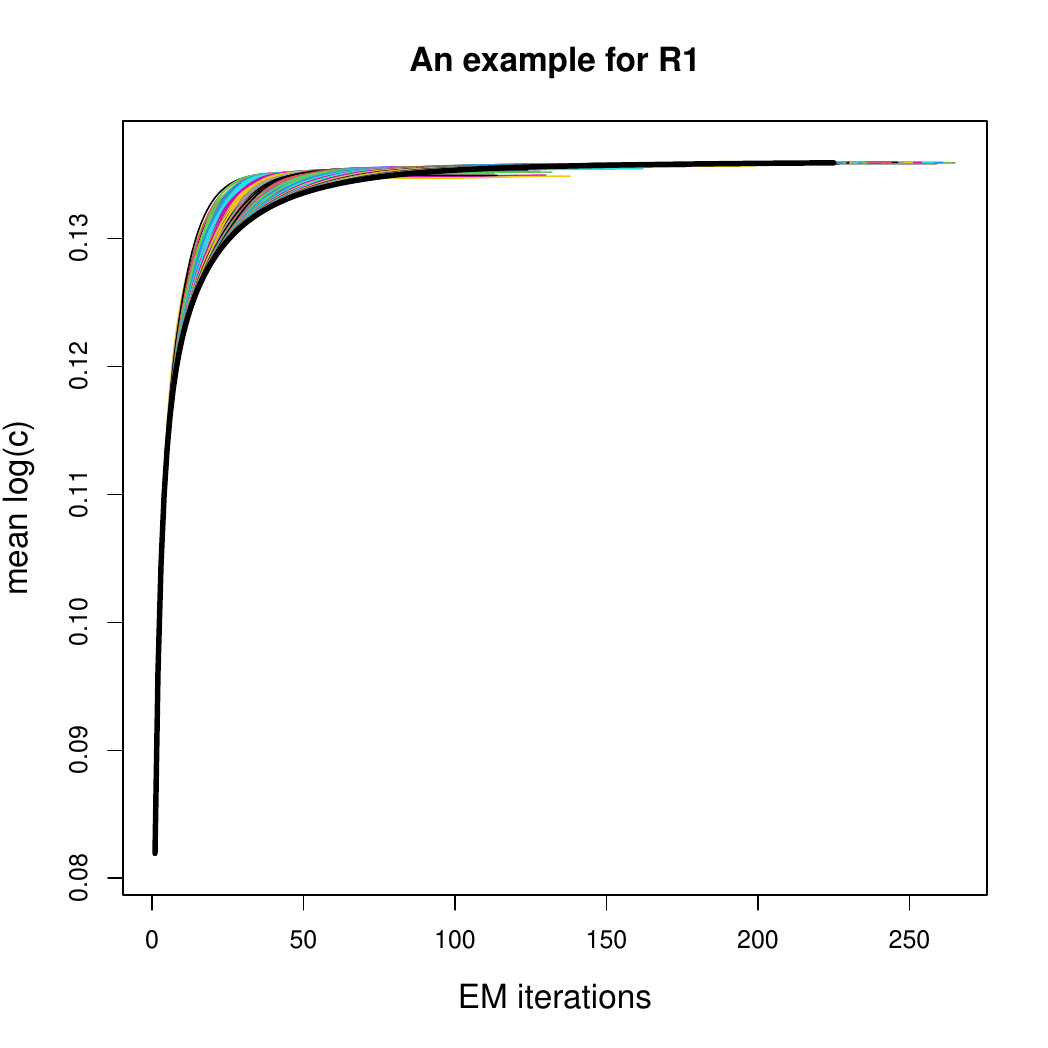} \\
      \includegraphics[width=0.5\linewidth]{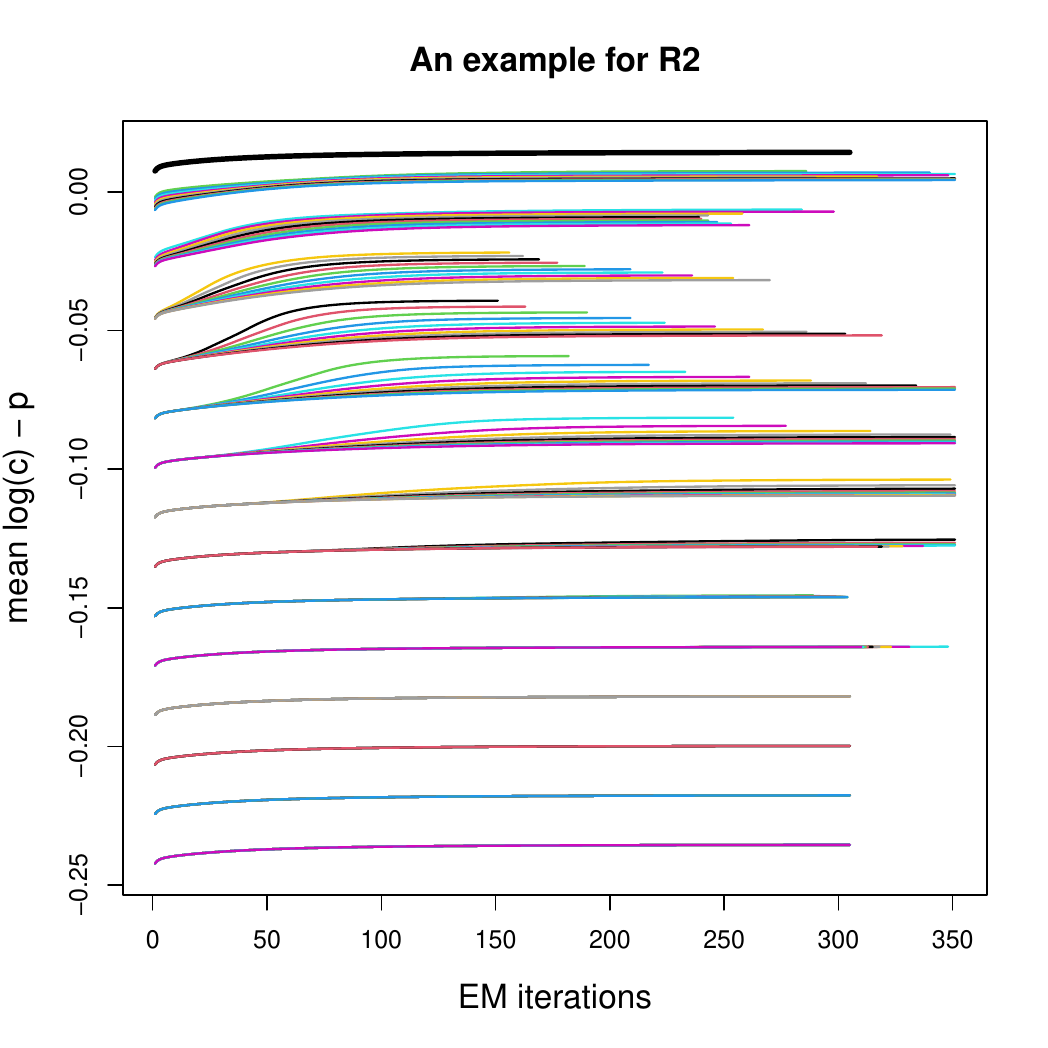} &
        \includegraphics[width=0.5\linewidth]{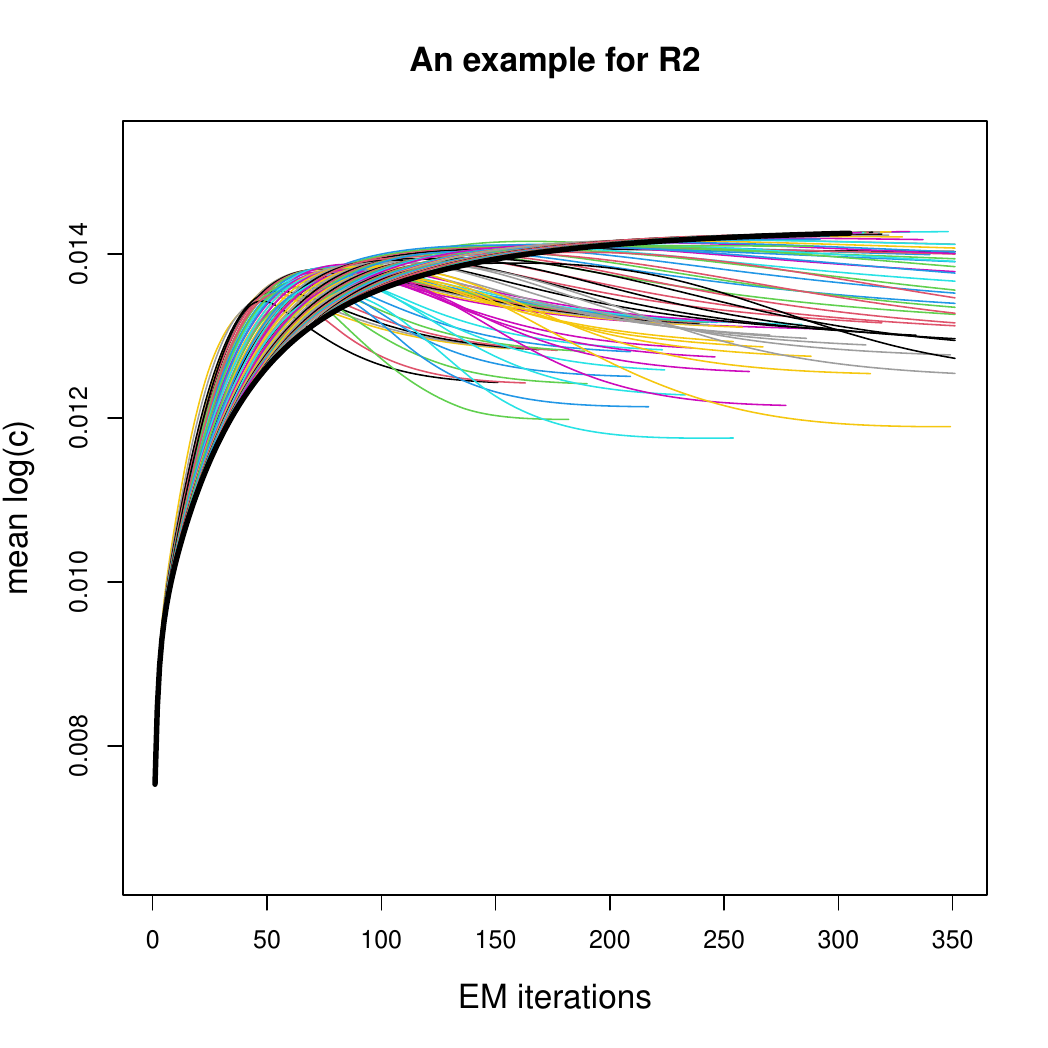} \\
      \includegraphics[width=0.5\linewidth]{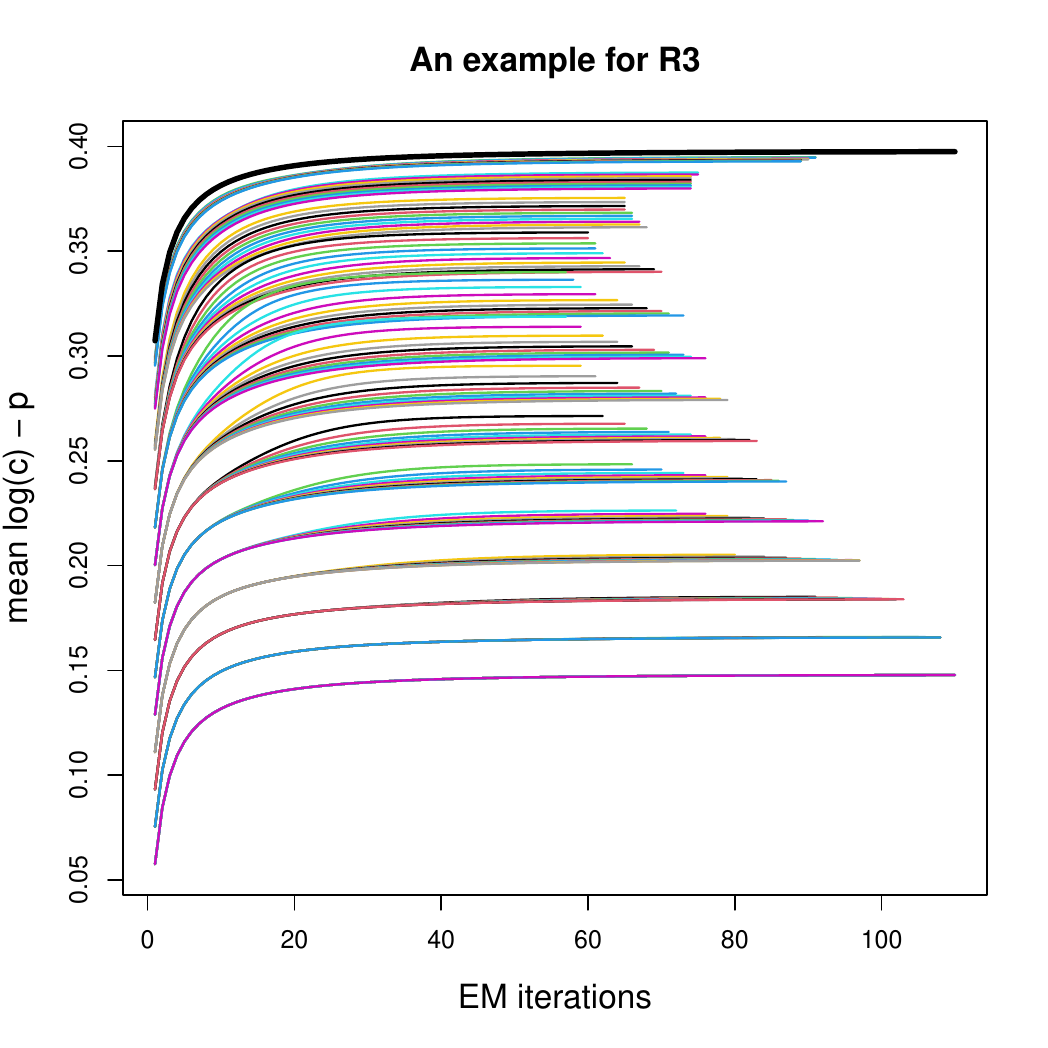}  &
        \includegraphics[width=0.5\linewidth]{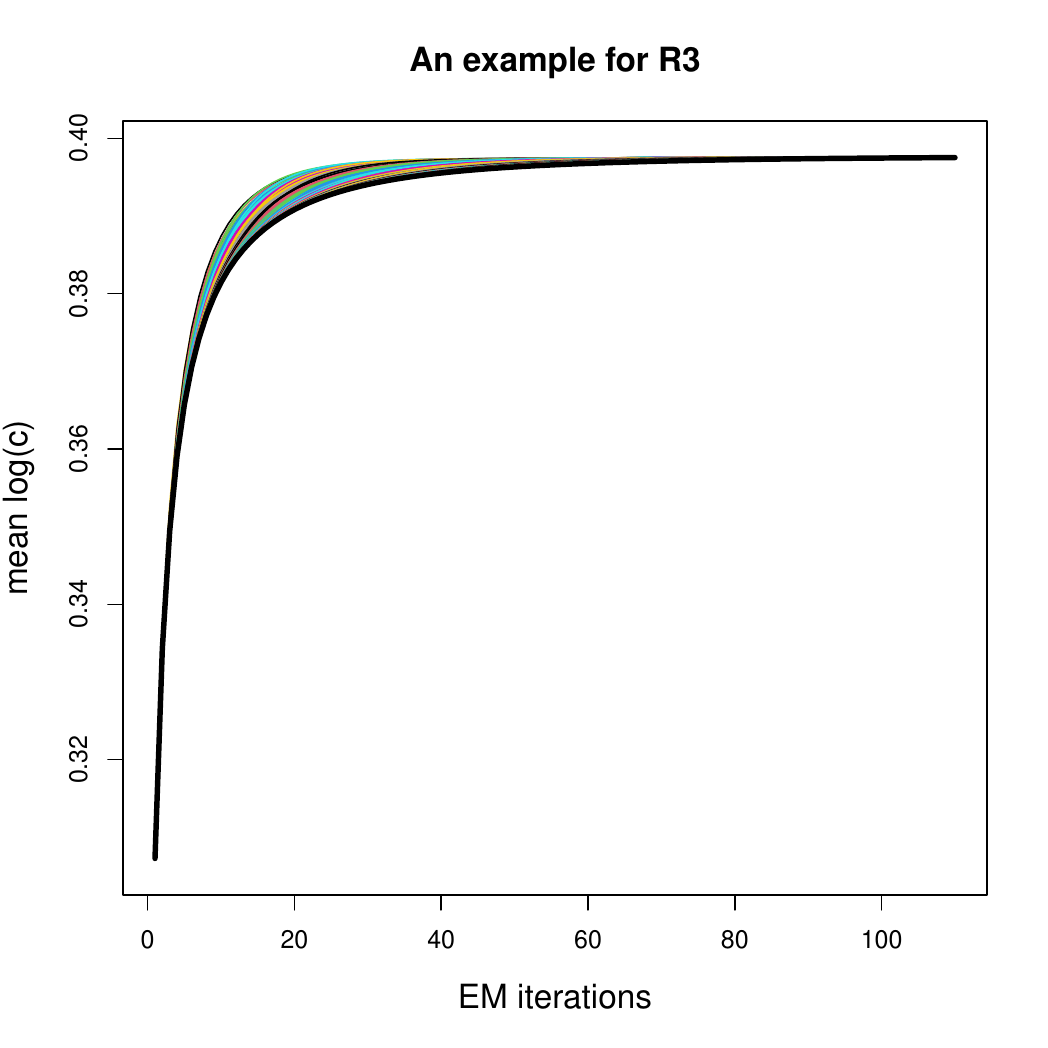} 
\end{tabular}
\end{center}
\caption{Horizontal axis is the number of EM iterations. 
Vertical axis is for 
(\ref{pll}) or (\ref{l*r}) of three data sets.
Left: convergence of (\ref{pll}) for different pairs of $(\alpha, \beta)$; Right: convergence of  (\ref{l*r}) for different pairs of $(\alpha, \beta)$.}
\label{fig:converg}
\end{figure}

The graphs on the left side of Figure \ref{fig:converg} show monotonically increasing convergence of the average penalized pseudo-log-likelihood functions (\ref{pll}) when the algorithm is applied to each data set.  As (\ref{pll}) also depends on the values of $(\alpha, \beta)$ then we have in each panel a group of 15 sets of penalized log-likelihood curves, and each group has 10 curves, for the $\alpha$'s and $\beta$'s, respectively.  

As expected, when $\alpha=0$, the log-likelihood function attains its highest value; and as $\alpha$ increases, the penalized log-likelihood function takes large decreases in value.  That is, each group of curves corresponds to a single value of $\alpha$, and increases in the value $\alpha$ leads to substantially smaller values of the penalized log-likelihood function.  For the cases in which $\alpha=0$ or $\alpha \ge \max\{ r_{k,\ell}\}$, since $L_p(R)= L_p^*(R) - \mbox{constant}$, we see that the top ($\alpha=0$) and bottom ($\alpha \ge \max\{ r_{k,\ell}\}$) lines in each left panel of Figure \ref{fig:converg} behave similarly.

For fixed $\alpha$, each group of curves indicate that changes in the value of $\beta$ lead only to minor changes in the values of the penalized log-likelihood function.  This phenomenon again implies that the choice of $\alpha$ is more crucial to the penalized log-likelihood function than the choice of $\beta$.  Further, it is also evident from these graphs that the EM algorithm maximizes the penalized log-likelihood in each case and without any difficulty.  

For the same data sets, the graphs on the right-hand side of Figure \ref{fig:converg} illustrate the behaviors of (\ref{l*r}), the mean of the pseudo-log-likelihood functions for the 150 pairs of $(\alpha, \beta)$. 
For $0 < \alpha < \max \{r_{k, \ell}\}, \ k=1,\ldots, m,\  \ell=1, \ldots, n$, we see that the EM method enables (\ref{l*r}) to attain its maximum value at early stages of convergence. However, the pseudo-log-likelihood function subsequently may decrease temporarily before attaining convergence.  Nevertheless the EM algorithm without penalization, i.e., for the case in which $\alpha=0$, continues to increase and convergence is attained perhaps at a slow pace, as shown in the thick black line.

\subsection{Choosing the tuning parameters (Simulation Study II)}
\label{subsec:simu_ab}

\begin{figure}[htbp]
\begin{center}
\begin{tabular}{c}
    \includegraphics[width=0.6\linewidth]{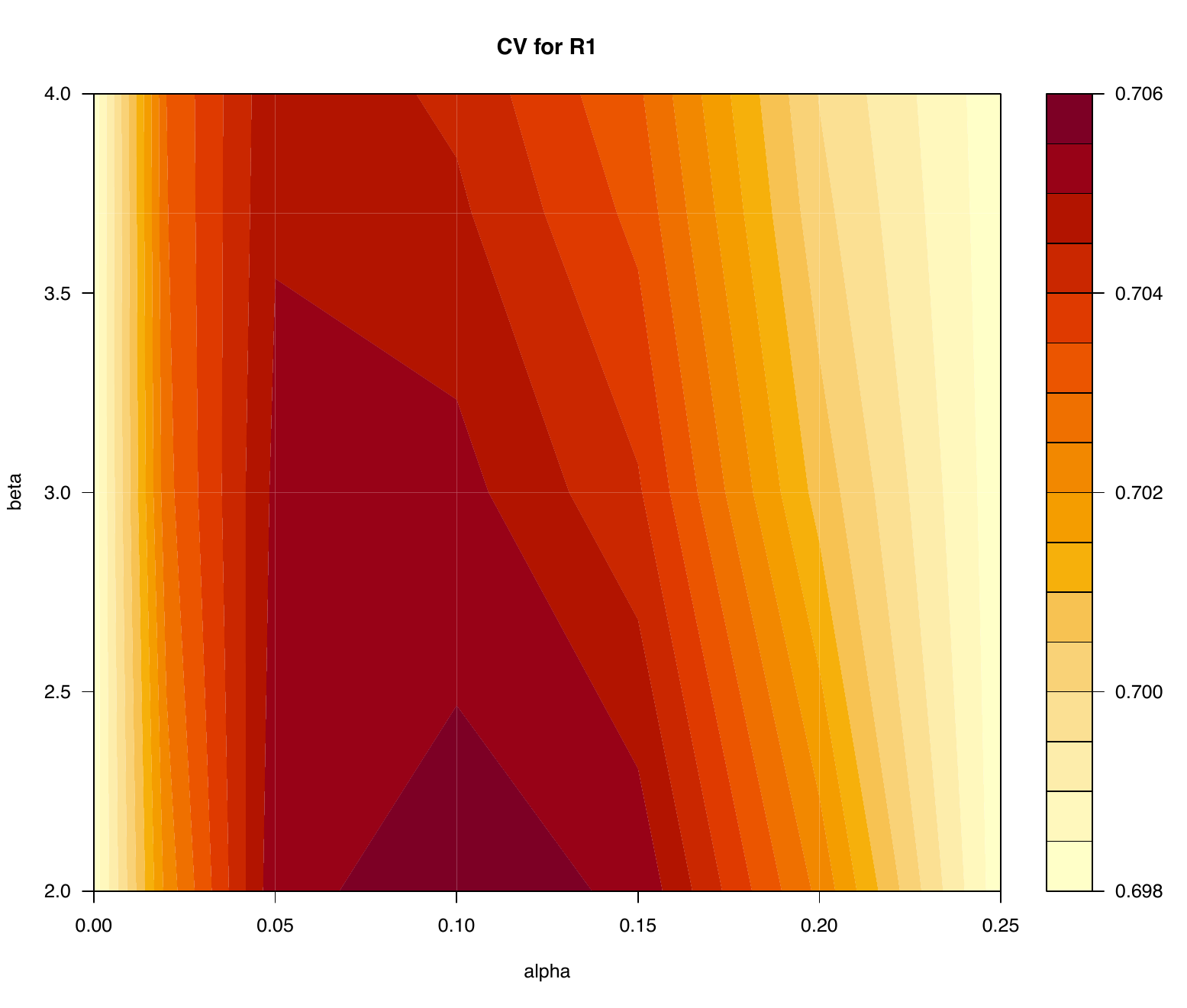}  \\
    \includegraphics[width=0.6\linewidth]{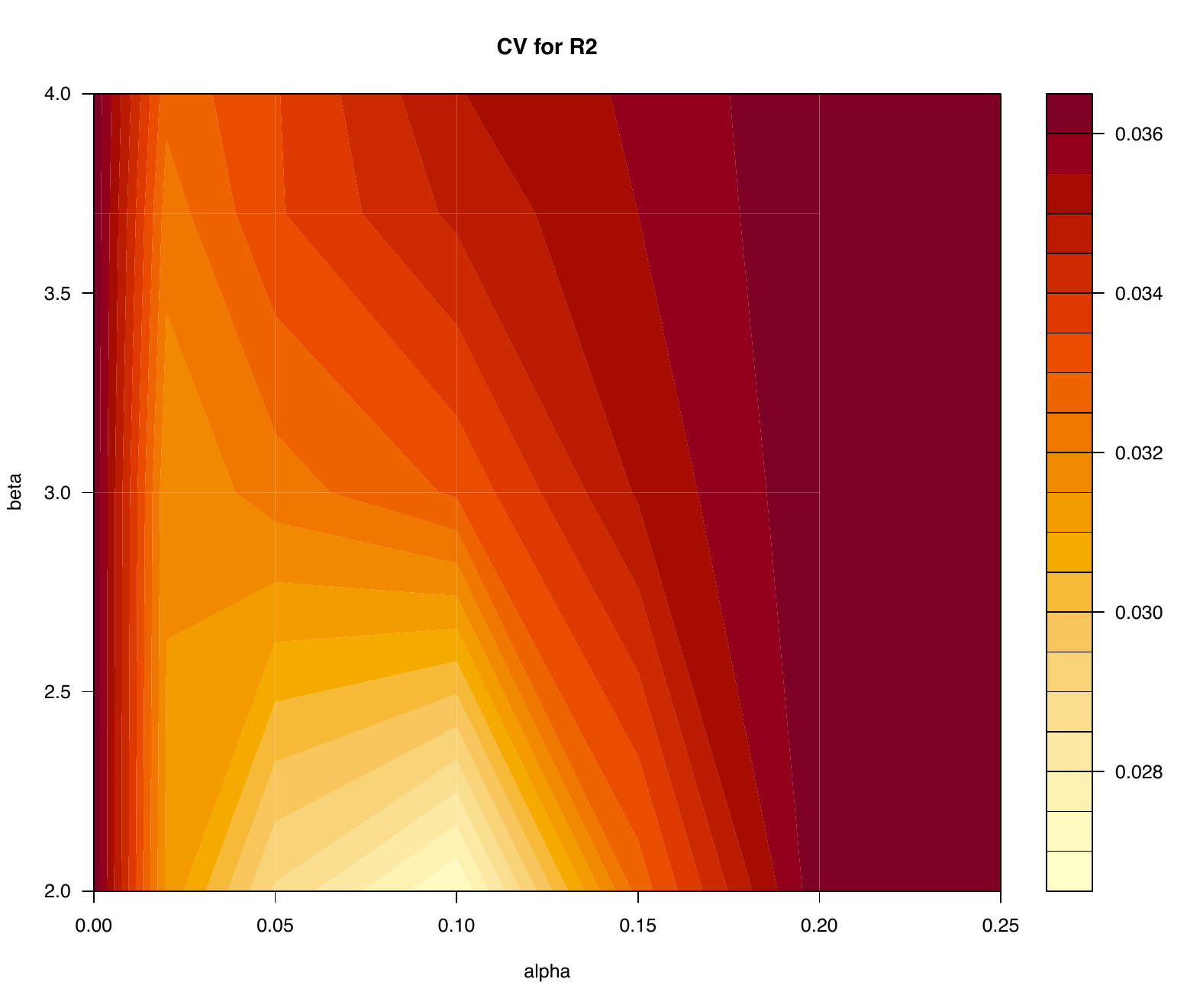} \\
    \includegraphics[width=0.6\linewidth]{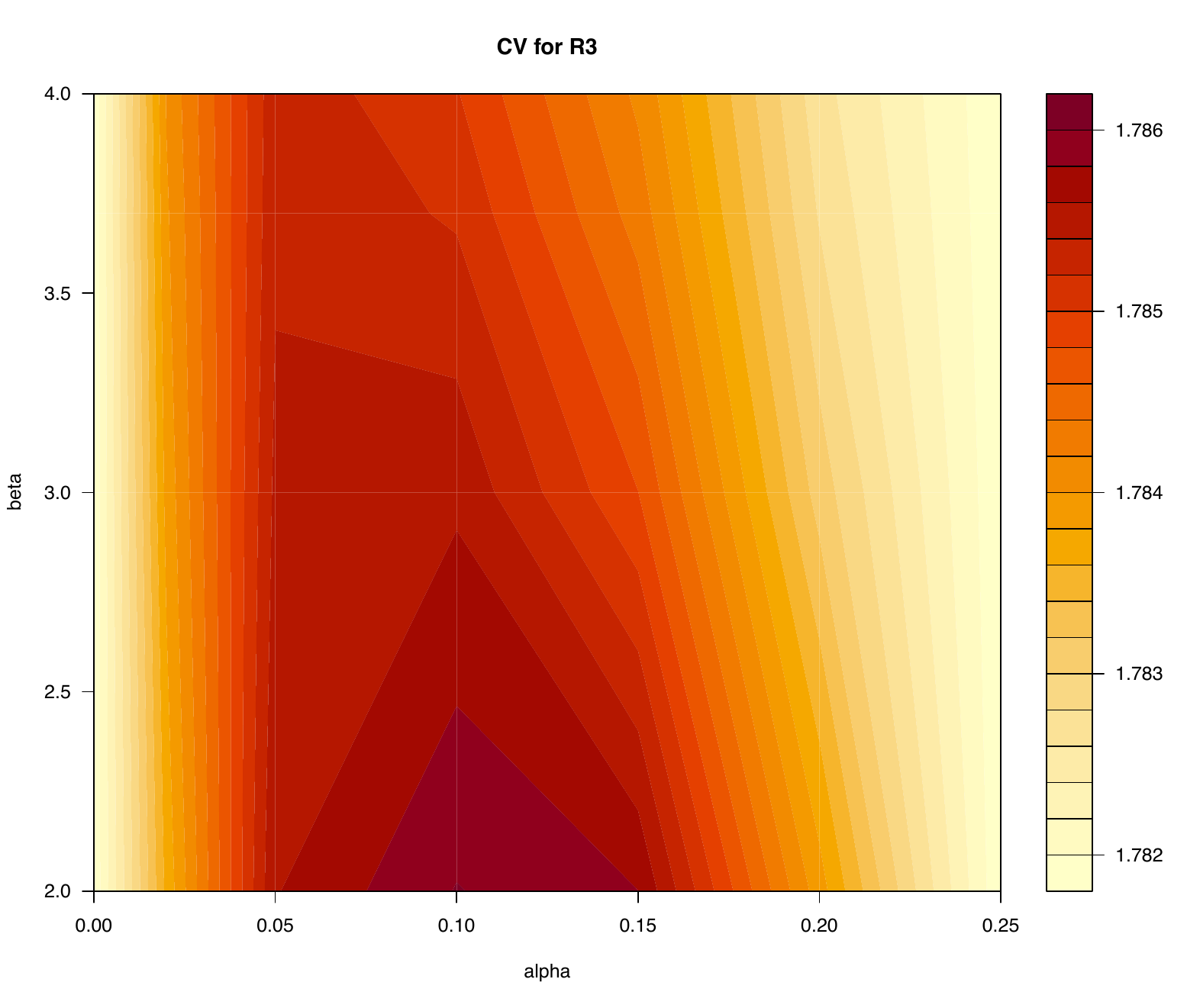}  
\end{tabular}
\end{center}
\caption{Horizontal axis is $\alpha \in \{ 0, 0.02, 0.05, \allowbreak 0.1, 0.15, 0.2, 0.25\}$; vertical axis is $\beta \in \{ 2, 3, 3.7, 4 \}$.
Contour plots of average $\overline{CV}(\alpha,\beta;m,n)$ in (\ref{cv_m}) for pairs of $(\alpha, \beta)$. }
\label{fig:cv_m}
\end{figure}

In the second simulation study, we evaluate the cross-validation method in Section \ref{sec_tuning_parameters} for the tuning parameters $(\alpha, \beta)$.  Because of the time-consuming nature of cross-validation, we consider reduced sets of $\alpha$ and $\beta$, with $\alpha \in \{ 0, 0.02, 0.05, 0.1, 0.15, 0.2, 0.25\}$ and $\beta \in \{ 2, 3, 3.7, 4 \}$.  For each $R_i$, $i=1,2,3$, we use the same $J=100$ sets of random data of size 1,000 as in the first simulation.  For each data set $j, \ j=1, \ldots J$, we calculate $CV_j(\alpha, \beta; m,n)$ by (\ref{cv}), and then we calculate the arithmetic mean of the $CV_j$, 
\begin{equation}
\overline{CV}(\alpha,\beta;m,n) = \frac{1}{J}\sum^J_{j=1} CV_j(\alpha, \beta;m,n).
\label{cv_m}
\end{equation}
The contour plots of $\overline{CV}(\alpha,\beta;m,n)$ are graphed in Figure \ref{fig:cv_m}.

Since a larger value of $\overline{CV}(\alpha,\beta;m,n)$ indicates superior model fit, 
we see that the results of Simulation Study II are consistent with the results of Simulation Study I. 
That is, in the top and bottom panels for $R_1$ and $R_3$, respectively, for moderate values of $\alpha$, such as $0.05 < \alpha < 0.2$, cross-validation results in larger values of $\overline{CV}$; as $\alpha$ approaches zero or larger than 0.2, we see that $\overline{CV}(\alpha,\beta;m,n)$ decreases.
This means that $\alpha \in (0.05, 0.2)$ provides better estimates for $R_1$ and $R_3$.
On the other hand, in the case of $R_2$, where no zero elements are contained in the parameter matrix, 
$\alpha=0$ and $\alpha > 0.2$ give larger values of average $CV$ than those in the center. 
This also suggests 
that no penalization is necessary in estimating $R_2$.
Therefore we infer from this study and the related graphs that the cross-validation method with (\ref{L*}) is effective for tuning parameter selection and is able to choose $(\alpha, \beta)$ accurately.

\subsection{Model selection (Simulation Study III)}
\label{subsec:simu_mn}

In this subsection, we use simulations to examine the performance of the cross-validation and pseudo-AIC approaches to determining the size of the parameter matrix $R$.  
For selecting $(m, n)$ we set $\alpha = 0$, for simplicity, and we also recall the monotonic increasing property of (\ref{l*r}) when $\alpha = 0$. With this choice of $\alpha$, the parameter $\beta$ becomes extraneous, so we will use the notation $CV(m,n)$ as shorthand for $CV(\alpha,\beta;m,n)$.
For each $R_i$, $i=1, 2, 3$, with $d=3$, and equally-spaced interior knots of B-spline basis functions,
we generate $J=100$ data sets, each of sample size 1,000 as before.

Using the $j$th data set, we calculate (\ref{cv}) for each pair of integers $(m,n)$; thus we obtain $CV_j(m, n)$, $j=1,\ldots, J$, and then we compute the mean and standard derivation of all $CV_j(m, n)$.
We also define $AIC_j(m,n)$, the value of the pseudo-AIC obtained by applying \eqref{aic} to the $j$th data set, $j=1,\ldots, J$, and then we calculate the mean and standard derivation of all $AIC_j(m, n)$.
The computed results for $m,n \in \{ 4,5,6,7,8\}$ are given in Tables \ref{cv1_mn}--\ref{aic3_mn}.

From the results given in these tables, we see that both the cross-validation and pseudo-AIC approaches are useful for selecting $(m,n)$.  For $R_1$ and $R_3$, with help of these methods, we easily detect the correct choices of $(m,n)$.  In the case of $R_2$, both methods prefer $(m,n)=(4,4)$; however the values of $CV$ and $AIC$ for the second-best model, $(m,n)= (4,5)$, are close to the best and are much closer than all others values of $(m,n)$.  Consequently, we 
may choose either $(m,n) = (4,4)$ or $(m,n)= (4,5)$ for $R_2$.

\begin{table}[htbp]
\begin{center}
\caption{Mean and standard deviation of $CV(m,n)$ for $R_1$}
\begin{tabular}{|c|c|c|c|c|c|c|}
\hline
$m\,\backslash\,n$ &  & 4 & 5 & 6 & 7 & 8 \\\hline  
\multirow{2}{1.5em}{4} & mean & 0.518 & \fbox{0.698} & 0.685 & 0.685 & 0.679 \\
& s.d. & 0.042 & 0.075 & 0.071 & 0.076 & 0.078\\
\hline
\multirow{2}{1.5em}{5} & mean &0.551 & 0.690 & 0.674 & 0.668 & 0.660\\
& s.d. & 0.048 & 0.075 & 0.077 & 0.077 & 0.078\\
\hline
\multirow{2}{1.5em}{6}  & mean & 0.550 & 0.679 & 0.664 & 0.655 & 0.647\\
& s.d. & 0.050 & 0.077 & 0.080 & 0.082 & 0.084\\
\hline
\multirow{2}{1.5em}{7}  & mean & 0.556 & 0.674 & 0.656 & 0.645 & 0.637\\
& s.d.  & 0.053 & 0.077 & 0.080 & 0.083 & 0.083\\
\hline
\multirow{2}{1.5em}{8}  & mean &0.551 & 0.668 & 0.646 & 0.635 & 0.623\\
& s.d. & 0.055 & 0.078 & 0.082 & 0.084 & 0.084\\
\hline
\end{tabular}
\label{cv1_mn}
\end{center}
\end{table}

\begin{table}[htbp]
\begin{center}
\caption{Mean and standard deviation of $CV(m,n)$ for $R_2$}
\begin{tabular}{|c|c|c|c|c|c|c|}
\hline
$m\,\backslash\,n$ &  & 4 & 5 & 6 & 7 & 8 \\\hline  
\multirow{2}{1.5em}{4} & mean & \fbox{$0.039$} & \fbox{$0.036$}  & $0.030$  & $0.023$  & $0.017$ \\
& s.d. & $0.028$ & $0.030$ & $0.032$ & $0.033$ & $0.033$ \\
\hline
\multirow{2}{1.5em}{5} & mean & $0.032$ & $0.024$  & $0.015$  & $0.006$ & $-0.003$ \\
& s.d. &  $0.030$ & $0.033$ & $0.036$ & $0.036$ & $0.038$ \\
\hline
\multirow{2}{1.5em}{6}  & mean &  $0.025$ & $0.014$  & $0.003$ & $-0.008$ & $-0.019$ \\
& s.d. & $0.030$ & $0.033$ & $0.035$ & $0.038$ & $0.040$ \\
\hline
\multirow{2}{1.5em}{7}  & mean & $0.020$ & $0.008$ & $-0.006$ & $-0.019$ & $-0.030$ \\
& s.d.  &  $0.032$ & $0.034$ & $0.036$ & $0.040$ & $0.041$ \\
\hline
\multirow{2}{1.5em}{8}  & mean & $0.014$ & $0.000$ & $-0.016$ & $-0.030$ & $-0.045$ \\
& s.d. & $0.036$ & $0.036$ & $0.040$ & $0.043$ & $0.045$ \\
\hline
\end{tabular}
\label{cv2_mn}
\end{center}
\end{table}


\begin{table}[htbp]
\begin{center}
\caption{Mean and standard deviation of $CV(m,n)$ for $R_3$}
\begin{tabular}{|c|c|c|c|c|c|c|}
\hline
$m\,\backslash\,n$ &  & 4 & 5 & 6 & 7 & 8 \\\hline  
\multirow{2}{1.5em}{4} & mean & 1.585 & 1.585 & 1.687 & 1.680 & 1.684\\
& s.d. &  0.081 & 0.081 & 0.102 & 0.098 & 0.103\\
\hline
\multirow{2}{1.5em}{5} & mean &1.585 & \fbox{1.781} & 1.774 & 1.772 & 1.765\\
& s.d. & 0.081 & 0.115 & 0.112 & 0.114 & 0.116\\
\hline
\multirow{2}{1.5em}{6}  & mean & 1.685 & 1.775 & 1.758 & 1.755 & 1.747\\
& s.d. & 0.099 & 0.112 & 0.114 & 0.115 & 0.115\\
\hline
\multirow{2}{1.5em}{7}  & mean & 1.678 & 1.771 & 1.754 & 1.747 & 1.738\\
& s.d.  & 0.097 & 0.115 & 0.115 & 0.116 & 0.115\\
\hline
\multirow{2}{1.5em}{8}  & mean & 1.680 & 1.764 & 1.745 & 1.735 & 1.726\\
& s.d. &  0.102 & 0.118 & 0.117 & 0.119 & 0.120\\
\hline
\end{tabular}
\label{cv3_mn}
\end{center}
\end{table}

\clearpage

\begin{table}[t]
\begin{center}
\caption{Mean and standard deviation of $AIC(m,n)$ for $R_1$}
\begin{tabular}{|c|c|c|c|c|c|c|}
\hline
$m\,\backslash\,n$ &  & 4 & 5 & 6 & 7 & 8 \\\hline  
\multirow{2}{1.5em}{4} & mean &  $-195.15$ & \fbox{$-261.39$} & $-254.19$ & $-254.17$ & $-250.97$ \\
& s.d. &  16.08 & 27.98 & 26.68 & 28.32 & 28.93\\
\hline
\multirow{2}{1.5em}{5} & mean & $-203.70$ & $-255.60$ & $-251.04$ & $-248.10$ & $-243.62$ \\
& s.d. & 18.16 & 27.63 & 28.78 & 29.00 & 29.03\\
\hline
\multirow{2}{1.5em}{6}  & mean &  $-201.15$ & $-251.18$ & $-245.44$ & $-241.23$ & $-235.69$ \\
& s.d. & 18.65 & 28.61 & 29.61 & 29.94 & 29.90\\
\hline
\multirow{2}{1.5em}{7}  & mean & $-200.61$ & $-246.12$ & $-239.36$ & $-233.97$ & $-227.18$ \\
& s.d.  &  19.67 & 28.67 & 29.21 & 29.77 & 29.80\\
\hline
\multirow{2}{1.5em}{8}  & mean &  $-196.64$ & $-240.96$ & $-232.95$ & $-226.38$ & $-218.10$ \\
& s.d. & 19.90 & 28.60 & 29.14 & 29.36 & 29.32\\
\hline
\end{tabular}
\label{aic1_mn}
\end{center}
\end{table}

\begin{table}[h]
\begin{center}
\caption{Mean and standard deviation of $AIC(m,n)$ for $R_2$}
\begin{tabular}{|c|c|c|c|c|c|c|}
\hline
$m\,\backslash\,n$ &  & 4 & 5 & 6 & 7 & 8 \\\hline  
\multirow{2}{1.5em}{4} & mean & \fbox{$-12.43$} & \fbox{$-11.38$} & $-8.13$ & $-4.93$ & $-1.39$ \\
& s.d. &  8.82 & 9.71 & 10.09 & 10.30 & 10.28\\
\hline
\multirow{2}{1.5em}{5} & mean &$-9.61$  & $-7.04$ & $-2.99$ & $1.50$ &  $5.91$ \\
& s.d. & 9.34 & 10.11 & 10.51 & 10.79 & 10.82\\
\hline
\multirow{2}{1.5em}{6}  & mean & $-6.27$ & $-2.69$ & $2.51$ & $8.19$ & $13.74$ \\
& s.d. & 9.54 & 10.29 & 10.79 & 11.19 & 11.50\\
\hline
\multirow{2}{1.5em}{7}  & mean &  $-2.87$  & $1.63$ & $8.04$ & $14.86$ & $21.46$ \\
& s.d.  &  9.84 & 10.50 & 10.84 & 11.29 & 11.43\\
\hline
\multirow{2}{1.5em}{8}  & mean &  $0.54$  & $6.19$ & $13.81$ & $21.88$ & $29.94$ \\
& s.d. &  10.34 & 10.86 & 11.65 & 11.97 & 12.40\\
\hline
\end{tabular}
\label{aic2_mn}
\end{center}
\end{table}

\begin{table}[h]
\begin{center}
\caption{Mean and standard deviation of $AIC(m,n)$ for $R_3$}
\begin{tabular}{|c|c|c|c|c|c|c|}
\hline
$m\,\backslash\,n$ &  & 4 & 5 & 6 & 7 & 8 \\\hline  
\multirow{2}{1.5em}{4} & mean &  $-615.98$ & $-610.00$ & $-647.05$ & $-638.55$ &  $-638.62$ \\
& s.d. &  32.40 & 32.40 & 40.05 & 38.44 & 40.32\\
\hline
\multirow{2}{1.5em}{5} & mean &  $-610.01$ & \fbox{$-686.51$}&  $-676.60$ & $-674.80$ & $-669.65$ \\
& s.d. & 32.40 & 44.68 & 43.24 & 44.69 & 44.88\\
\hline
\multirow{2}{1.5em}{6}  & mean & $-646.23$ & $-676.93$ & $-671.97$ & $-666.42$ & $-659.84$ \\
& s.d. &  39.14 & 43.39 & 45.02 & 45.21 & 45.04\\
\hline
\multirow{2}{1.5em}{7}  & mean &  $-638.16$ & $-674.47$ & $-666.53$ & $-659.31$ & $-651.77$ \\
& s.d.  &  37.94 & 44.82 & 45.18 & 45.31 & 45.33\\
\hline
\multirow{2}{1.5em}{8}  & mean &  $-637.29$ & $-669.25$ & $-659.59$ & $-651.06$ & $-642.17$ \\
& s.d. &   39.62 & 45.65 & 45.56 & 46.05 & 46.09\\
\hline
\end{tabular}
\label{aic3_mn}
\end{center}
\end{table}

\clearpage

\subsection{Comparison of the B-spline and Bernstein copulas (Simulation Study IV)}
\label{subsec_5}

Using a simulated 3-dimensional data set, we compare the B-spline and Bernstein copulas.  The data are generated in the same way as in \cite{Dou-etal16}, Section 3.3.  That is, we first generate $(u_{1,i}, u_{2,i}, u_{3,i}), i=1, \ldots, N$, from a trivariate Baker distribution with copula density 
\begin{equation}
c(u_1, u_2, u_3)=n_1 n_2 n_3 \sum^{n_1}_{k_1=1}\sum^{n_2}_{k_2=1}\sum^{n_3}_{k_3=1} r_{k_1, k_2, k_3} \prod^3_{j=1} b_{k_j-1, n_j-1}(u_j),
\label{c_3d}
\end{equation}
where 
\[
b_{k,n}(u)= \dbinom{n}{k}u^k(1-u)^{n-k}, \quad u \in [0,1],
\]
and the parameter $R=(r_{k_1, k_2, k_3})$ is designed as
\[
r_{k_1, k_2, 1}= \dfrac{1}{2n_1 n_2} \ \  (\mbox{for \,  all\ }k_1, k_2 ), \quad r_{k_1, k_2, 2}=
\begin{cases}
\dfrac{1}{2n_1}, & (\mbox{if}\ k_1=k_2),\\
0, &  (\mbox{if}\ k_1 \neq k_2),
\end{cases}
\]
with $(n_1,n_2,n_3) = (20,20,2)$.

Next, we use $\Phi^{-1}(\cdot)$, the quantile function of the standard normal distribution, to define 
\[
x_i= \Phi^{-1}(u_{1,i}), \quad y_i= \Phi^{-1}(u_{2,i}), \quad z_i= \Phi^{-1}(u_{3,i}), \quad i=1, \ldots, N.
\]
This converts each uniform marginal distributions of $(x_i, y_i, z_i)$ to $N(0,1)$, the standard normal distribution.   

\begin{figure}[htbp]
\begin{center}
\begin{tabular}{ccc}
    \includegraphics[width=0.38\linewidth]{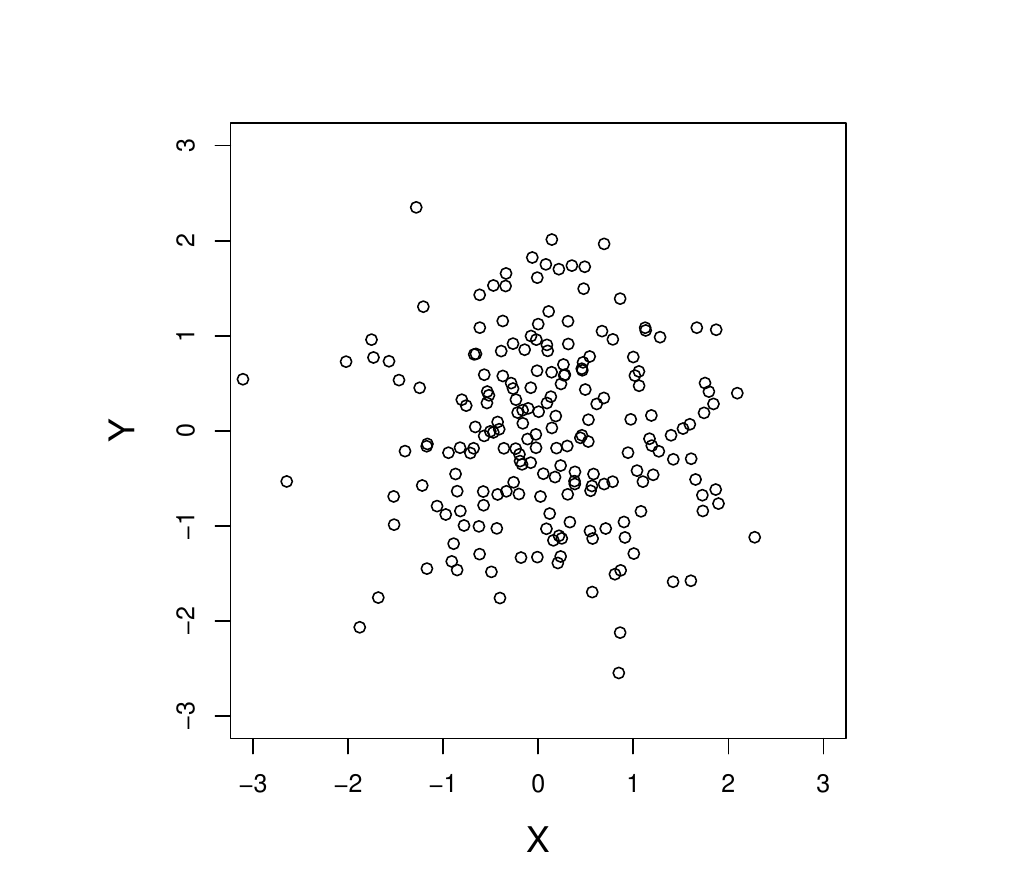}  &  \hspace{-12mm}
      \includegraphics[width=0.38\linewidth]{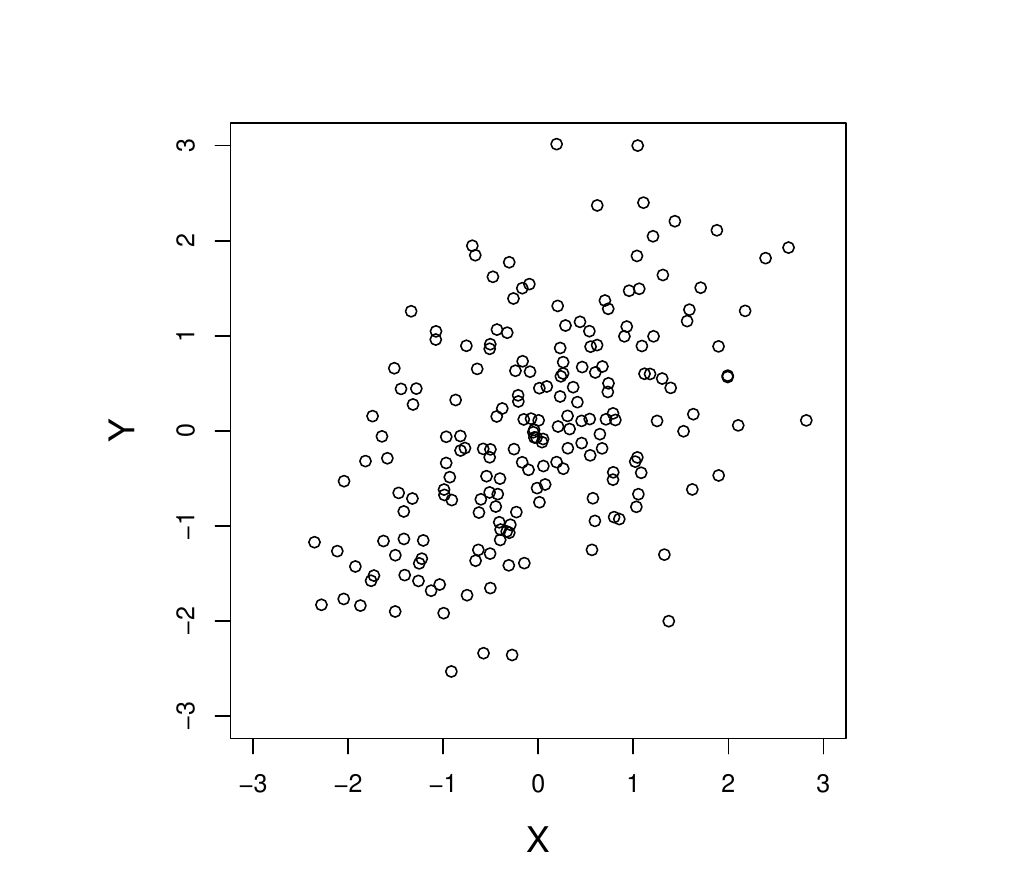}  &  \hspace{-12mm}
      \includegraphics[width=0.38\linewidth]{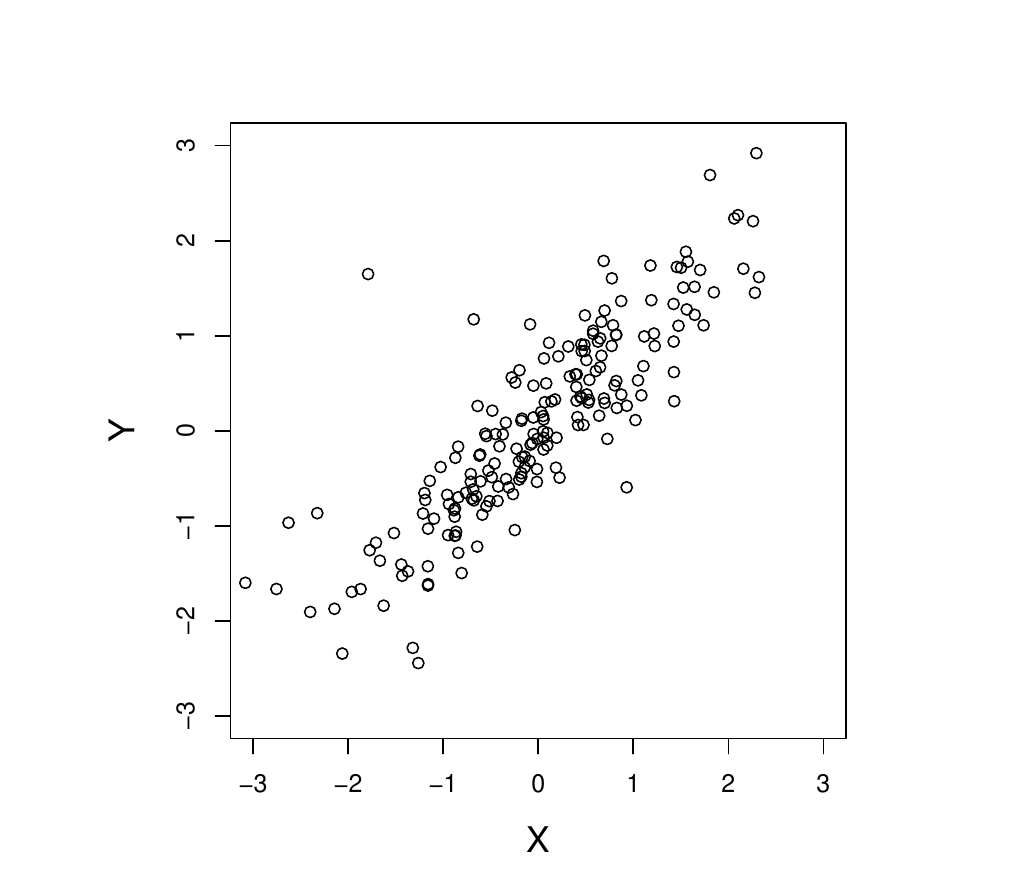} \\
    \includegraphics[width=0.32\linewidth]{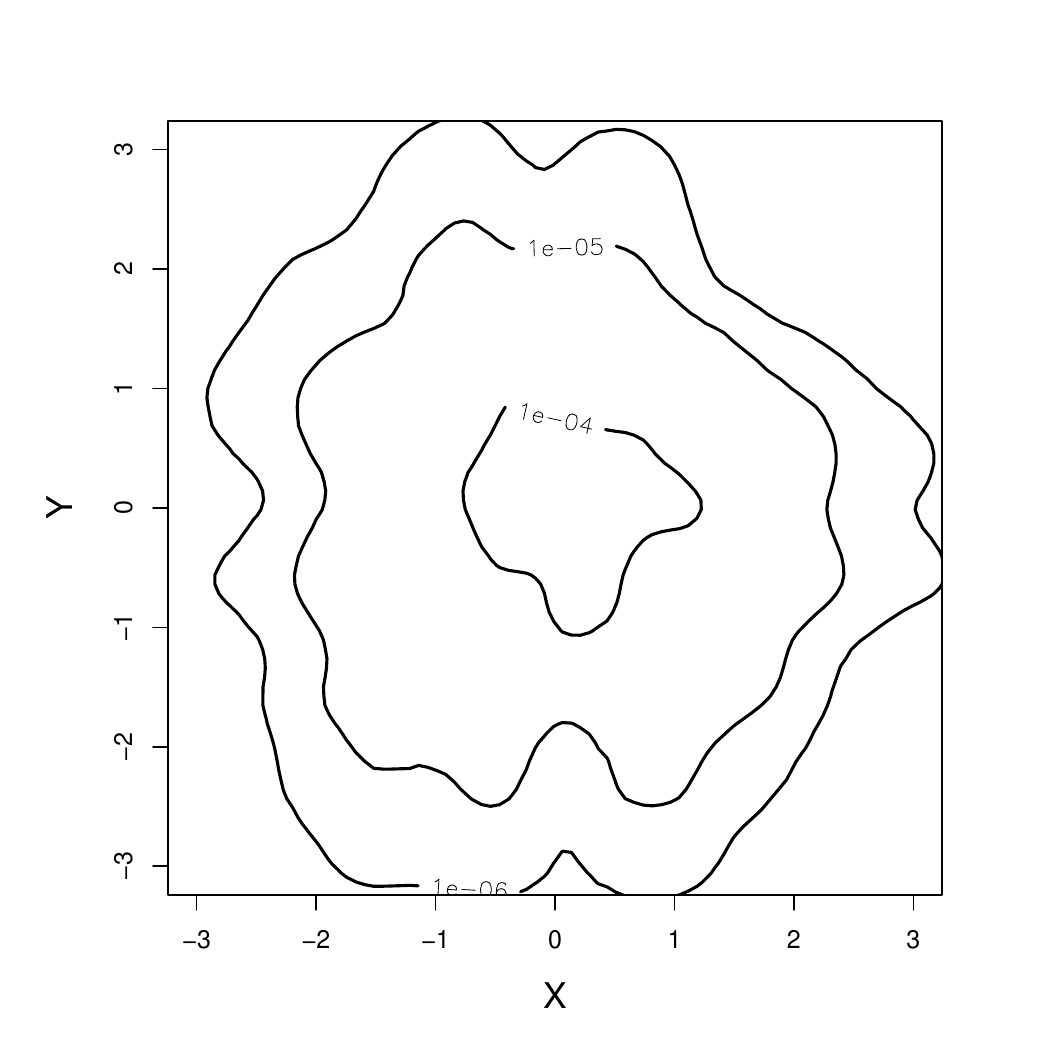}  &  \hspace{-12mm}
      \includegraphics[width=0.32\linewidth]{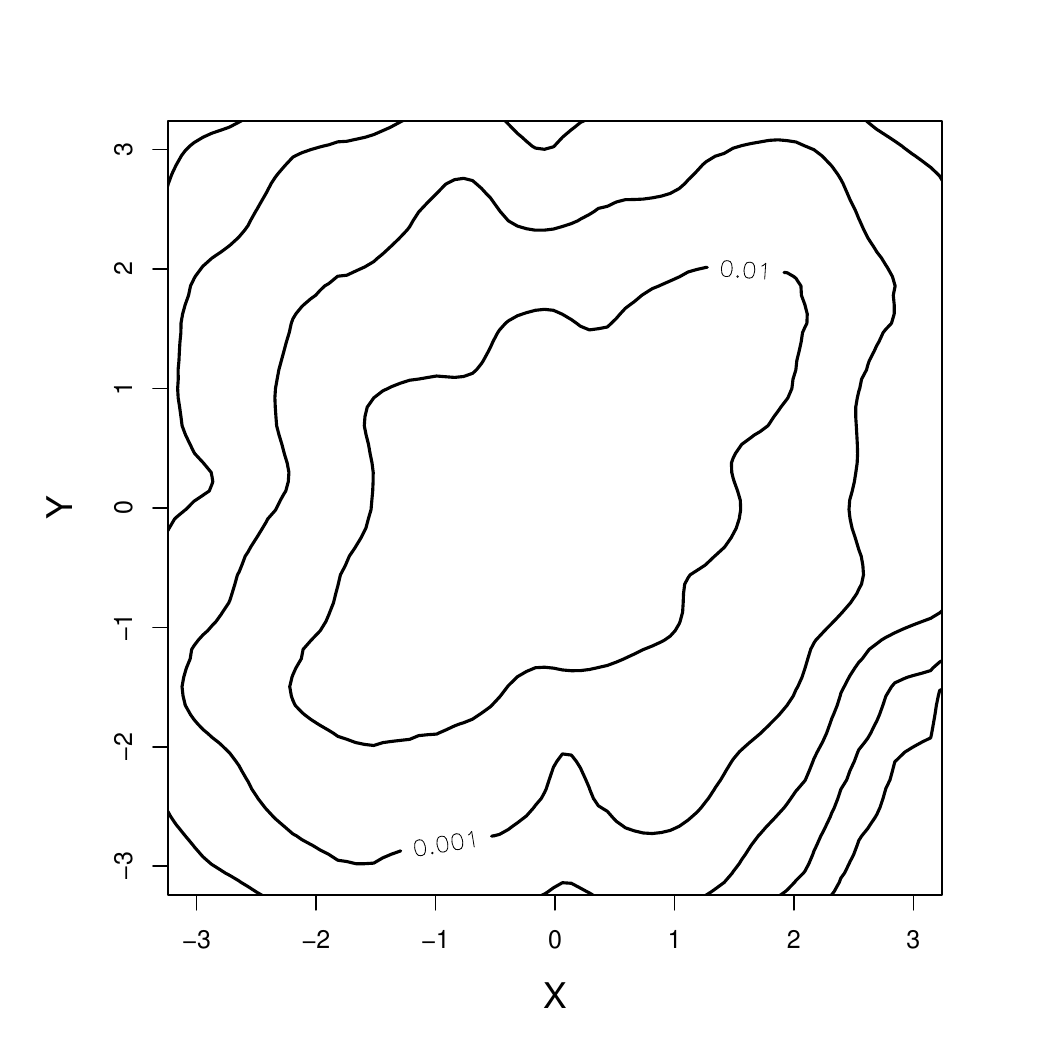}  &  \hspace{-12mm}
      \includegraphics[width=0.32\linewidth]{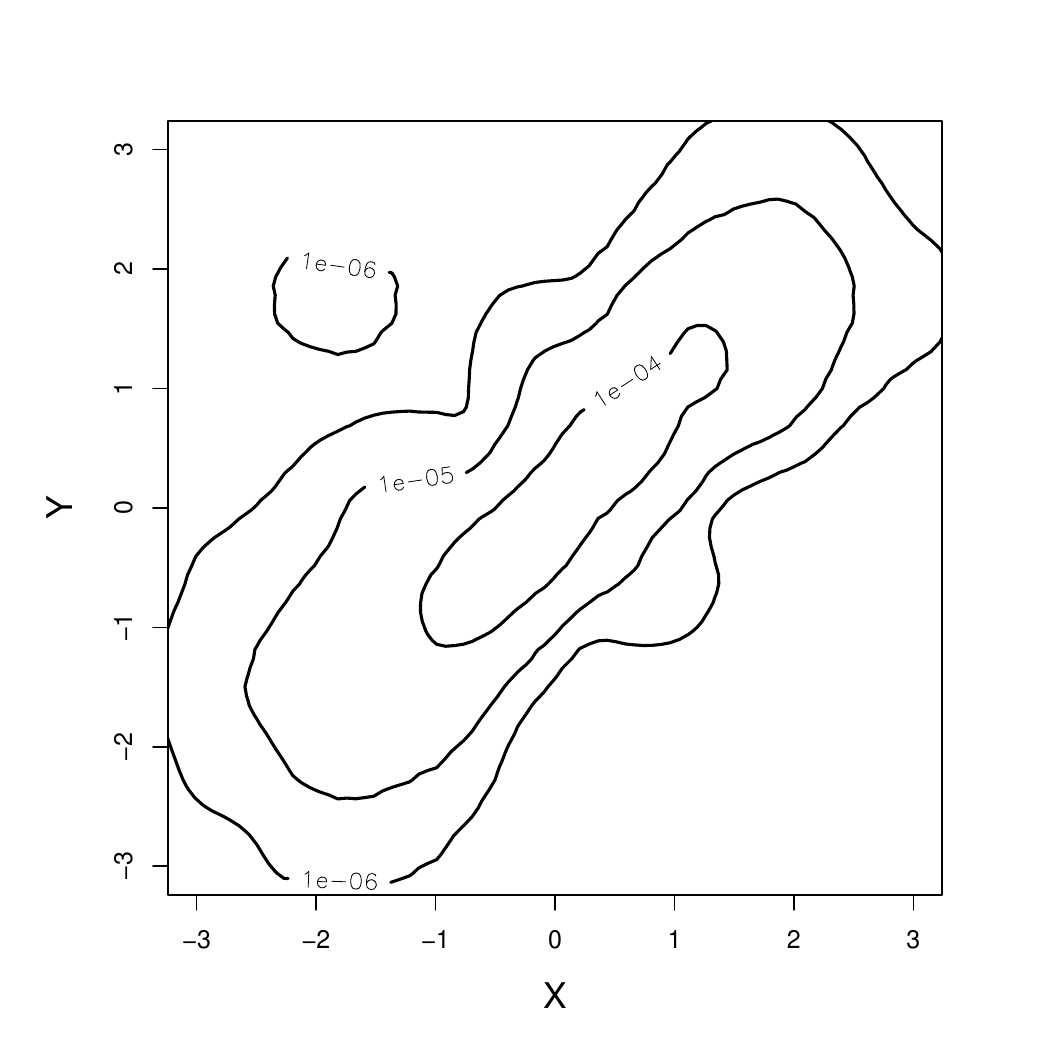} \\ 
    \includegraphics[width=0.32\linewidth]{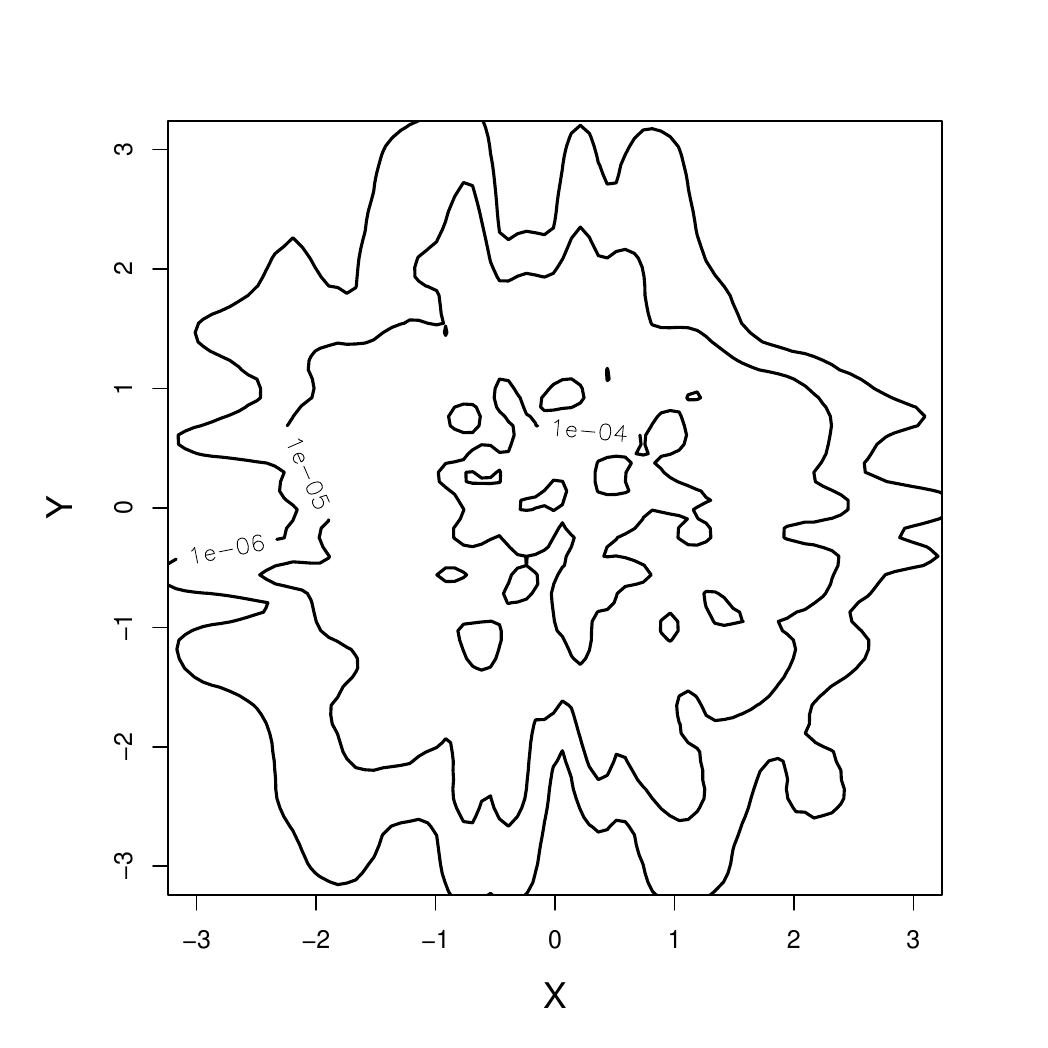}  & \hspace{-12mm}
      \includegraphics[width=0.32\linewidth]{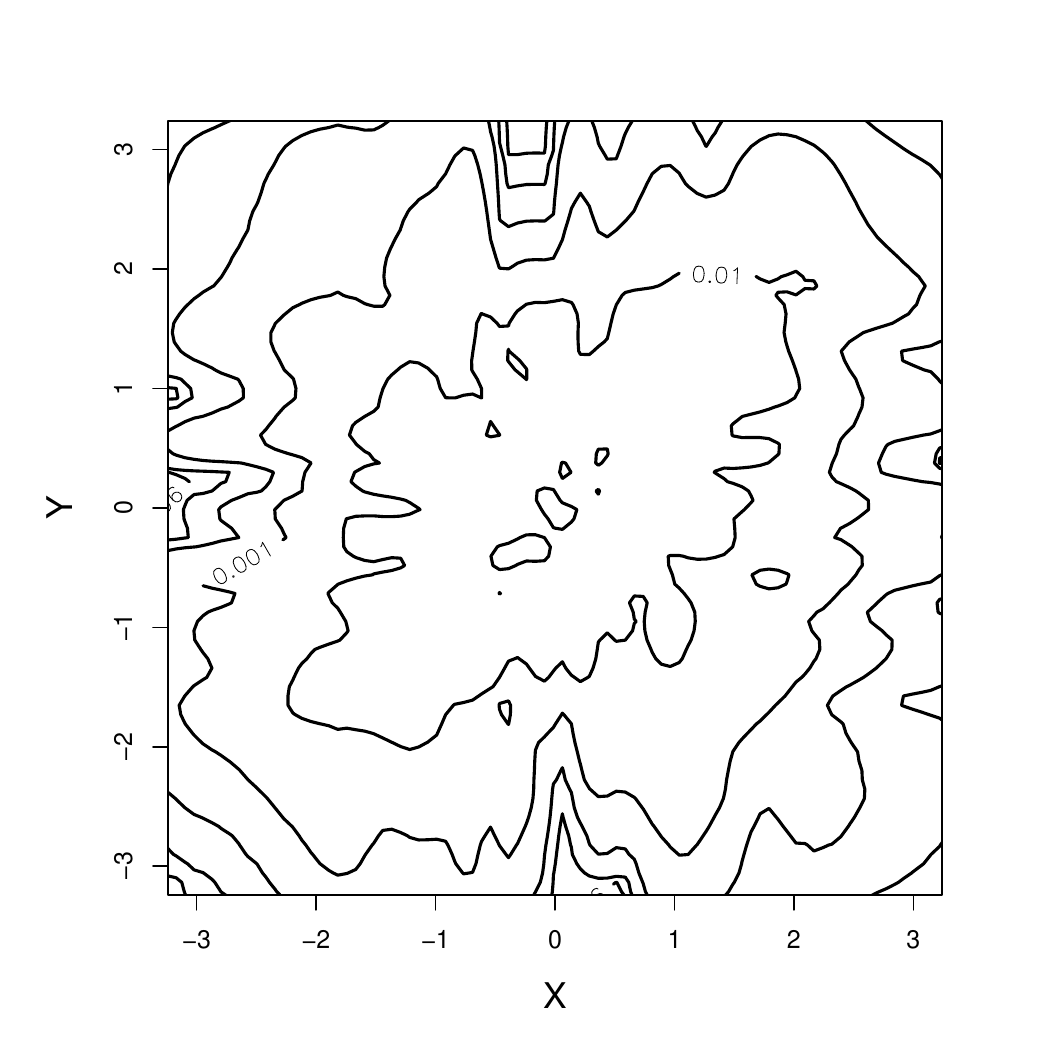}  & \hspace{-12mm}
      \includegraphics[width=0.32\linewidth]{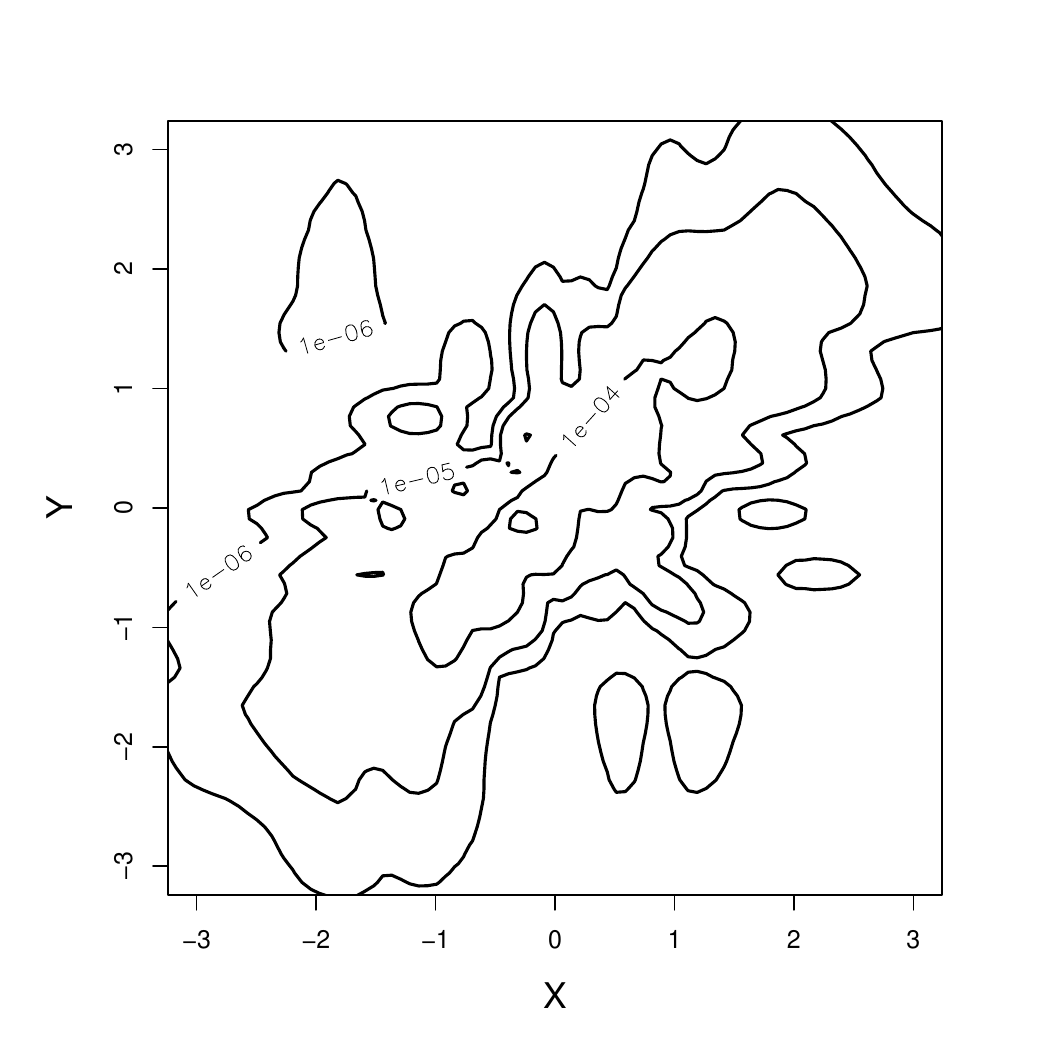} \\
          \includegraphics[width=0.32\linewidth]{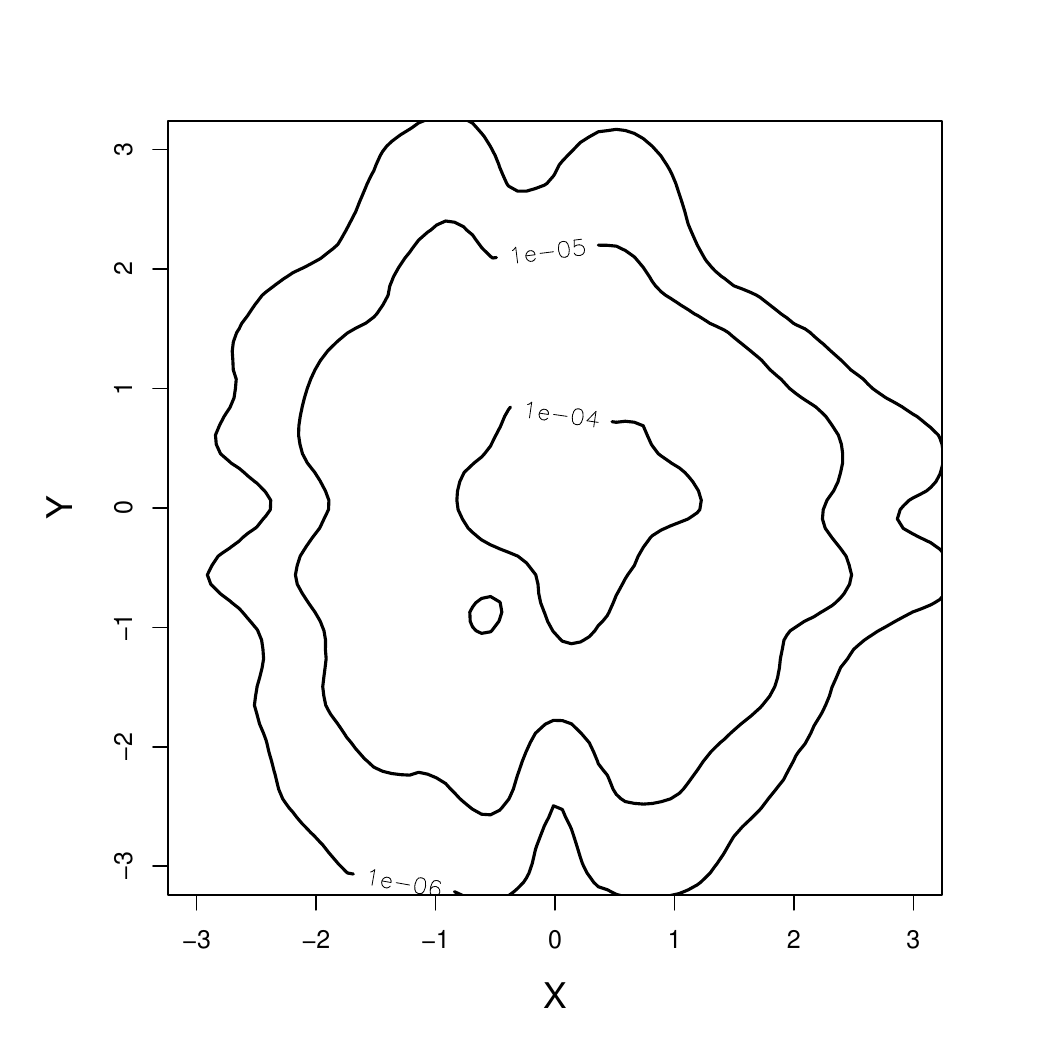}  & \hspace{-12mm}
      \includegraphics[width=0.32\linewidth]{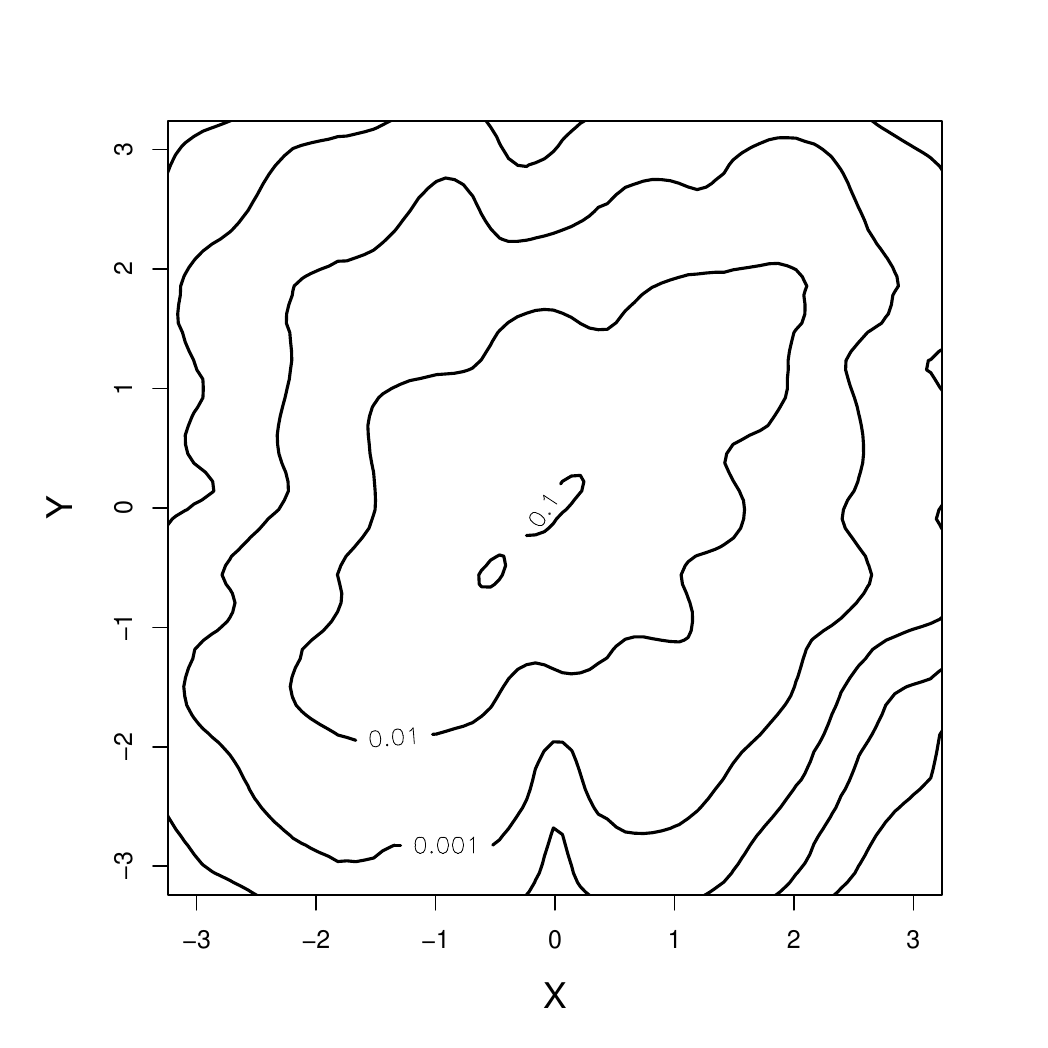}  & \hspace{-12mm}
      \includegraphics[width=0.32\linewidth]{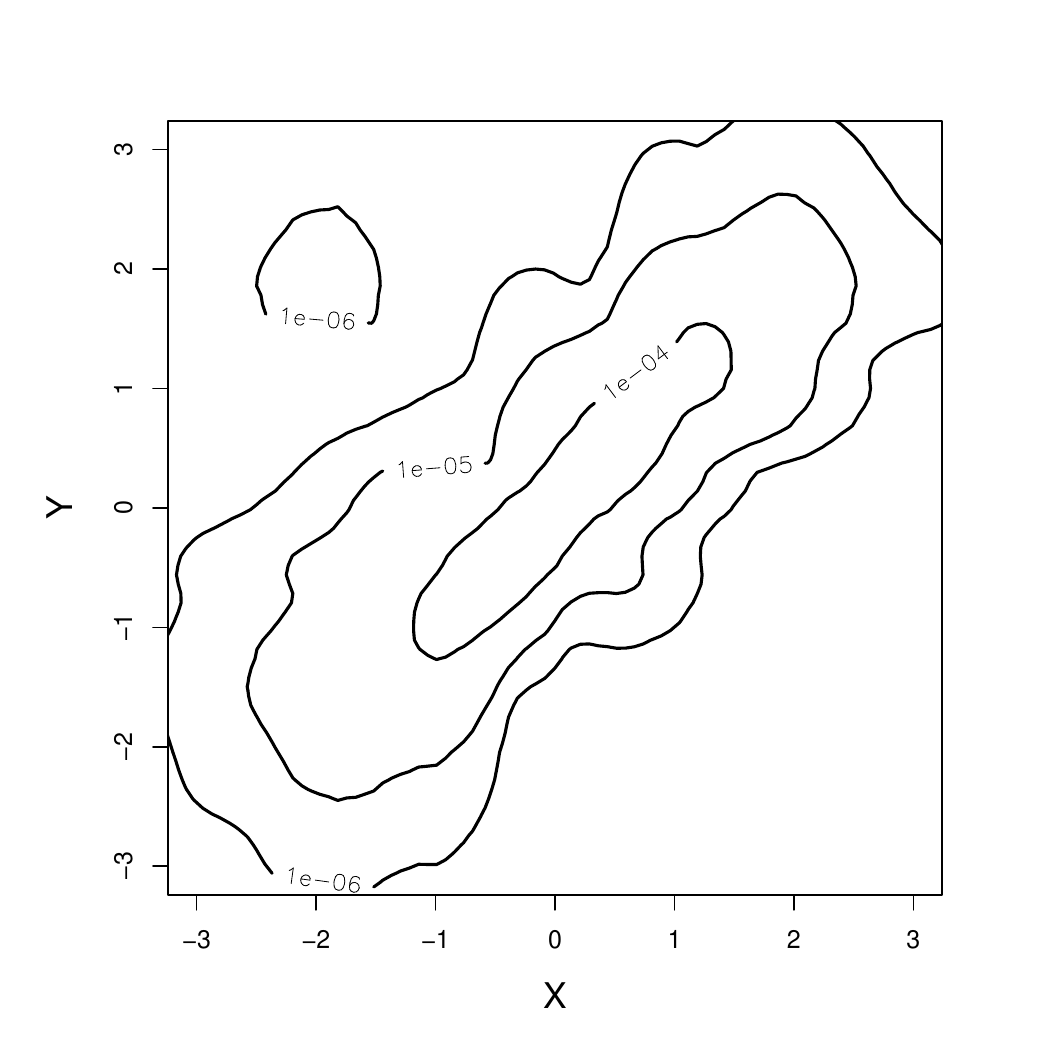} \\
\end{tabular}
\end{center}
\caption{
Scatterplots for the stratified data (first row), and their estimated contour plots using the Bernstein copula with parameter size $20 \times 20 \times 2$ (second row), the B-spline copula with parameter size $20 \times 20 \times 2$ (third row), and the B-spline copula with parameter size $10 \times 10 \times 2$ (fourth row). In all cases, the SCAD penalty with tuning parameters ($\alpha=0.01, \beta=2.25$) is used.}
\label{fig:simu3dp}
\end{figure}

The sample size of the simulated trivariate data set is chosen to be $N=2,000$.  In the first row of Figure \ref{fig:simu3dp}, we provide scatterplots of $X$ vs. $Y$ for small values of $Z$, $Z \le \Phi^{-1}(0.1)$, (in column 1), moderate values of $Z$, $\Phi^{-1}(0.45) \le Z \le \Phi^{-1}(0.55)$, (in column 2), and large values of $Z$, $Z \ge \Phi^{-1}(0.9)$, (in column 3).  It is evident that the correlation between $X$ and $Y$ increases with increasing values of $Z$.

To construct Figure \ref{fig:simu3dp}, we use the EM algorithm with the SCAD penalty and tuning parameters ($\alpha=0.01, \beta=2.25$).  
With parameter size $20 \times 20 \times 2$, the contour plots of the joint density estimated by the Bernstein copula are shown in the second row of Figure \ref{fig:simu3dp}.
The results of using B-spline copulas with parameter size $20 \times 20 \times 2$, and $10 \times 10 \times 2$ are given in the third row and fourth row of Figure \ref{fig:simu3dp}, respectively.
In the B-spline copulas, the B-spline functions are defined with equally spaced interior knots, and the degrees of the B-spline functions are set as $d_1=3, d_2=3$, and $d_3=1$ for $u_1, u_2, $ and $u_3$, respectively.

We see that with parameter size $20 \times 20 \times 2$, the Bernstein copula provides a good fit for the data.  With similarly-sized parameters, the B-spline copula returns more detailed contour plots. With a smaller size ($10 \times 10 \times 2$) of parameter, the results of the B-spline copula are as good as those obtained by the Bernstein copula. This implies that the B-spline copula, even with fewer parameters, can provide good estimates for the joint density function.

With regard to the simulations depicted in Figure \ref{fig:simu3dp}, a reviewer noted that when the performances of the B-spline and the Bernstein copulas are compared then there arises an overfitting problem when the B-spline copula with parameter size $20 \times 20 \times 2$ is used.  We believe that, for sample sizes such as $N = $2,000 and when the Bernstein copula has a large parameter size, such as $20 \times 20 \times 2$, overfitting is likely to arise since the B-spline copula requires the use of interior knots to define the B-spline functions, whereas the Bernstein copula does not require any such knots.

Although it is tempting to compare the results of the various estimated models visually, as given in Table \ref{fig:simu3dp}, it is better to use their (estimated) mean-square error for such comparisons.  To that end, we calculated the estimated mean-square errors as follows.  

For the simulated data $(x_i, y_i, z_i), i=1,\ldots, 2000$, we estimate the corresponding joint density function $\widehat{h}(x_i,y_i,z_i)$ using the three models, {\it viz.}, the Bernstein copula and the B-spline copula each with parameter size $20 \times 20 \times 2$, and the B-spline copula with parameter size $10 \times 10 \times 2$.  In each case, the density estimator is 
\[
\widehat{h}(x_i, y_i,z_i)= \widehat{c} \big( \widehat{F}_X(x_i), \widehat{F}_Y(y_i), \widehat{F}_Z(z_i); \widehat{R} \big) \widehat{f}_X(x_i) \widehat{f}_Y(y_i) \widehat{f}_Z(z_i),
\]
where the marginal density functions are estimated by the kernel method.  Using (\ref{c_3d}), the true values of the joint density function are given by 
\[
h(x_i, y_i,z_i)= c \big( \Phi(x_i), \Phi(y_i), \Phi(z_i); R \big) \, \phi(x_i) \, \phi(y_i) \, \phi(z_i),
\]
where $ \Phi(\cdot)$ and $ \phi(\cdot)$ are, respectively, the cumulative distribution function and the probability density function of the standard normal distribution, $N(0,1)$.  Then the mean squared error, 
\[
\mathrm{MSE}(\widehat{h}) = \frac{1}{N} \sum^N_{i=1} \big[\widehat{h}(x_i,y_i,z_i) - h(x_i,y_i,z_i)\big]^2
\]
can be calculated, and its values are displayed in Table \ref{table.mse}.

\def\Topspc{\rule{0pt}{10pt}}
\def\Btmspc{\rule[-6pt]{0pt}{0pt}}
\begin{table}[t]
\begin{center}
\caption{MSE comparisons of the joint densities in Simulation Study IV}
\begin{tabular}{|l|c|c|c|}
\hline 
\Topspc \Btmspc Copula & Size of $R$ & Number of interior knots & $\mathrm{MSE}(\widehat{h})$ \\ 
\hline 
\Topspc \Btmspc Bernstein & $20 \times 20 \times 2$ & $(0,0,0)$   & $3.872255e-05$ \\ 
\hline 
\Topspc \Btmspc B-spline  & $20 \times 20 \times 2$ & $(16,16,0)$ & $9.13041e-05$ \\ 
\hline 
\Topspc \Btmspc B-spline  & $10 \times 10 \times 2$ & $(6,6,0)$   & $5.279504e-05$ \\ 
\hline
\end{tabular}
\label{table.mse}
\end{center}
\end{table}

\section{An illustrative example}
\label{sec_example}

\begin{figure}[htbp]
\begin{center}
\includegraphics[width=0.56\linewidth]{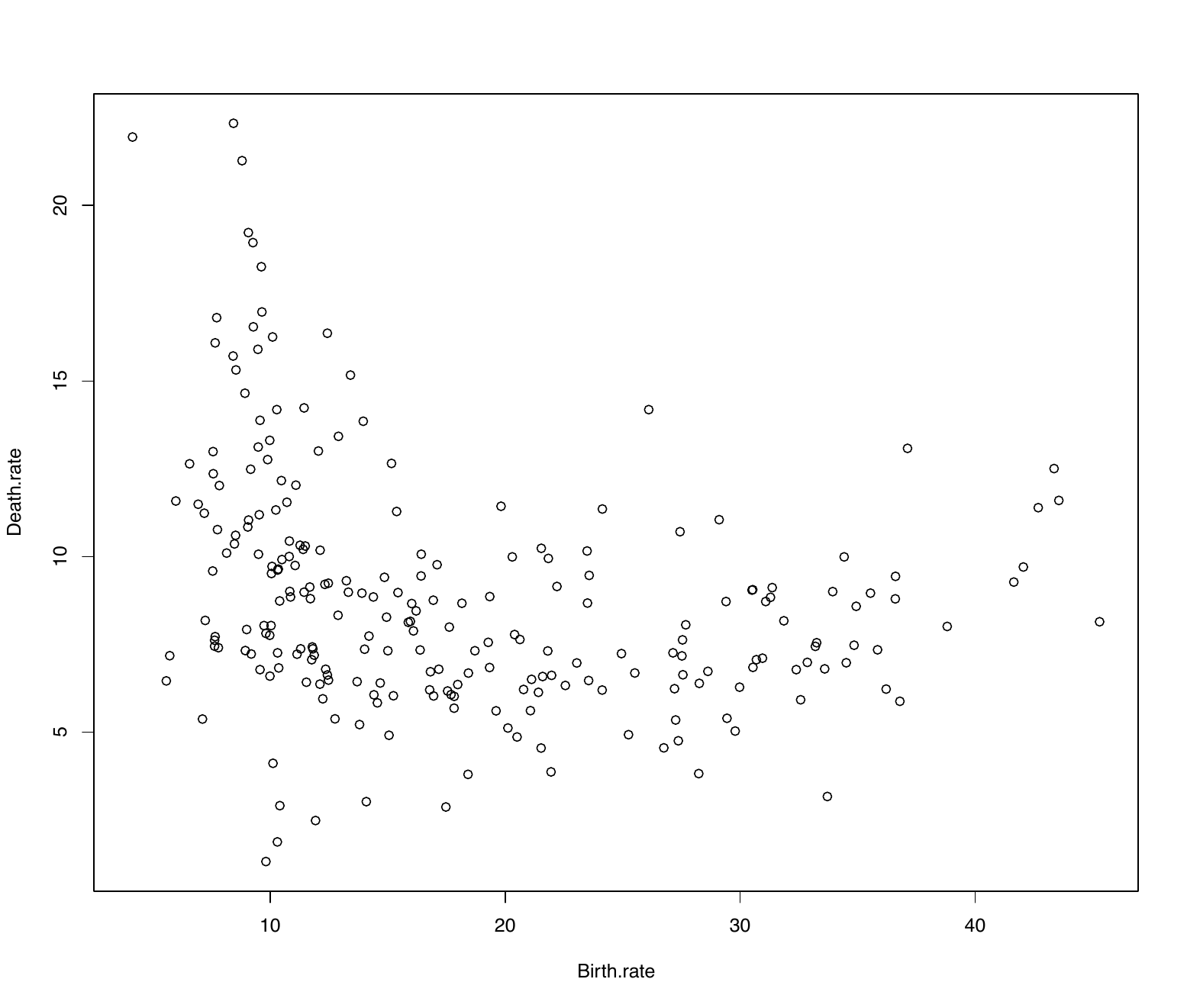} 
\end{center}
\caption{Scatterplot of the data set, birth and death rate.}
\label{fig:bdr}
\end{figure}

\begin{figure}[htbp]
\begin{center}
\begin{tabular}{cc}
\includegraphics[width=0.48\linewidth]{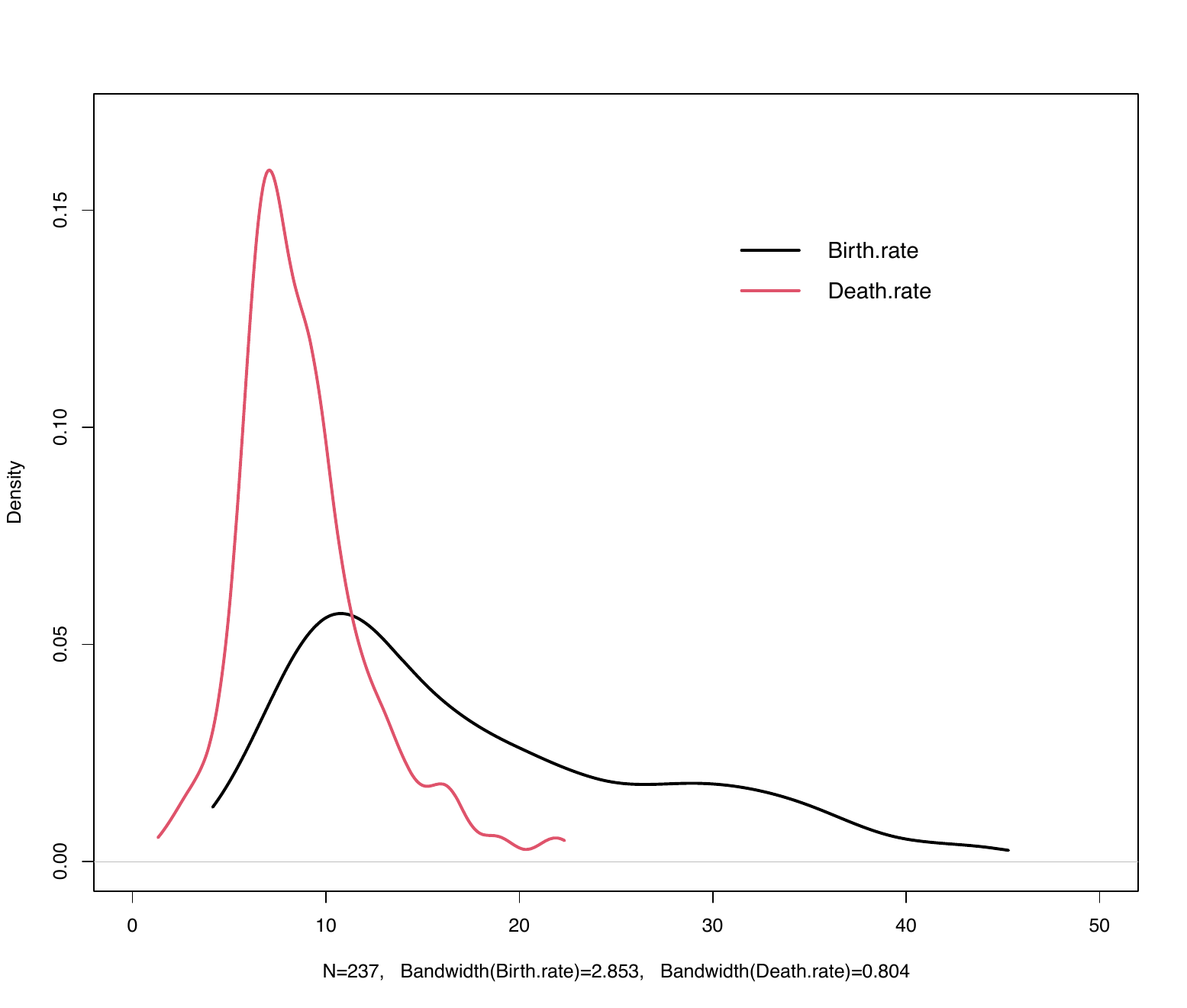} &
  \includegraphics[width=0.48\linewidth]{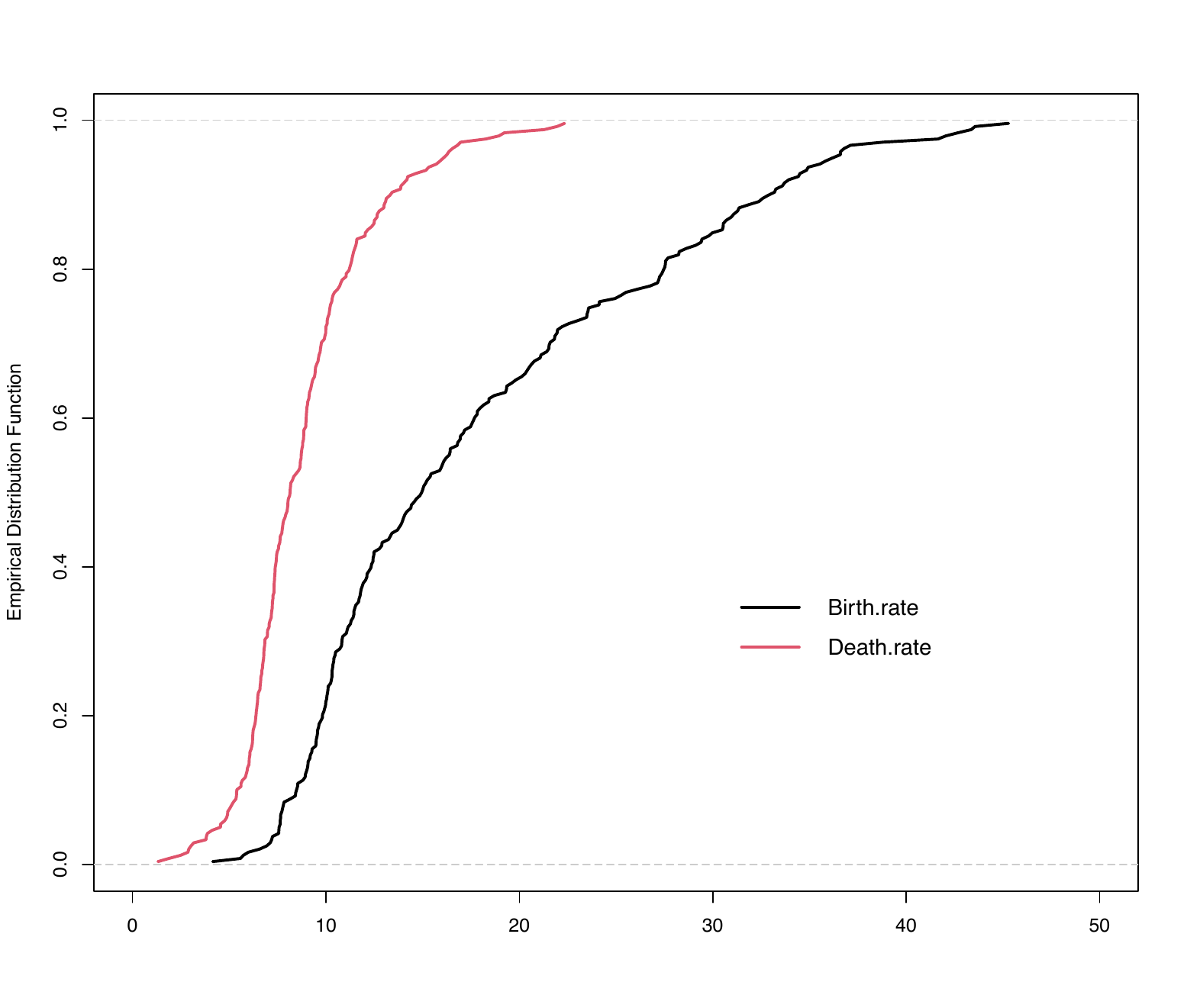} 
\end{tabular}
\end{center}
\caption{Marginal density functions and empirical cumulative distribution functions of the birth and death rates.}
\label{fig:pdf_cdf}
\end{figure}

This section presents an application of the proposed methods using birth and death rate data available at the website of ``\href{https://ourworldindata.org/}{Our World in Data}.'' 
The data for year 2021 in Figure \ref{fig:bdr} pertain to 237 countries, and both the birth and death rates are given per 1,000 people of each country's population.  The marginal densities and distribution functions of the birth and death rates are estimated by the kernel method and by the empirical cumulative distribution function, and the estimates are graphed in Figure \ref{fig:pdf_cdf}.

\begin{table}[htbp]
\begin{center}
\caption{AIC values for $(m,n)$}
\begin{tabular}{|c|c|c|c|c|c|}
\hline  
$m\,\backslash\,n$ & 4 & 5 & 6 & 7 & 8 \\\hline  
4 & $-39.77$ & $-36.99$ & $-34.94$ & $-32.16$ & $-28.76$ \\
5 & \fbox{$-40.30$} & $-39.15$ & $-34.61$ & $-28.51$ & $-26.26$ \\
6 & $-40.00$ & $-34.63$ & $-29.02$ & $-22.00$ & $-16.96$ \\
7 & $-36.92$ & $-31.62$ & $-24.33$ & $-16.56$ & $-11.68$ \\
8 & $-32.69$ & $-25.74$ & $-16.97$ &  $-7.48$ & $-2.69$ \\
\hline
\end{tabular}
\label{tab:aic}
\end{center}
\end{table}

\begin{table}[htbp]
\begin{center}
\caption{CV values for $(m,n)$}
\begin{tabular}{|c|c|c|c|c|c|}
\hline  
$m\,\backslash\,n$ & 4 & 5 & 6 & 7 & 8 \\\hline  
4 & 0.457 & 0.445 & 0.447 & 0.440 & 0.435 \\
5 & \fbox{0.525} & 0.521 & 0.496 & 0.451 & 0.469 \\
6 & 0.501 & 0.480 & 0.419 & 0.380 & 0.408 \\
7 & \fbox{0.526} & 0.442 & 0.404 & 0.332 & 0.449 \\
8 & 0.508 & 0.431 & 0.385 & 0.272 & 0.386 \\
\hline
\end{tabular}
\label{tab:cv}
\end{center}
\end{table}

By calculating the pseudo-AIC and carrying out cross-validation for the size $(m,n)$ of the parameter matrix, we obtain the results in Tables \ref{tab:aic} and \ref{tab:cv}.  We observe that the pseudo-AIC attains its minimum at $(m,n)= (5,4)$.  However the cross-validation method causes us to hesitate because it provides two competitive larger values at $(m,n)= (7,4)$ and $(m,n)= (5,4)$.  

\begin{figure}[t!]
\begin{center}
 \includegraphics[width=0.56\linewidth]{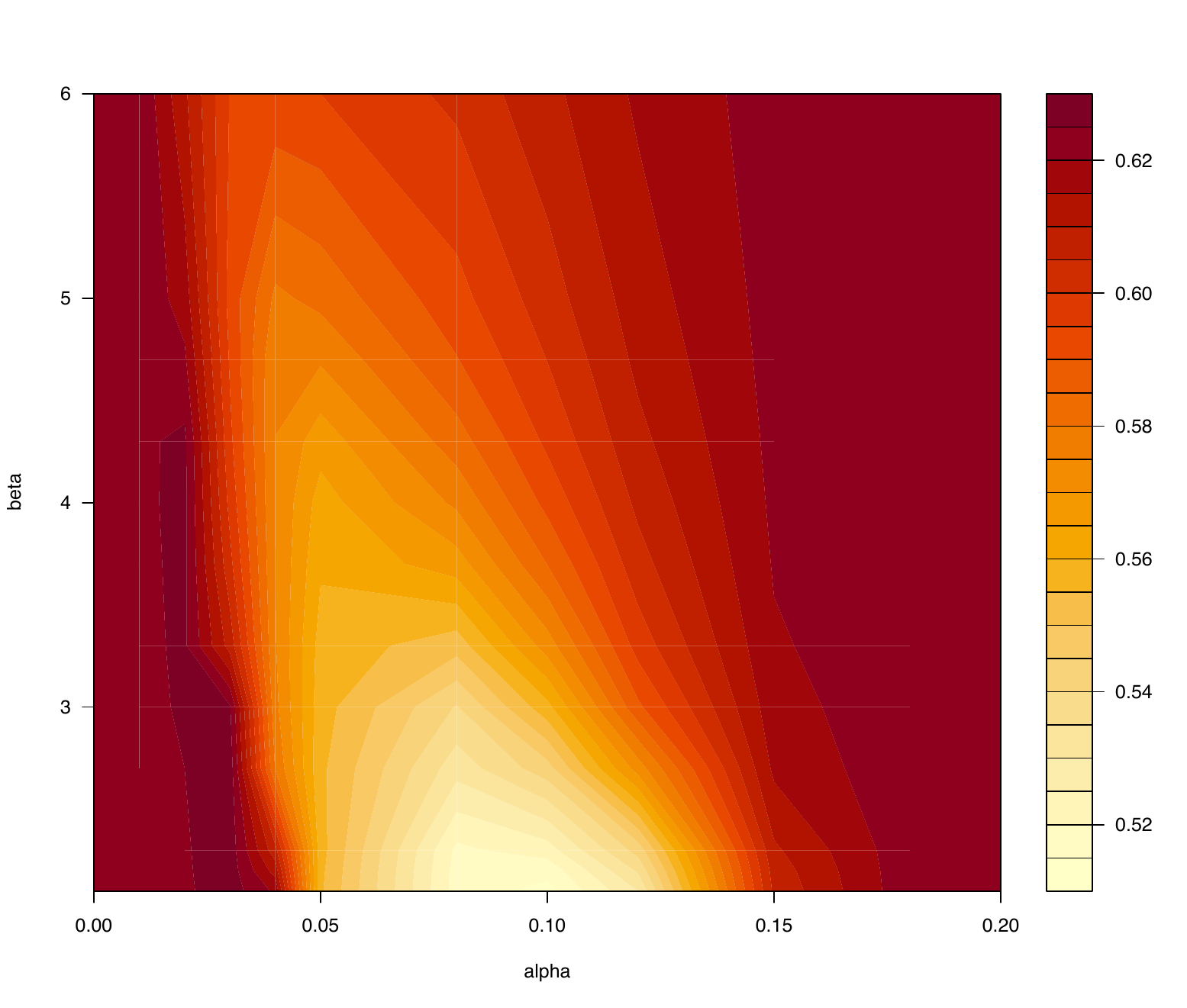} 
\end{center}
\caption{$CV(\alpha, \beta;m=5,n=4)$ of a 5-fold cross-validation for the birth and death rate data.}
\label{fig:cv2_bd}
\end{figure}

Thus, let us consider the case $(m,n)=(5,4)$.  We depict in Figure \ref{fig:cv2_bd} the results of a five-fold cross-validation study to choose the tuning parameters, i.e., with $M=5$ in (\ref{cv}); from that study, we 
find that $(\widehat{\alpha}, \widehat{\beta})= (0.03, 2.7)$ from the combinations of $(\alpha, \beta)$ where 
\begin{align*}
\alpha &\in \{ 0, 0.01, 0.02, 0.03, 0.04, 0.05, 0.08, 0.1, 0.12, 0.15, 0.18, 0.20 \} \\
\beta  &\in \{ 2.1, 2.2, 2.3, 2.7, 3.0, 3.3, 3.7, 4.0, 4.3, 4.7, 5.0, 6.0 \}.
\end{align*}

\begin{figure}[htbp]
\begin{center}
\begin{tabular}{cc}
\includegraphics[width=0.48\linewidth]{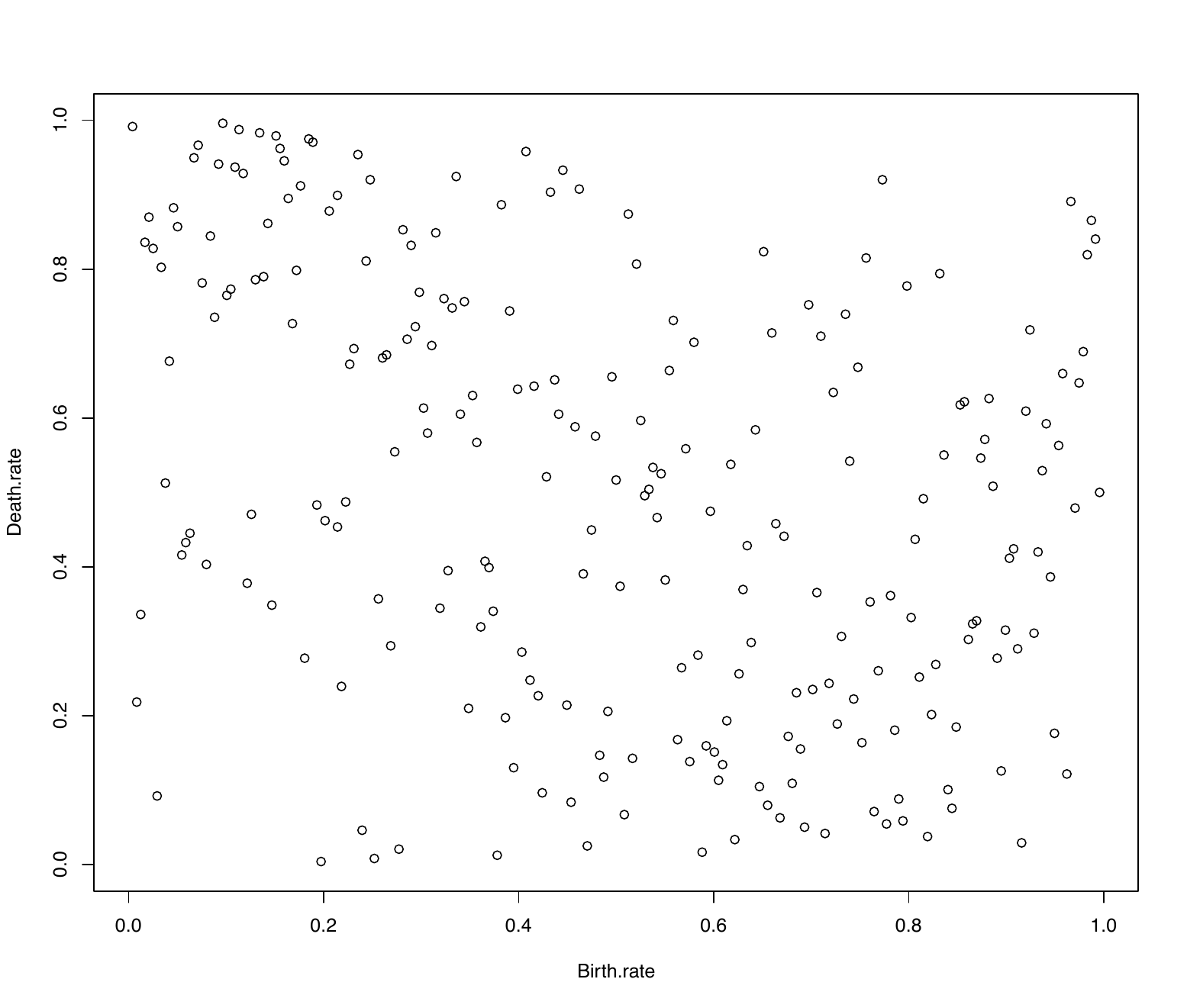} &
  \includegraphics[width=0.48\linewidth]{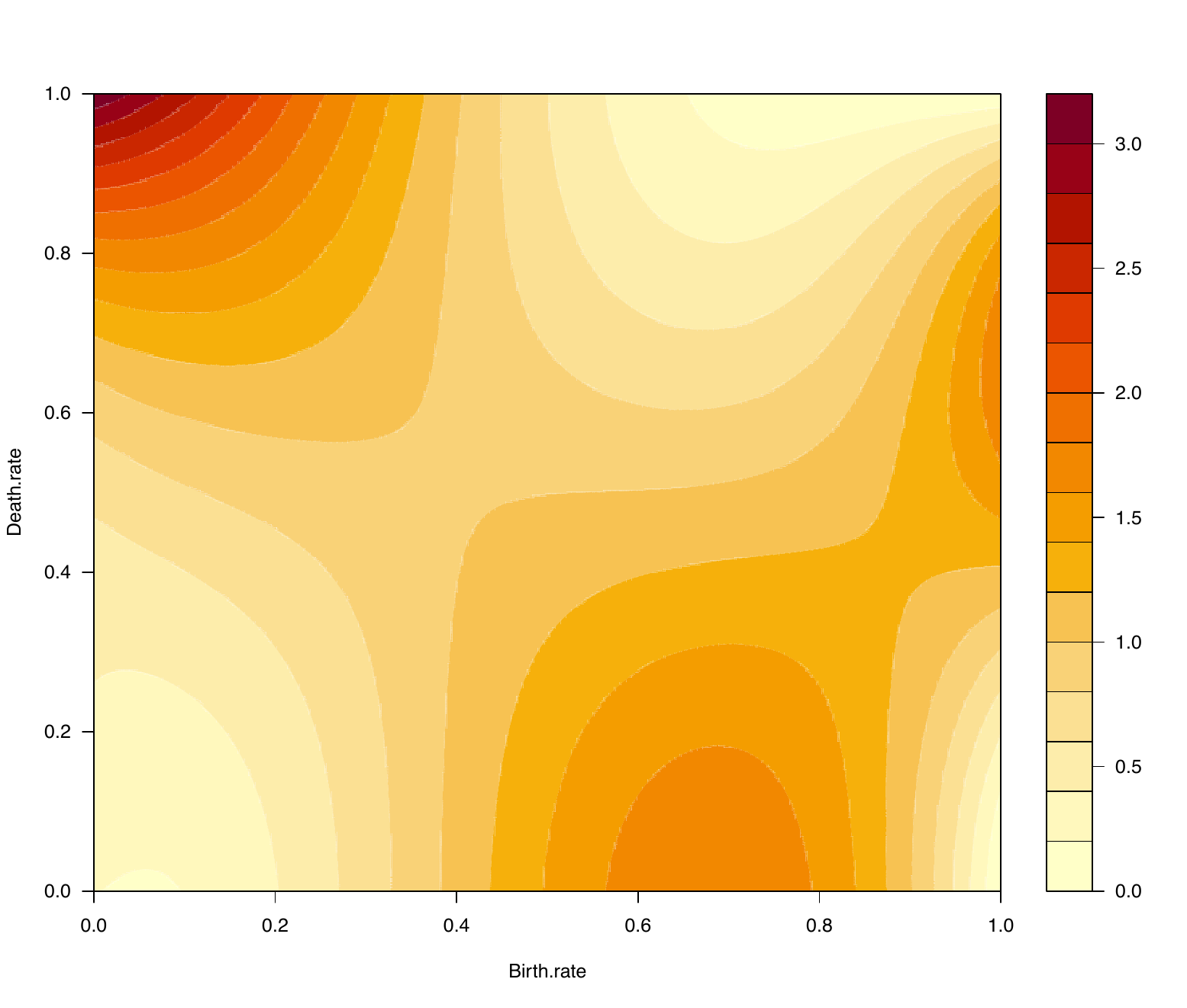} 
\end{tabular}
\end{center}
\caption{Scatterplot of the values of $(\widehat{u}_t,\widehat{v}_t)$ (left) and a filled contour plot of the estimated copula density (right).}
\label{fig:cuv}
\end{figure}

\begin{figure}[htbp]
\begin{center}
\begin{tabular}{cc}
\includegraphics[width=0.48\linewidth]{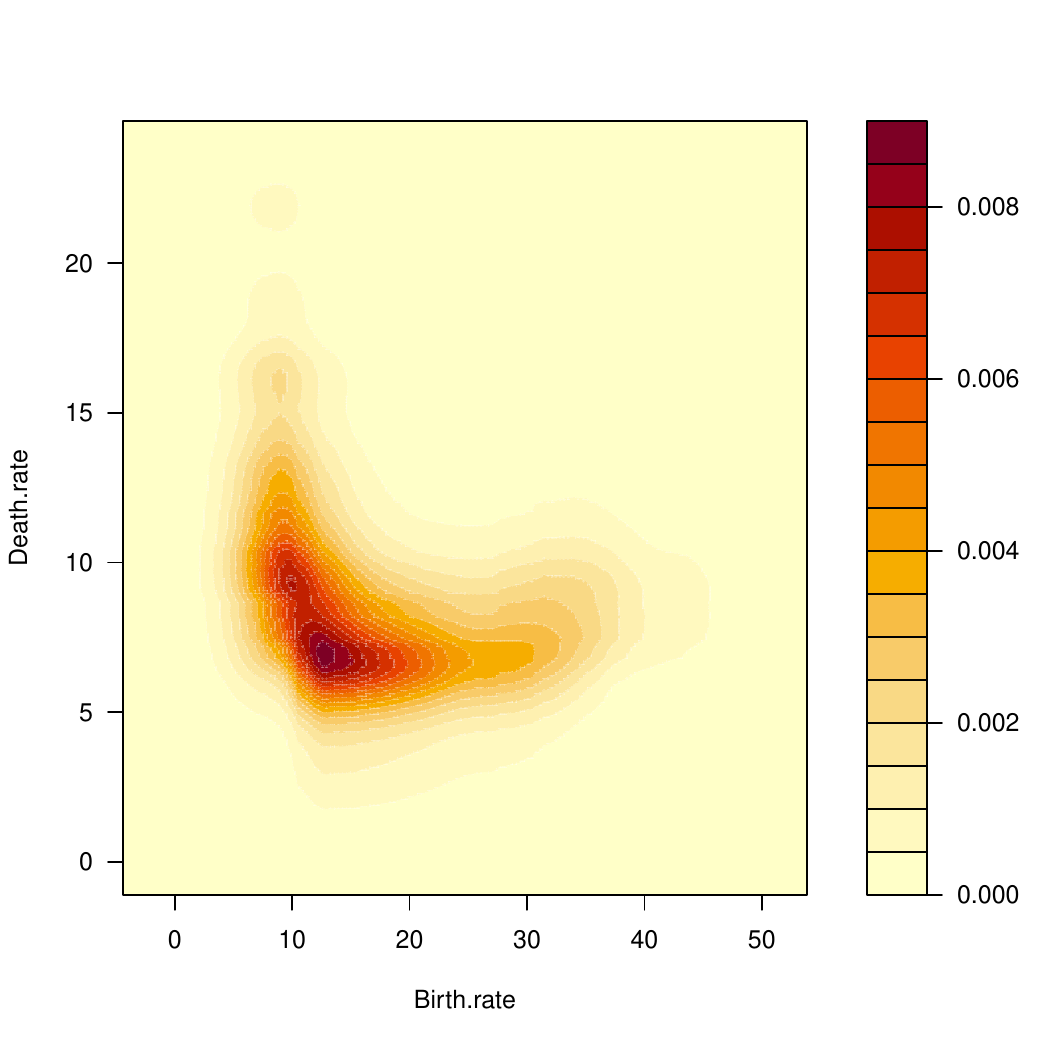} &
  \includegraphics[width=0.48\linewidth]{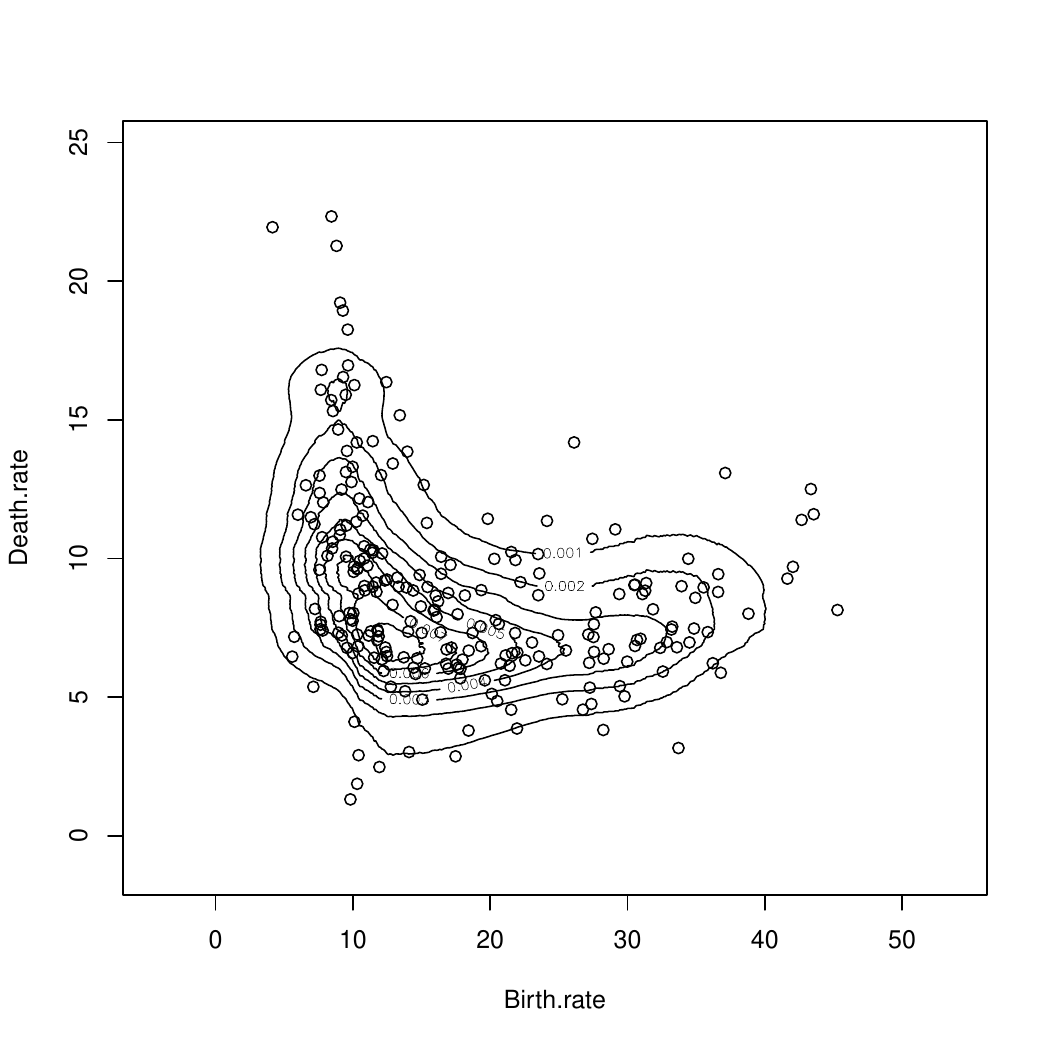} 
\end{tabular}
\end{center}
\caption{Joint density function of the data.}
\label{fig:h_xy}
\end{figure}

The EM algorithm for the penalized pseudo-likelihood function provides for $R$ the estimate 
\[
\widehat{R}=
\begin{pmatrix}
0.007 & 0.019 & 0     & 0.099 \\
0     & 0     & 0.099 & 0.150 \\
0.110 & 0.111 & 0.029 & 0 \\
0.133 & 0.117 & 0     & 0 \\
0     & 0.003 & 0.122 & 0 
\end{pmatrix}.
\]
We present in Figure \ref{fig:cuv} a scatterplot of the values of $(\widehat{u}_t,\widehat{v}_t)$, $t=1,\ldots, 237$, calculated by (\ref{eq_ut_vt}), and a filled contour plot of the estimated copula density.  Also, the estimated joint density of the data is shown in Figure \ref{fig:h_xy}.

\section{Discussion}
\label{sec_discussion}

In this article, we have proposed the EM algorithm for penalized pseudo-likelihood objective functions to estimate the parameter matrix of the B-spline copula.
To determine the size of the parameter matrix and the tuning parameters, we provided a new model of $CV$. By means of simulations, we see that the 
penalized method achieves good performance when the parameter matrix is sparse.  
We have observed that the non-penalized EM algorithm is superior when there are no zeros in the parameter matrix, and in that case, by setting the tuning parameter $\alpha$ equal to $0$, we can change the EM algorithm to being non-penalized.  We also showed that the cross-validation method can choose appropriate tuning parameters for the penalty function and further, that the cross-validation and pseudo-AIC approaches are useful in model selection. 

Additionally, by comparing the performance of the B-spline and Bernstein copulas, we see that the B-spline copula can provide more accurate estimates of the joint density function; moreover,  it still can provide good results with fewer parameters.  
Therefore we conclude that the B-spline copula can be considered a more efficient tool for modelling multivariate distributions.

\bmhead{Acknowledgments}

The authors are grateful to the reviewers and the editors for their careful review and constructive comments which significantly helped improve this article.  
The authors also thank Benjamin Poignard of Osaka University for valuable discussion and advice on this work during the Workshop on Copula Theory at the Institute of Statistical Mathematics, 2022.
This work was supported by Waseda University Grants for Special Research Projects (2022R-048).

\begin{appendices}

\section{Proofs of Propositions \ref{prop_Rinf} and \ref{prop_monot}}
\label{sec_appendix}

We first provide the following notation.  Let $\bm{w}$ denote the complete data; that is, $\bm{w}$ consists of the pseudo-observations 
 $(\widehat{\bm{u}}, \widehat{\bm{v}})=(\widehat{u}_t, \widehat{v}_t)_{t=1,\ldots,N}$ (referred to as the incomplete data), and the label $\tau_t\in\{1,\ldots,m\}\times\{1,\ldots,n\}$, $t=1,\ldots,N$, indicating the B-spline basis to which the $t$-th individual belongs (referred to as the missing data).  

The pseudo-likelihood function for the given incomplete data is 
\[
g(\widehat{\bm{u}}, \widehat{\bm{v}}; \!R)= \left\{ \prod^N_{t=1} \bigg[\sum^m_{k=1} \sum^n_{\ell=1} r_{k,\ell} \phi_{k,m}(\widehat{u}_t) \psi_{\ell,n}(\widehat{v}_t) \bigg] \right\}^{1/N},
\]
and the penalty function is 
$$
\tilde{p}(R) = \sum_{k=1}^m \sum_{\ell=1}^n p(r_{k,\ell}; \alpha, \beta).
$$
Then the penalized pseudo-likelihood is 
\begin{align}
\tilde{g}(\widehat{\bm{u}}, \widehat{\bm{v}};\!R) & = g(\widehat{\bm{u}}, \widehat{\bm{v}}; R) \, \exp(-\tilde{p}(R)) \nonumber \\
& =\bigg[ \prod^N_{t=1} \sum^m_{k=1} \sum^n_{\ell=1} r_{k,\ell} \phi_{k,m}(\widehat{u}_t) \psi_{\ell,n}(\widehat{v}_t)  \bigg]^{1/N} \exp\bigg[- \sum_{k,\ell} p(r_{k,\ell}; \alpha, \beta) \bigg]. \nonumber
\end{align}
Denote by $g_c(\bm{w};\!R)$ the likelihood function for the full data set; then the likelihood of the complete data is 
\[
\tilde{g}_c(\bm{w}; \!R)= g_c(\bm{w}; R)\, \exp(-\tilde{p}(R)).
\]
The conditional likelihood of $\bm{w}$ given $(\widehat{\bm{u}}, \widehat{\bm{v}})$ is obtained as
\[
k(\bm{w} \vert \widehat{\bm{u}}, \widehat{\bm{v}};\!R)= \frac{\tilde{g}_c(\bm{w}; R)}{\tilde{g}(\widehat{\bm{u}}, \widehat{\bm{v}}; R) } = \frac{g_c(\bm{w}; R)}{g(\widehat{\bm{u}}, \widehat{\bm{v}}; R)}.
\]
The penalized pseudo-log-likelihood of the incomplete data $L_p(R)$ in (\ref{pll}) can be rewritten as 
\begin{align}
L_p(R) &= \log g(\widehat{\bm{u}}, \widehat{\bm{v}};R)- \tilde{p}(R)  \nonumber \\
&= \log \tilde{g}(\widehat{\bm{u}}, \widehat{\bm{v}}; R)  \nonumber \\
&= \log \tilde{g}_c(\bm{w};R)-\log k(\bm{w} \vert \widehat{\bm{u}}, \widehat{\bm{v}}; R).
\label{lpr}
\end{align}
Let $R^{(s)}$ be the parameter matrix used in the $s$-th iteration of the EM algorithm, $\tilde{Q}(R; R^{(s)})$ be the expectation of the penalized pseudo-log-likelihood of the complete data for $R^{(s)}$, and $H(R; R^{(s)})$ be the expectation of conditional log-likelihood of the complete data given $(\widehat{\bm{u}}, \widehat{\bm{v}})$ and $R^{(s)}$.
Taking expectations in \eqref{lpr}, we obtain
\begin{align*}
L_p(R) & = E_{R^{(s)}}\left[ \log \tilde{g}_c(\bm{w}; R) \vert (\widehat{\bm{u}}, \widehat{\bm{v}}) \right] - E_{R^{(s)}} \left[ \log k (\bm{w} \vert (\widehat{\bm{u}}, \widehat{\bm{v}}); R) \vert (\widehat{\bm{u}}, \widehat{\bm{v}}) \right] \nonumber \\
& = \tilde{Q}(R; R^{(s)})- H(R; R^{(s)}).
\end{align*}

\begin{proof}[Proof of Proposition \ref{prop_Rinf}]
Let $R^{(\infty)}$ be the maximizer of $\tilde{Q}(R; R^{(s)})$. Then 
\begin{align*}
0 & = \frac{\partial}{\partial R}\tilde{Q}(R; R^{(\infty)})\bigg\vert_{R= R^{(\infty)}} \\
&= E_{R} \left[ \frac{\partial}{\partial R} \log \tilde{g}_c(\bm{w}; R) \vert (\widehat{\bm{u}}, \widehat{\bm{v}}) \right]\bigg\vert_{R= R^{(\infty)}}\\
& = E_{R} \left[ \frac{\partial}{\partial R} \log \left\{ \tilde{g}(\widehat{\bm{u}}, \widehat{\bm{v}}; R) + \log k(\bm{w} \vert \widehat{\bm{u}}, \widehat{\bm{v}}; R)\right\} \vert (\widehat{\bm{u}}, \widehat{\bm{v}}) \right] \bigg\vert_{R= R^{(\infty)}}\\
& = \frac{\partial}{\partial R} \log \tilde{g}(\widehat{\bm{u}}, \widehat{\bm{v}}; R) \bigg\vert_{R= R^{(\infty)}} + E_R  \left[ \frac{\partial}{\partial R} \log k(\bm{w} \vert \widehat{\bm{u}}, \widehat{\bm{v}}; R) \vert (\widehat{\bm{u}}, \widehat{\bm{v}})  \right] \bigg\vert_{R= R^{(\infty)}}.
\end{align*}
The second expression in the above equation is 
\begin{align*}
E_R \bigg[ \frac{\partial}{\partial R} \log & k(\bm{w} \vert \widehat{\bm{u}}, \widehat{\bm{v}}; R) \vert (\widehat{\bm{u}}, \widehat{\bm{v}}) \bigg] \bigg\vert_{R= R^{(\infty)}}\\
= & \int_{\mathcal{W}(\widehat{\bm{u}}, \widehat{\bm{v}})} \frac{\partial}{\partial R} \log k(\bm{w} \vert \widehat{\bm{u}}, \widehat{\bm{v}}; R) \cdot  k(\bm{w} \vert \widehat{\bm{u}}, \widehat{\bm{v}}; R) \, dw \bigg\vert_{R= R^{(\infty)}}\\
= & \frac{\partial}{\partial R} \int_{\mathcal{W}(\widehat{\bm{u}}, \widehat{\bm{v}})} k(\bm{w} \vert \widehat{\bm{u}}, \widehat{\bm{v}}; R) \, dw  \bigg\vert_{R= R^{(\infty)}} = \frac{\partial}{\partial R} 1 = 0.
\end{align*}
Hence, 
\[
 \frac{\partial}{\partial R}L_p(R) \bigg\vert_{R= R^{(\infty)}} = \frac{\partial}{\partial R} \log \tilde{g}(\widehat{\bm{u}}, \widehat{\bm{v}}; R)\bigg\vert_{R= R^{(\infty)}}=0.
\]
The proof now is complete.  
\end{proof}

\begin{proof}[Proof of Proposition \ref{prop_monot}]
For any $s$-th and $(s+1)$-th iterations of the EM algorithm, we consider the difference of their penalized pseudo-log-likelihoods 
\begin{align*}
L_p&(R^{(s+1)}) - L_p(R^{(s)}) \\
&= \big[ \tilde{Q}(R^{(s+1)}; R^{(s)}) - \tilde{Q}(R^{(s)}; R^{(s)}) \big] - \big[ H(R^{(s+1)}; R^{(s)})- H(R^{(s)}; R^{(s)}) \big].
\end{align*}
Note that the first term on the right hand side satisfies
\[
\tilde{Q}(R^{(s+1)}; R^{(s)})- \tilde{Q}(R^{(s)}; R^{(s)}) \ge 0,
\]
because the EM algorithm is designed to achieve a larger log-likelihood than the previous iteration.  Also, the second term equals
\begin{align*} 
H(&R^{(s+1)};R^{(s)})- H(R^{(s)}; R^{(s)}) \\
 &= E_{R^{(s)}}\left[ \log k(\bm{w} \vert \widehat{\bm{u}}, \widehat{\bm{v}}; R^{(s+1)}) \vert (\widehat{\bm{u}}, \widehat{\bm{v}}) \right] -E_{R^{(s)}}\left[ \log k(\bm{w} \vert \widehat{\bm{u}}, \widehat{\bm{v}}; R^{(s)}) \vert (\widehat{\bm{u}}, \widehat{\bm{v}}) \right]\\
 &= E_{R^{(s)}}\left[ \log \frac{k(\bm{w} \vert \widehat{\bm{u}}, \widehat{\bm{v}}; R^{(s+1)})}{k(\bm{w} \vert \widehat{\bm{u}}, \widehat{\bm{v}}; R^{(s)})}\bigg\vert (\widehat{\bm{u}}, \widehat{\bm{v}}) \right].
\end{align*}
Since the logarithm function is concave then, by applying Jensen's inequality, we obtain 
\begin{align*} 
H(R^{(s+1)}; R^{(s)}) &- H(R^{(s)}; R^{(s)}) \\
&\le \log E_{R^{(s)}} \left[ \frac{k(\bm{w} \vert \widehat{\bm{u}}, \widehat{\bm{v}}; R^{(s+1)})}{k(\bm{w} \vert \widehat{\bm{u}}, \widehat{\bm{v}}; R^{(s)})}\big\vert (\widehat{\bm{u}}, \widehat{\bm{v}}) \right] \\
 =&  \log \int_{\mathcal{W}( \widehat{\bm{u}}, \widehat{\bm{v}})} \frac{k(\bm{w} \vert \widehat{\bm{u}}, \widehat{\bm{v}}; R^{(s+1)})}{k(\bm{w} \vert \widehat{\bm{u}}, \widehat{\bm{v}}; R^{(s)})} k(\bm{w} \vert \widehat{\bm{u}}, \widehat{\bm{v}}; R^{(s)}) \, dw\\
 = & \log \int_{\mathcal{W}( \widehat{\bm{u}}, \widehat{\bm{v}})} k(\bm{w} \vert \widehat{\bm{u}}, \widehat{\bm{v}}; R^{(s+1)}) \, dw= \log 1=0.
\end{align*}
Therefore, we obtain 
\[
L_p(R^{(s+1)}) \ge L_p(R^{(s)}),
\]
and since $s$ was chosen arbitrarily then we have proved that the algorithm always results in monotonically increasing values of $L_p(R^{(s)})$.  
\end{proof}

\clearpage

\section{Rejection sampling for generating random data from the B-spline copula}
\label{sec_sampling}

\begin{algorithm}
\caption{Generate random data from the B-spline copula}
\label{alg:rv}
\begin{algorithmic}[1]
\State 
Set the degree of the B-spline basis functions to equal $d$, the size of $R$ to equal $m\times n$, and choose equally-spaced interior knots.  This determines the basis functions as well as the $q_{k,m}$ and $q^*_{\ell,n}$ for each argument $(u,v)$ of the copula $c(u,v;\!R)$, $u \in [0,1]$, $v \in [0,1]$. 
\State 
Choose a parameter matrix $R$ that satisfies the conditions in (\ref{R}).
\State 
Calculate $c_{\max} := \max_{u,v} c(u,v;\!R)$, the maximum value of $c(u,v;\!R)$ on $[0,1]\times[0,1]$, and then the scaling parameter is $K = 1/c_{\max}$.
\State 
Generate independent observations, $U$ and $V$, both from the Unif$(0,1)$ distribution and calculate $c(U,V;\!R)$.
\State 
Compute $K \, c(U,V;\!R)$, the rate of choosing such random pairs $(U,V)$.
\State 
Generate a new random number $S \sim$ Unif$(0,1)$.
\State 
If $S \le K \, c(U,V;\!R)$, then accept $(U,V)$ as a pair of random numbers of $c(u,v;\!R)$; otherwise reject $(U,V)$ and repeat Steps 4--7.
\end{algorithmic}
\end{algorithm}

\section{Small-sample behavior of the pseudo-MLE}
\label{sec_small_pseudo_mle}

Tsukahara (2005, Theorem 1) established the consistency of the pseudo-MLE, thereby settling its large-sample behavior.  In response to a comment from a reviewer, we investigated the small-sample behavior of the pseudo-MLE in our context by carrying out simulations at sample sizes far smaller than those values, $N=1000$, that were used in Subsection \ref{subsec_1}.  

\begin{figure}[htbp]
\begin{center}
\begin{tabular}{cc}
    \includegraphics[width=0.5\linewidth]{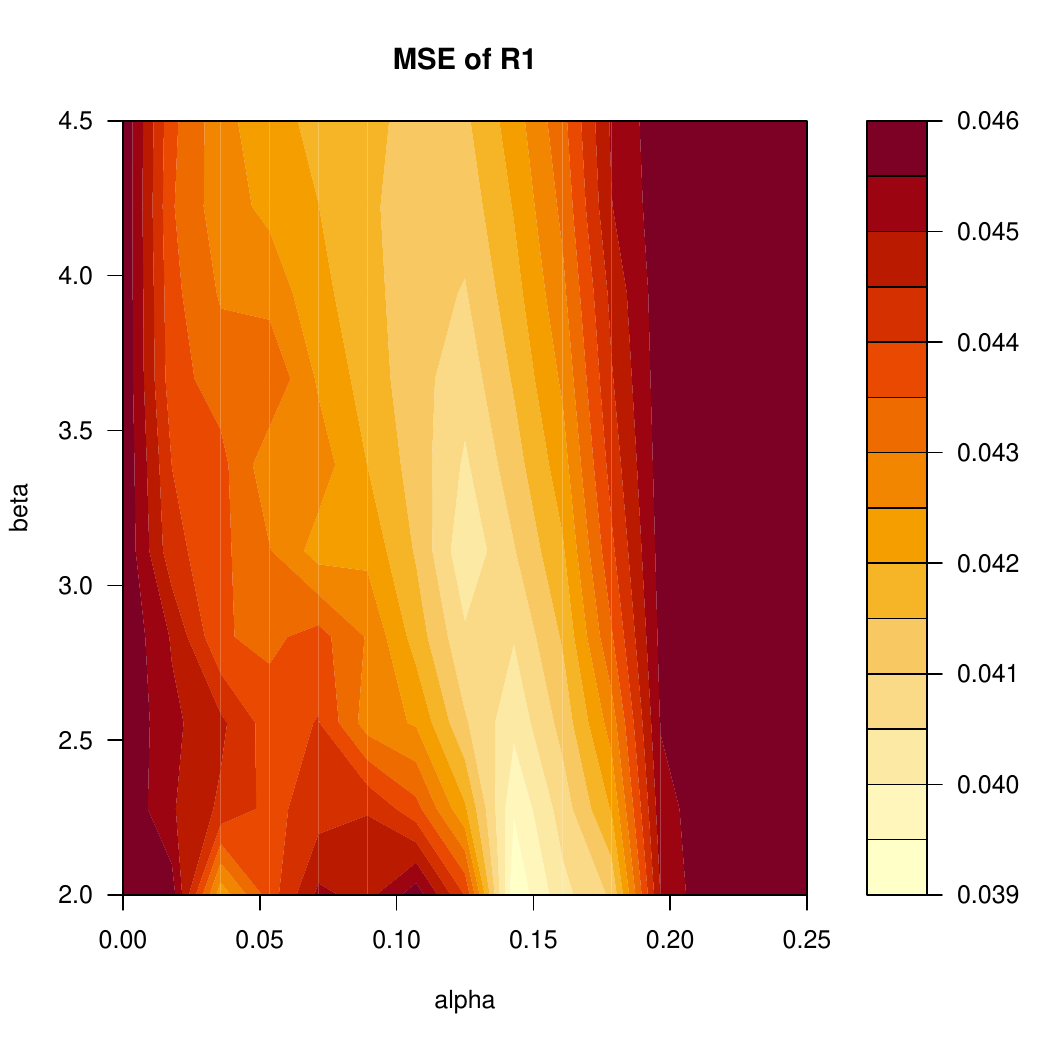} &
        \includegraphics[width=0.5\linewidth]{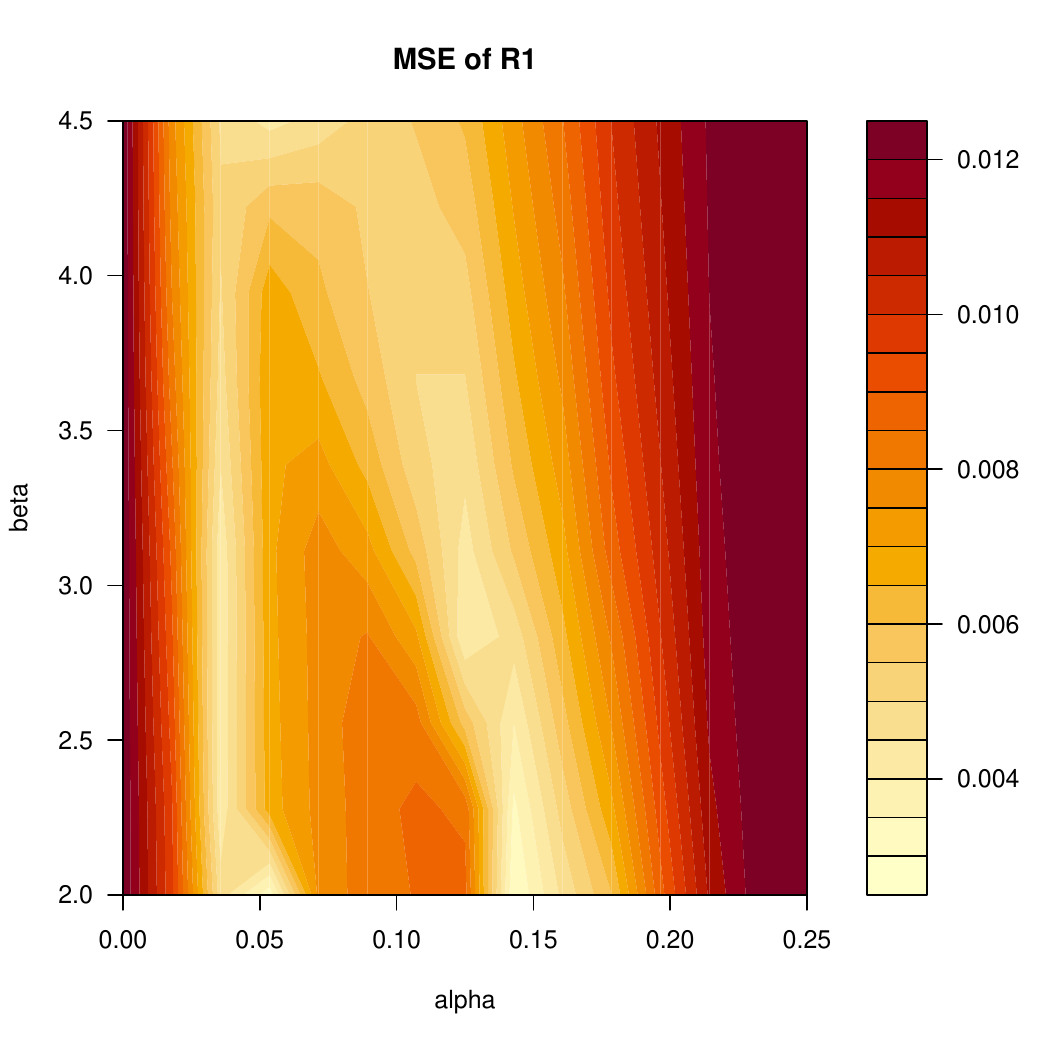} \\
            \includegraphics[width=0.5\linewidth]{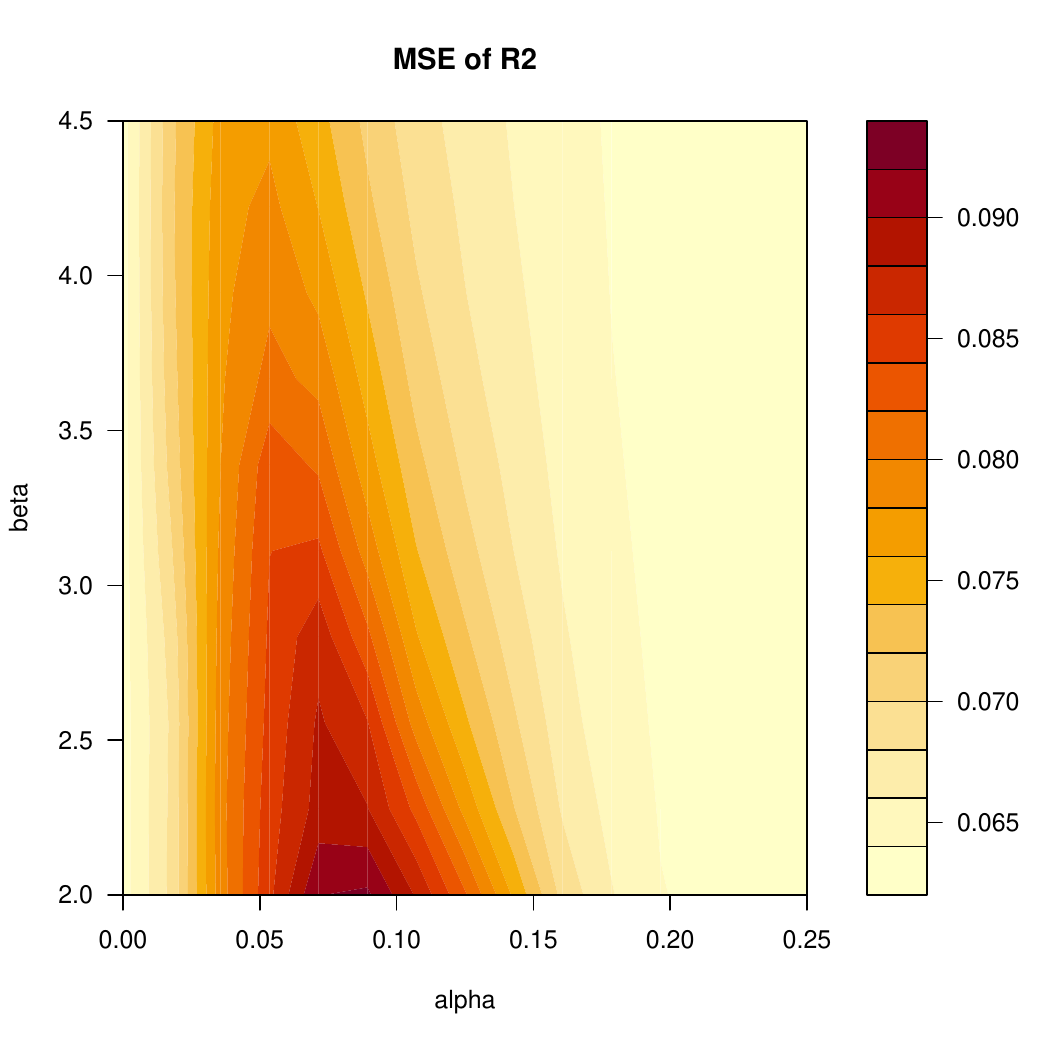} &
        \includegraphics[width=0.5\linewidth]{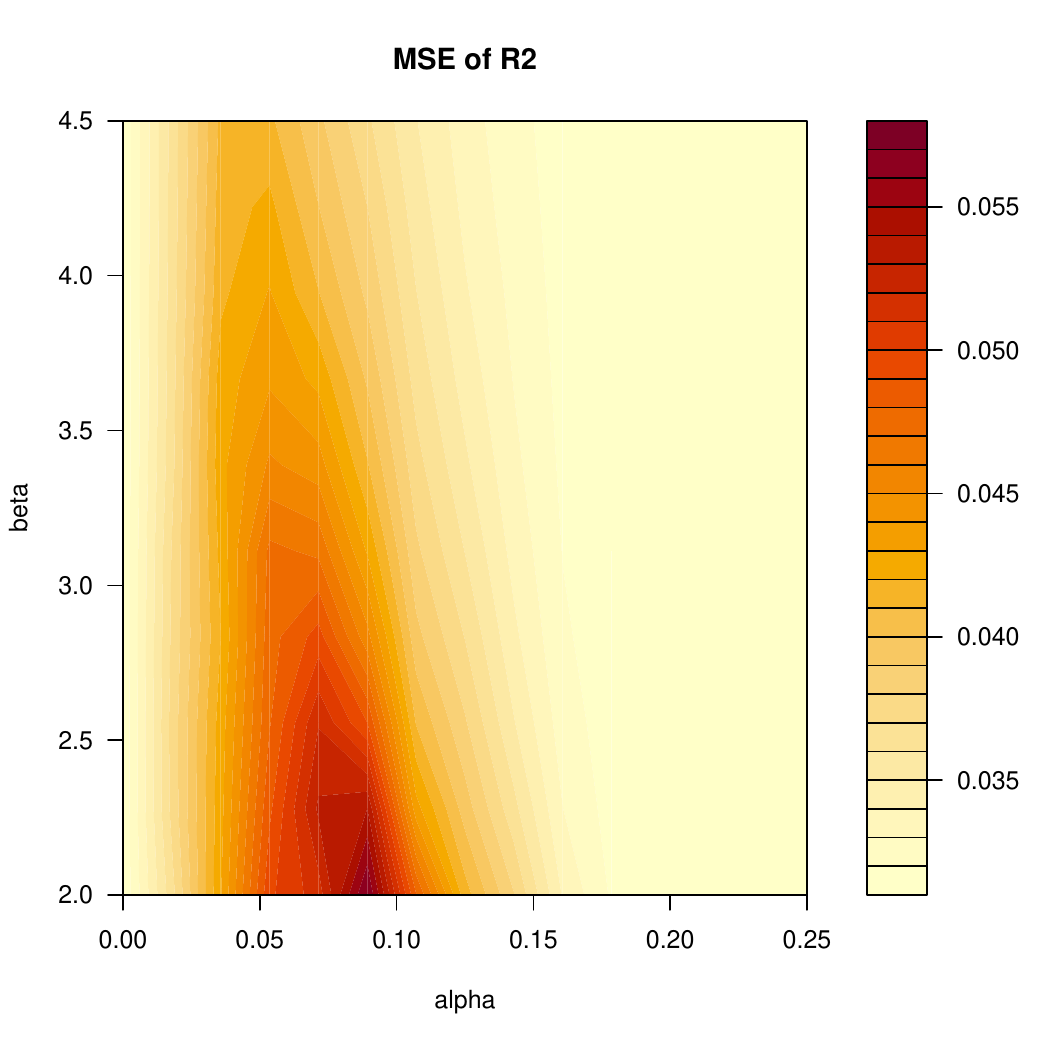} \\
            \includegraphics[width=0.5\linewidth]{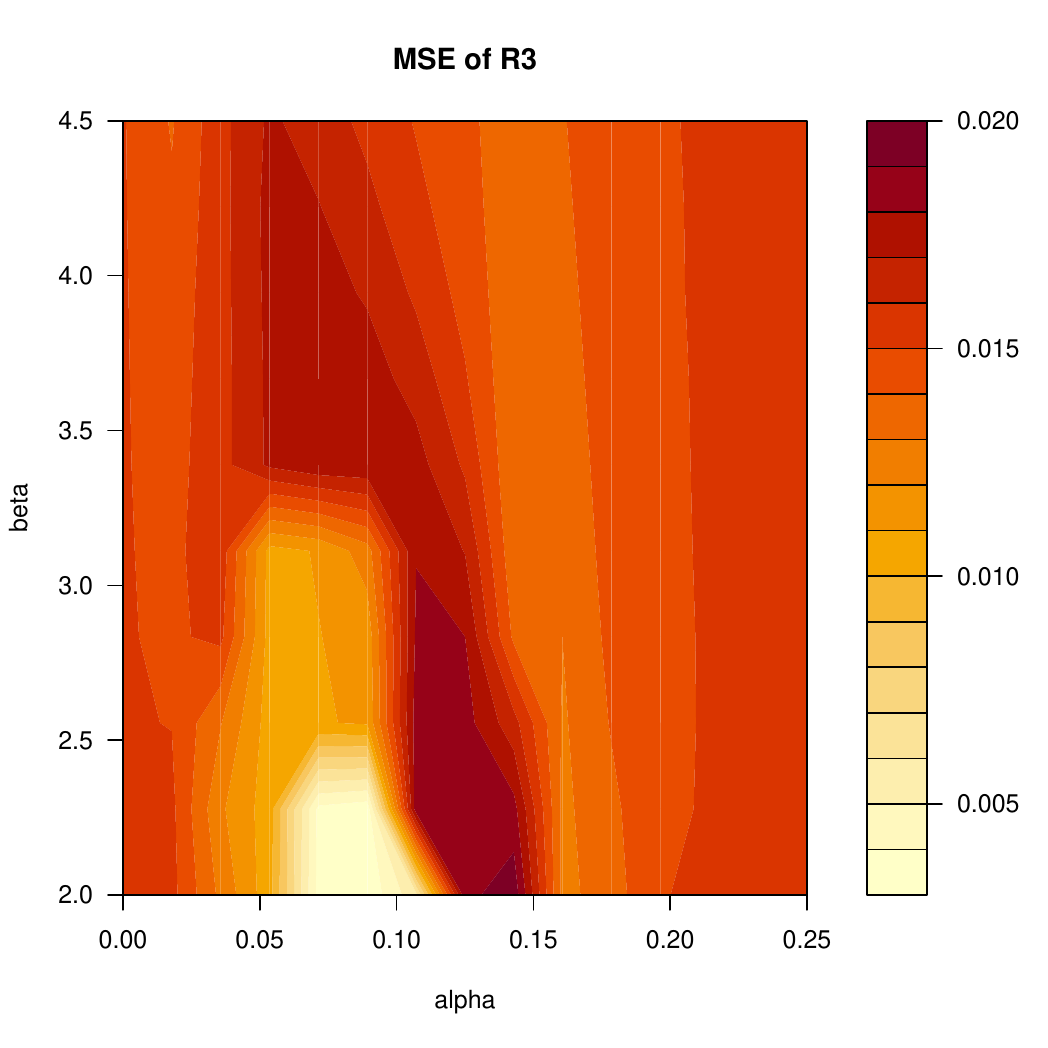} &
       \includegraphics[width=0.5\linewidth]{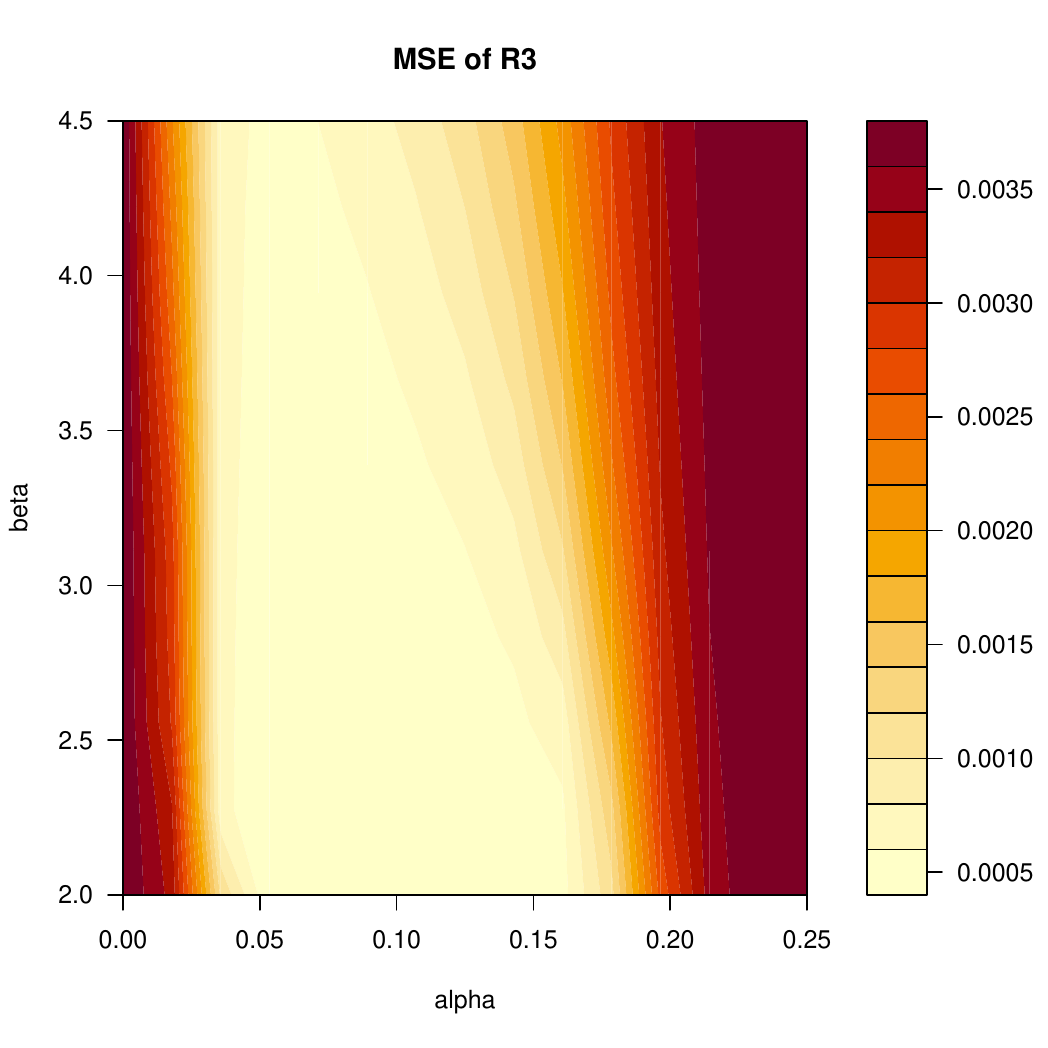} 
\end{tabular}
\end{center}
\caption{Horizontal axis is $\alpha \in [0,0.25]$; vertical axis is $\beta \in [2, 4.5]$. Contour plots of MSEs in (4.1) of $\widehat{R}_1$,$\widehat{R}_2$ , and $\widehat{R}_3$ for $(\alpha, \beta)$ obtained from $J=30$ data sets of sample sizes $N=100$ (left) and $N=300$ (right).}
\label{fig:mse_smallN}
\end{figure}

Specifically, with $J=30$ data sets of sizes $N=100$ and $N=300$, respectively, we calculated the MSE in \eqref{mse} for $R_1$, $R_2$ and $R_3$.  The contour plots in Figure \ref{fig:mse_smallN} of the MSEs of $\widehat{R}_1$, $\widehat{R}_2$, and $\widehat{R}_3$, for various values of $(\alpha,\beta)$, are shown in the first, second, and third rows, respectively.  

In Figure \ref{fig:mse_smallN}, the graphs in the first column correspond to the sample size $N=100$, and those in the second column correspond to $N=300$.

When Figure \ref{fig:mse_smallN} is compared with Figure 2, we can see that the images in Figure C1 become increasingly similar to those in Figure 2.  Also the computed values of the MSE decrease in size when the sample size increases.  This validates that the proposed method works properly, even for smaller sample sizes, and it demonstrates that pseudo-MLE is able to attain near-efficiency at relatively smaller sample sizes.

\end{appendices}

\clearpage

\bibliographystyle{sn-apacite}
\bibliography{sn-bibliographyEMBC}

\end{document}